\documentclass[fonts]{asd}

\usepackage{moreverb}

\usepackage[breaklinks,colorlinks,bookmarksopen,bookmarksnumbered,linkcolor=ICSTblue,citecolor=blue,urlcolor=ICSTblue]{hyperref}
\usepackage{breakurl}
\usepackage{doi}

\usepackage{float}
\usepackage{stfloats}
\usepackage[caption=false,labelfont=sf,textfont=footnotesize]{subfig}
\usepackage{picins}
\usepackage{array}
\usepackage{array}
\usepackage{hyperref}

\newcommand\BibTeX{{\rmfamily B\kern-.05em \textsc{i\kern-.025em b}\kern-.08em
T\kern-.1667em\lower.7ex\hbox{E}\kern-.125emX}}

\def\volumeyear{2014}

\journalname{Internet of Things}
\articletype{Research Article}
 \def\volumeyear{2018}
 \def\volumenumber{04}
  \def\issuemonth{10}
  \def\issuenumber{16}
  \def\articleid{XX}
  \def\copyrightholder{Jobish John\textit{ et al.}, licensed to ICST}
  \copyrightnote{This is an open access article distributed under the terms of the Creative Commons Attribution license (\url{http://creativecommons.org/licenses/by/3.0/}), which permits unlimited use, distribution and reproduction in any medium so long as the original work is properly cited.}
  \def\articledoi{10.4108/eai.13-7-2018.158420}
\received{19 July 2018}
  \accepted{21 September 2018}
  \published{31 October 2018}

\begin{document}

\runningheads{J. John et al.}{Design and Implementation of a Wireless Sensor Network for Agricultural Applications}

\title {Design and Implementation of a Wireless Sensor Network for Agricultural Applications}

\author{Jobish John\affil{1}\fnoteref{1}, Gaurav S. Kasbekar\affil{1}, Dinesh K. Sharma\affil{1}, V. Ramulu\affil{2}, Maryam Shojaei Baghini\affil{1}}

\address{\affilnum{1}Department of Electrical Engineering, Indian Institute of Technology Bombay, Powai - 400076, Mumbai, India\\
\affilnum{2}Water Technology Center, Professor Jayashankar Telangana State Agricultural University, Rajendranagar - 500030, Hyderabad, India}

\abstract{

We present the design and implementation of a shortest path tree based, energy efficient data collection wireless sensor 
network to sense various parameters in an agricultural farm using in-house developed low cost sensors. Nodes follow a 
synchronized, periodic sleep-wake up schedule to maximize the lifetime of the network. The implemented network consists 
of 24 sensor nodes in a 3 acre maize farm and its performance is captured by 7 snooper nodes for different data collection 
intervals: 10 minutes, 1 hour and 3 hours.  The almost static nature of wireless links in the farm motivated us to use the 
same tree for a long data collection period (3 days). The imbalance in energy consumption across nodes is observed to be 
very small and the network architecture uses easy-to-implement protocols to perform different network activities including 
handling of node failures. We present the results and analysis of extensive tests conducted on our implementation, 
which provide significant insights.
}


\tnotetext[1]{Part of this paper was presented at NCC'15~\cite{john2015design}.}

\fnotetext[1]{Corresponding author.  Email: \email{jobish.john@ee.iitb.ac.in}}

\maketitle

\section{Introduction}

\label{intro}

Wireless sensor network (WSN) is one of the emerging technologies, which finds application in a variety of fields such as environmental
and health monitoring, battle field surveillance, and industry process control \cite{chong2003sensor}.
Sensor networks consist of sensor nodes, which are usually deployed in an ad-hoc manner and they
self-organize and coordinate among themselves to perform a sensing task. The design of a WSN mainly focuses on extending the lifetime of
the system since the sensor nodes work on battery. In contrast, energy constraints are secondary criteria to the traditional wireless networks 
like cellular networks \cite{akyildiz2002survey}. The architecture of WSN
should be chosen in such a way that the network will be efficient in terms of energy consumption and should yield maximum lifetime 
for the network, while maintaining the required level of reliability for the data packets \cite{zheng2009wireless}. 

Sensor networks can play an integral part in providing solutions for different types of applications which can be broadly categorized as tracking, event detection, 
and periodic monitoring applications. The wireless network architecture and  its design criteria heavily depends on the respective application.

We focus on monitoring applications, where a particular environment is monitored periodically. There are several deployment 
studies which are reported in the literature for monitoring applications like structural health monitoring \cite{pakzad2008design},
habitat monitoring \cite {mainwaring2002wireless},
environment monitoring \cite {lombardo2017wireless} \cite {lazarescu2013design} \cite{yang2010design}, volcano monitoring \cite {werner2006fidelity},
forest surveillance \cite{greenorbs} etc. 

In India, agriculture is one of the sectors which uses a lot of water resources. 
India has 4\%  of the world's fresh water resource, out of which 80\% is used in agriculture \cite{VibhaDhawan}.
Moreover, India is the largest user of ground water for irrigation \cite{fishman2015can}.
Proper irrigation and water management can lead to an improvement in crop productivity as well as saving of water \cite{mashnik2017increasing}. With this motivation,
we have designed and developed an affordable agri sensor network system which can help farmers to irrigate their fields properly by
measuring multiple soil parameters like soil moisture, soil temperature, atmospheric temperature and relative humidity.

Soil moisture, soil temperature, atmospheric temperature and relative humidity are the major parameters which play a crucial role in the field of 
precision agriculture. Monitoring of these parameters is essential to enhance the crop productivity through irrigation management 
and by applying fertilizers in proper time intervals.
High soil temperature destroys the crops and low temperature prevents the roots from absorbing water from
the field. Similarly, it is important to measure the soil moisture at regular intervals because low 
moisture adversely affects the crops. Relative humidity (RH) is another parameter which indirectly 
affects photosynthesis and the plant growth. High RH reduces carbon dioxide uptake in the plants \cite{palaparthy2013review}. 
Atmospheric temperature is another important parameter which affects plant development and productivity, \emph{e.g.}, pollination is a 
temperature sensitive phenomenon. Hence, monitoring temperature becomes beneficial for the deployment of some agricultural 
strategies \cite {hatfield2015temperature}.

In \cite{ojha2015wireless}, the major practices followed in the agricultural domain with the help of sensor networks are outlined.
Many researchers have reported sensor network implementations for agricultural applications.
The performances of random and grid topologies of sensor networks for precision agriculture are compared in \cite{keshtgary2012efficient}
through simulations in the OPNET simulator. They measure the performance in terms of parameters such as delay and throughput.
An irrigation management system is described in \cite{mafuta2013successful} that uses a star network of Waspmote nodes from Libelium.
A ZigBee based star WSN deployment for irrigation is presented in \cite{zhou2009wireless} and \cite{gutierrez2014automated},
while \cite {kim2008remote} describes a Bluetooth based sensor network with a star architecture for irrigation control.
Star architecture is one of the very simple network architectures in which there is a direct wireless connection between one of nodes
(called the center) and every other node; however, multi-hop communication architectures are required for covering large farms.
In this paper, we present the design and implementation of a WSN with a multi-hop architecture, which can be used to cover
large farms.

As \cite{raman2008censor} points out, there exists a gap between protocol or network architecture designs proposed in the theoretical research literature and those actually used in practical sensor network implementations.  
In this paper, we seek to close this gap, by both designing protocols and a network architecture, and practically implementing a sensor network which can measure various parameters like
soil moisture, soil temperature, atmospheric temperature and relative humidity at various locations in the field, which can help farmers
to optimally irrigate their crops. We propose an energy efficient synchronized tree based data collection approach to collect data from
various sensor nodes. Depending on the sensed information, proper irrigation management actions can be accomplished. Our main focus is on the 
networking aspects of the data collection part and hence, we do not address the irrigation control part. We present the results of extensive tests conducted on our implementation and their analysis in this paper, which provide insight.  The major contributions of this 
paper are as follows:

\begin{enumerate}
 \item {We present a synchronized tree based network architecture which can be used for data collection in deployment environments
 which do not have too much link variation, as is typically the case for most agricultural farms.}
 \item {The effect of link dynamics in the agricultural environment is studied and is taken into account while deciding the network architecture.}
 \item {The nodes in the network are kept time synchronized with each other for data collection by energy efficient approaches.
 In particular, all the nodes follow a periodic sleep-wake-up schedule, in which they alternate between \emph{sleep state}, in which
 battery energy is conserved, and \emph{wake-up state}; also, the sleep and wake-up intervals of all the nodes are synchronized.}
 \item {Low complexity methods to handle unexpected node failures are included in the design.}
 \item {The transmissions in the network  are well characterized by analyzing the various activities in the network using powerful snooper nodes.}
\item {A simple approach to find the remaining capacity from the battery voltage is introduced.}
 \item {The energy expenditure profiles of nodes of different traffic loads are well captured and 
	our results show several trends that are often overlooked in energy optimization design techniques for
	outdoor low duty cycled data collection applications.}
\end{enumerate}

The rest of this paper is organized as follows. Section~\ref{RelatedWork} gives a review of sensor network deployment studies focused on 
wireless network analysis.
Section \ref{problem_definition} describes our problem statement, and Section \ref{Network_architecture} explains the network architecture.
The implemented network and experimental methodology are described in Section \ref{experimental_methodology}.
Section \ref{motivations_architecture} describes the motivations which led to the design of the proposed network architecture and 
Section \ref{field_experiments} details the field experiments that we conducted and provides an analysis of the results. We conclude the paper in 
Section \ref{conclusion}.

\section{Related Work}
\label {RelatedWork}

WSN is a topic that has been well researched over the past two decades and there are several works in the literature which
mainly concentrate on protocol designs for a particular layer in the layered network design approach. 
For example, \cite{zheng2009wireless} classifies Medium Access Control (MAC) protocols into 
contention-based \cite{van2003adaptive,lu2004adaptive,el2003wisemac}, 
contention-free \cite{rajendran2006energy,sohrabi2000protocols},and hybrid \cite{rhee2008z,ahn2006funneling} protocols.
Most of these protocols are designed to address one or more of the general concerns in WSNs such as achieving energy efficiency,
reducing  idle listening, maintaining good throughput, acceptable latency and fairness etc.

In \cite{zheng2009wireless},  routing protocols for WSNs have been classified into different categories such as location-aided protocols, 
data centric protocols, mobility based protocols, multi-path based protocols, quality of service based protocols, heterogeneity based 
protocols etc.
Also there are several protocols in the literature  which concentrate on a single aspect of WSN design such as neighbour discovery, 
broadcasting techniques or time synchronization techniques where the authors try to improve the network performance with respect to some particular metrics. For example, \cite{sun2014energy} reviews various Neighbour Discovery Protocols (NDP) and most of them are evaluated with respect to their 
duty cycle and discovery latency. Taxonomies of time synchronization protocols are discussed in \cite{TPSN,FTSP}.

Most of the above protocols, which are designed for a particular networking activity, are evaluated theoretically and/ or via simulations.
Also, the above papers focus on a specific networking task such as neighbour discovery, time synchronization etc and do not address the 
design or implementation of a complete sensor network system. In contrast, in this paper, we present the design and implementation of 
a complete WSN for agricultural application.

Next, we present a brief review of some of the sensor network deployment studies reported in the research literature with a focus on the wireless 
networking aspects. GreenOrbs \cite{greenorbs} is a large scale deployment of 330 TelosB nodes in a wild forest to sense the forest canopy
using sensors for temperature, humidity, light etc., where synchronized data collection is  
performed using the collection tree protocol (CTP) \cite {gnawali2009collection}, one of the defacto routing protocols for sensor network deployments.
They observe that a small number of nodes bottleneck the network and there is a reduction in the number of 
data packets reaching the sink node as the hop count increases. 
Network concurrency plays a major role compared to environment dynamics and an event based routing scheme is suggested.
CitySee \cite{liu2013citysee} is another deployment of 1200 nodes to monitor mainly the $CO_2$ levels in the environment and it also employs CTP 
as the routing protocol.

Habitat monitoring \cite {mainwaring2002wireless} is one of the earliest implementations of a sensor network which uses a tiered architecture
to collect data. Sensorscope \cite {barrenetxea2008sensorscope} is a sensor network deployment in Switzerland for environmental monitoring
with sensors for 7 different parameters including soil water content. For routing the data packets, each node chooses one of its good quality
neighbor nodes at random so that this approach will eliminate the cost to maintain a backbone tree structure and reduce the imbalance of 
traffic load among nodes. Using the deployment, experimental results of 16 solar powered sensor stations in an area of 2500 $m^2$ are presented. 
VigilNet \cite{vicaire2009achieving} is another 200 node deployment mainly focusing on power management strategies for military surveillance
application. The implementation of a 100 node sensor network to monitor temperature and humidity in a potato farm is detailed in 
\cite{langendoen2006murphy} along with their field experiences.
It uses MintRoute \cite{woo2003taming} protocol which is a spanning tree based
approach and is configured to collect 1 data packet from each node once every 10 minutes.
Several tree based data collection approaches in sensor networks are discussed in \cite{woo2003taming} along with their performance comparisons
where routing action is performed using link quality estimation and
neighborhood management.

In India, generally most of the farmers hold an agricultural farm of a few acres in size. Hence, in this paper,  we do not consider 
sensor network implementations of 100's or 1000's of nodes. Also, assuming some levels of spatial similarities in the field conditions and 
to provide a cost 
effective solution, the network under consideration is not a dense network. These facts lead to the requirement of a sensor network of a few 10's 
of sensor nodes where a centralized approach can provide a good solution. We focus on the design and implementation of such a sensor network in this paper.

We now compare our deployment with the afore mentioned literature. Most of the deployments are not for agricultural applications, whereas we detail the design and performance evaluation
of a sensor network for outdoor agricultural application. Sensorscope \cite {barrenetxea2008sensorscope} uses battery voltage as a 
measure of available energy while 
no other implementation takes into account energy consumption at a node by node level. In contrast, in this paper, we provide a simple approach
to find out the remaining battery capacity of a sensor node from its battery voltage, and use the battery capacity as a measure of the available energy.  Handling of node failures is not addressed in 
\cite{mainwaring2002wireless} and \cite {barrenetxea2008sensorscope}, whereas in this paper, low complexity methods to handle node failures are included in the design. The data collection protocol, CTP, used in \cite{greenorbs} and \cite{liu2013citysee}
uses adaptive beaconing
to characterize the network topology changes which requires several packet transmissions; such transmissions are an overhead and are not required in our implementation since link qualities do not change rapidly in the environment we consider.  Also, we use an optimal tree for data collection in our implementation; to the best of our knowledge, no implementation in prior work uses an optimal tree. Even though \cite{langendoen2006murphy} details a sensor network implementation in a potato farm, no test results obtained from the implementation are provided. In contrast, we present the results of extensive tests conducted on our implementation and their analysis in this paper, which provide insight. 

\section{Network Model and Problem Definition}
\label{problem_definition}

The network consists of $n$ wireless sensor nodes, placed arbitrarily in any environment which is expected to have low wireless link variation,
for low duty cycled data collection. We do not make any assumptions about the node placement except the fact that 
all the nodes in the network together form a connected graph. 
The aim is to design an energy efficient data collection scheme which takes into account
the remaining energy of nodes (to provide a longer lifetime) and the link dynamics (to provide reliability to data packets).
The scheme can be tuned for any medium scale wireless network (consisting of 10's of nodes) in any environment where energy harvesting 
may or may not be feasible. 

In this work we implement the proposed architecture for an agricultural application where
each node has sensors for measuring soil moisture, soil temperature, atmospheric temperature and  relative humidity. 
Each node periodically senses all the parameters and reports them to a sink node in the network
through a data collection tree. We use TelosB  motes \cite{sky2007ultra} which are programmed using the TinyOS  platform \cite{tinyos}.
It uses the CC2420 radio \cite{cc2420datasheet} for wireless data transfer.

\section{Network Architecture}
\label{Network_architecture}

In sensor networks which are deployed for any environment monitoring applications such as agricultural farm  monitoring, \emph{convergecast} is the most common operation~\cite{fastcollection}.
That is, data
from all the individual sensor nodes are collected at a sink node via transmissions along the edges of a tree.

\begin{figure}[h]
 \centering
\includegraphics[scale=0.40]{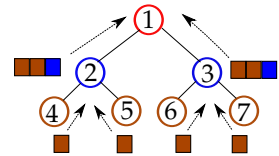}
 \caption{Data collection along a convergecast tree}
 \label{Datacollection_convergecasttree}
\end{figure}

There are three approaches which are widely used for data collection in wireless sensor network applications.
In the first approach, each sensor node sends its data packet (which contains its sensed data) to its parent and each parent node relays the received packets from each of its children as separate packets towards the sink node along the data collection tree. Also, the sensor data generated by a parent node itself is forwarded as a separate message along the tree.
There are approaches which employ aggregation methods to reduce the number of message transmissions occuring in the network. 
In one such method, each node receives data (e.g., 2 bytes) from all of its children and applies an aggregation technique
like taking average, minimum, maximum etc. on the collected dataset (received data as well as its own generated data) and forwards the aggregated
result (also 2 bytes in the above example) to its parent. This method is used mainly in dense deployments to sense 
parameters which have high spatial correlation. In another such method, each node collects the sensor data from its children and its own sensed data, concatenates them to form a single data packet and forwards it to its parent node as shown in Fig. \ref{Datacollection_convergecasttree}.

Our sensor network is not a dense deployment of nodes since it focuses on covering a large area with a small number of nodes so that the network will 
be a cost-effective solution to monitor the soil moisture and other parameters. Hence, we are assuming that all the sensor readings are equally important 
and are required to be transmitted to the sink node. So the data collection in our network is of the type shown in Fig. \ref{Datacollection_convergecasttree}.

In Section~\ref{SSC:overview}, we provide an overview of the architecture of our network, and in subsequent subsections we provide details on how various operations are performed in this architecture. 

\subsection{Overview of the network architecture}
\label{SSC:overview}
We have performed several experimental measurements which are detailed in Section \ref{motivations_architecture} and the insights gained from these
measurements lead us to use the network architecture shown in Fig. \ref{Datacollection_convergecasttree} for data collection.

\begin{figure}[h]
 \centering
\includegraphics[scale=0.5]{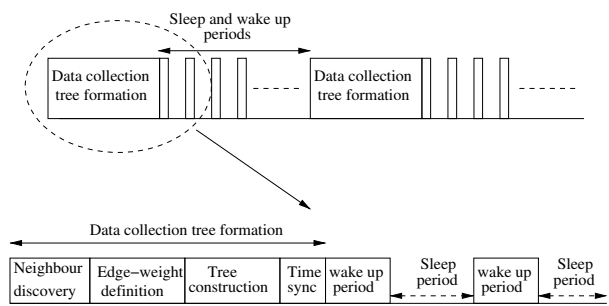}
 \caption{Sequence of operations occurring in the network}
 \label{periodic_operations_toplevel}
\end{figure}

Data collection in sensor networks mainly consists of two steps-- building a backbone structure (tree) to route the data packets
from each node to the sink node followed by scheduling of the data transmissions from the nodes.
The various operations occurring in the network are shown in Fig. \ref{periodic_operations_toplevel}. Periodically, the ``data collection tree formation'' phase is executed. Each ``data collection tree formation'' phase includes various stages like ``neighbour discovery'', where each node in the network finds its neighbours,  
followed by assignment of ``edge weights'', in which each node assigns a cost or weight to its neighbours depending upon the 
 remaining battery capacity of the nodes as well as the quality of the link connecting them. Then the list of neighbours and edge weights of each node are
transferred to the sink node through ``flooding''. The sink node builds the data collection tree from the collected information and
synchronizes the clocks of all the nodes along the edges of the tree. Once a time synchronized tree structure is constructed, each node periodically reports its
data to the sink node in a time-slotted manner.

To increase the lifetime of the network, the sensing nodes follow a periodic sleep and wake up schedule (see Fig. \ref{periodic_operations_toplevel}). 
The nodes are in energy saving sleep mode for most of the time in comparison with other stages
(for example, one sleep period = 58 minutes, one wakeup/ active \mbox {period = 2 minutes} and the total time between consecutive
``data collection tree formation'' phases = 3 days). During the wake up period, each node senses all the parameters: soil moisture,
soil temperature, atmospheric temperature and relative humidity and transfers the data to the
sink node along a convergecast tree to minimize the energy consumption as explained in Section \ref{tree_construction}.
As mentioned earlier,  sensor nodes are in the sleep state for most of the time. When a node becomes active (wake up), to initiate the transmission of a 
data message, the destination node should also come into active state from sleep mode.
To enable this, a time synchronization mechanism is periodically executed by the network to synchronize the clocks of all the nodes
as explained in Section \ref{timesync}.

Over a period of time, the nodes nearer to the sink node will have more battery drainage since they have to relay a lot of data coming from the 
nodes which are at a lower level in the tree. If proper precautions are not taken, the nodes nearer to the sink node will die (run out of energy) at an earlier 
stage than the nodes in the bottom levels. To ensure that the energy consumption occurs roughly uniformly at all the nodes in the network,
the convergecast tree is periodically recomputed (tree computations are performed during the ``data collection tree formation'' phases in Fig. \ref{periodic_operations_toplevel}); also, during each computation of the convergecast tree, nodes with a large amount of battery 
energy are preferentially assigned a large number of children in the tree and vice versa.
Section \ref{tree_construction} explains the construction of a covergecast tree. 

As mentioned earlier, the qualities of the links connecting different pairs of nodes are taken into account in the edge weight
calculations while constructing the data collection tree. In the rest of this subsection, we discuss how the link qualities 
are estimated in our implementation and the impact of their time variation on the network architecture. 
In \cite{baccour2012radio}, various link quality estimation techniques for  sensor networks are detailed. Unreliability in 
low power wireless links
mainly comes from three factors-- the environment, interference and the hardware platform. There are various software and 
hardware based approaches to characterize the quality of a wireless link. There are several contradicting observations also about the link
estimation. Received Signal Strength Indicator (RSSI) and Link Quality Indicator (LQI) 
are two hardware measures which can be used for capturing the link dynamics \cite{cc2420datasheet}. Several works have been reported in this regard.
RSSI is suggested as a better indicator of link dynamics in \cite{RSSI_underappreciated}.
The variation of RSSI as the distance between the transmitter and receiver varies is analyzed in \cite{greenorbs} and the
effect of temperature on RSSI is studied in \cite{temp_effects_RSSI}. \emph{Our experimental results (see Section \ref{motivations_architecture}) show that 
RSSI is a good indicator to capture the link dynamics between the nodes; hence, 
we use RSSI to estimate link qualities in this paper.}

Also, our experimental results (detailed in Section \ref{motivations_architecture}) reveal that the \emph{wireless link qualities 
in an agricultural farm are quite stable; hence, in our network architecture, we compute data collection trees infrequently, \emph{e.g.}, once every $3$ days.}
We now explain why the link qualities are stable. First, there are not many dynamics in the farm. Second, there does not exist 
much interference in open agricultural farms due to WiFi or Bluetooth signals which operate in the same 2.45GHz spectrum where our TelosB radio
operates (802.15.4).

\subsection{Neighbour discovery}
Since the placement of nodes in the field is considered as arbitrary, none of the nodes have any information about any other nodes 
in the network.
Hence, as a part of building the data collection tree, each node needs to discover its neighbours. 
Once the nodes are turned ON, each node enters into the neighbour discovery stage and 
tries to find out its 1-hop neighbours. The sink node initiates the exchange of Neighbour Discovery Messages (NDM) 
by broadcasting its message. Upon reception of a NDM, each node in the network initiates the broadcasting of its own messages.
While installing and setting up the network for the first time, we have to make sure that the sink node is turned ON last.
This will help to bring all the nodes into neighbour discovery stage almost concurrently providing a 
loose synchronized start of operation for the network which makes the actual field implementation easier.

A NDM is transmitted periodically for a specified number of times with the maximum transmission power \footnote{In our experimental evaluation, each node sends 60 NDM messages in 5 minutes, i.e., one message in every 5 seconds}.
The nodes receiving the NDM update its neighbour list with the source identifier (id), the remaining battery capacity (which is included in each NDM) and the received signal strength from the received message. The received signal strength 
indicator (RSSI) is a measure of strength of the received radio signal of the packet. Note that RSSI is a function of the distance 
between the nodes, shadow and multipath fading, and typically decreases as the distance between the nodes increases.
If the source id is already present in the neighbour table, the receiving node updates its neighbour table with the average RSSI from the respective source id.

To minimize collisions, the periodic NDM transmission at each node is initiated after a random delay. At the end of this phase each node has an updated neighbour list consisting of its 1-hop neighbours and the 
 edge-weight to each node in this list. The calculation of the edge-weight is detailed in Section \ref{edgewtcalculation}.

Neighbour discovery is performed periodically as a part of the ``data collection tree formation'' phase as shown in Fig. \ref{periodic_operations_toplevel}, 
in order to take into account the changes in the neighbours of a particular node which can happen either due to the 
addition of a new node into the network or due to the removal of a particular node because of various reasons
such as wireless communication failure, complete drainage of battery etc.

\subsection{Edge weight assignment}
\label{edgewtcalculation}
This phase assumes that each node has knowledge of its 1-hop neighbours. The aim of this phase is to assign an edge weight to 
all the communication links which exists between any pair of 
nodes in the network. The edge weight between a pair of nodes is defined in such a way that it captures the two node's current energy levels
as well as the link dynamics between them.

Each node uses a lithium ion battery as the source of energy and it is recharged using solar energy. Details about the sensor node are 
described in Section \ref{internal_structure}.
The current battery voltage can be used to measure a node's remaining battery capacity. Thus we define the edge weight between two nodes $u,v$ 
as: 
\begin{equation}
 e_{uv} = f(RC_u,RC_v) + g(avg RSSI)
 \label{edgeweighteqn}
\end{equation}
where $f$ is a function which contributes a cost term to $e_{uv}$ based on the remaining battery capacities $(RC_u,RC_v)$ of nodes $u$ and $v$ and
$g$ is a function which provides a cost term to $e_{uv}$ depending on the link quality between $u$ and $v$.
The lower the remaining battery capacity, the higher is the cost  provided by the function $f$ and the function $g$ provides a higher cost for
links which have lower avgRSSI\footnote{The question of how to choose the functions $f$ and $g$ in order to obtain the maximum lifetime tree is not in the scope of this work and is a direction for future research. For the current implementation, we have choosen $f = k_1 \times \left[\dfrac{1}{RC_u} + \dfrac{1}{RC_v} \right]$
and $g = k_2 \times avg RSSI$ where $k_1$ is positive and  $k_2$ and avgRSSI are negative. In particular, the values used to obtain the results provided in this paper 
are $k_1 = 5\times2200$ and $k_2 = -5$.}.
The data collection tree is constructed in such a way that edges with low edge weights are preferentially selected as part of the tree (see Section~\ref{tree_construction}); hence, edges corresponding to high battery capacities and high RSSI are preferred.

 \begin{figure}[h]
 \centering
\includegraphics[scale=0.7]{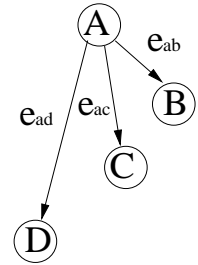}
 \caption{Node A and its neighbours with the respective edgeweights. From node A, the distance to node D $>$ the distance to
  node C $>$ the distance to node B and hence, $e_{ad}>e_{ac}>e_{ab}$ assuming all the nodes have the same stored energy.}
 \label{neighbour}
\end{figure}
 
For example as shown in Fig. \ref{neighbour}, nodes B, C, D are neighbours of node A in the increasing order of distance from node A. Node A
can calculate the the edgeweight $e_{ab}$ using (\ref{edgeweighteqn}) with the help of $RC_B$ and $avgRSSI$ at A from B which was 
computed at A during the neighbour discovery phase. In the same fashion node A finds the edge weights to all of its neighbours.
Every node finds its remaining battery capacity from the current battery voltage as explained in 
Section~\ref{BVtoRC}.

\subsection{Network information to the sink node through flooding}
\label{flood_sink}

\begin{figure}
 \centering
\includegraphics[scale=0.55]{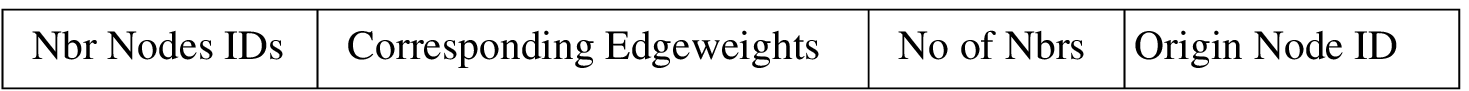}
 \caption{Neighbour broadcast message format}
 \label{NBM}
\end{figure}

At this stage, each node has details about its neighbouring nodes and the corresponding edge weights
which need to be transported to the sink node to build the
data collection tree. This is done through simple flooding \cite{floodingcomparison}:
each node in the network broadcasts its neighbour-list and corresponding edge weights, each
recipient of a broadcast packet re-broadcasts it and so on, until the sink
node receives the packet. Note that by \emph{broadcast}, we mean that the message is transmitted with a particular address so that every
other node in the transmission range of the sender receives it.
At the end of this flooding process, the sink
node has complete information about the topology and edge weights in the
entire network. Each node's broadcast message (Neighbour Broadcast Message, NBM) has a format as shown in Fig. \ref{NBM}.

Flooding is one of the most energy consuming operations in the tree building phase since it requires
a lot of message transmissions. To ensure reliable transfer of neighbour details of each node to the sink,
and to reduce the number of transmissions, the following mechanisms are employed.

\begin{enumerate}
 
 \item {A node makes sure that each NBM broadcasted by it reaches all of its neighbours through ACKs 
	(generated from application layer (with link layer ack enabled for this ack, that is; an ack for ack) since link 
	 layer ACKs are not supported for broadcast messages in tinyos).}
 
\item {Each node keeps a queue of fixed length to store NBMs. When a node receives NBMs from different nodes, it queues up the messages and rebroadcasts them at a later point in time.} 

\item {Also each node keeps a signature of each of the recent NBMs (Origin ID) it has forwarded~\footnote{Each node keeps the signatures of NBMs until
the neighbour details are transferred to the sink node.} to 
	avoid rebroadcasting of the same NBMs  a node may receive from its different neighbours.}
	
\item {Each node except the sink node has a NBM to pass to the sink node through flooding.  
      Therefore, there can be a worst case scenario of N-1 floods initiated at the same time in the network. 
      To reduce the congestion, each node first puts its NBM at the front of its queue and a timer is set
      to fire after a random time. The node in which the timer fires first, initiates the flooding process.
      The neighbouring nodes hearing this message send ACKs to the sender and add the received message to their send queue for rebroadcast.}
\end{enumerate}

Once a node finishes broadcasting all the messages in its send queue, and has not received any NBM from any of its neighbours 
for a particular timeout time, it implies that the flooding stage is over. At the end of this flooding process, the sink node has 
complete information about the network topology and edge weights.

\subsection {Tree construction phase}
\label{tree_construction}
\begin{figure}
 \centering
\includegraphics[scale=0.6]{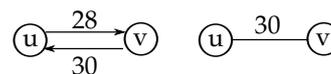}
 \caption{Undirectional assignment of edgeweights}
 \label{edgealignment}
\end{figure}

Once complete information about the network is available with the sink node, different data collection trees 
can be built in the network graph to collect data from the individual nodes, \emph{e.g.}, shortest path tree (SPT), minimum spanning tree (MST) etc~\cite{cormen2009introduction}. Before running any algorithm to construct a tree, the edges between the nodes are made 
undirectional by selecting the higher of the forward and backward weight  as the new edge weight as shown in Fig. \ref{edgealignment}; this is because, we indirectly represent the cost of message transfer as
the edge weight.

We use Dijkstra's algorithm \cite{Dijkstra} with the cost of each edge $(u,v)$ equal to $\max(e_{uv},e_{vu})$ (see Section \ref{edgewtcalculation}) 
to find the data collection tree.
Dijkstra's algorithm is used to find the shortest path from every node in
the network to the sink node. The union of all the shortest paths results
in a shortest path tree rooted at the sink node.
If the edge weight equals the energy needed for unit packet transfer, it is shown in \cite{SPTpaper} that,
for raw-data covergecast, in which the entire sensed data
needs to be sent to the sink node without fusion at intermediate nodes,
routing the packets through the shortest path tree minimizes the total
energy across all nodes, needed to deliver the packets to the sink node.
In our algorithm, since the edge weight is also dependent on the residual battery energies of nodes, 
the rate of energy consumption at different nodes is more uniform.

Once the data collection tree is built, the sink node constructs a message called ``Connection Detail Message (CDM)'' which
contains a list of the parent of each node in the data collection tree. The sink node finds all its children and forwards the CDM message to each of them.
On reception of the CDM, each node finds its parent node as well as all its children nodes and forwards the CDM to each of its children nodes. In this
manner, each node gets the information about its parent and children in the data collection tree.

\subsection{Time synchronization}
\label{timesync}

The sensor nodes, which are in sleep state for most of the time, periodically wake up, sense various parameters and transmit the data 
to their respective parent nodes in the data collection tree (see Fig. \ref{periodic_operations_toplevel}). To have a coordinated sleep and wake up schedule between the transmitting and the 
receiving node, time synchronization between the nodes is essential.
Since time accuracy to within fractions of seconds is generally acceptable in sensor networks, we can use any lightweight 
(in terms of energy) protocol.

Every node uses an oscillator of specified frequency to increment a register counter in hardware which is considered as its local hardware time (H) \cite{karl2007protocols}.
The oscillator has a small random variation from its specified frequency which varies with respect to time and is known as 
\emph{drift} \cite{karl2007protocols}. Due to this drift, two nodes of the same specified frequency will have different rate of increase for their time resulting in an 
error in time measurement called \emph{clock skew} error \cite{karl2007protocols}.
Also, each node in the network is switched ON at different times resulting in an \emph{offset/ phase shift} \cite{karl2007protocols}. Considering all these factors, for a node $i$,
we can represent its software logical time at a real time $t$ as $L_i(t) = \theta_i H_i(t) + \phi_i$, where $H_i(t)$ represents the 
local hardware time of node $i$ at time $t$ \cite{karl2007protocols}. We try to adjust the values of 
$\theta_i$ (clock skew) and $\phi_i$ (offset) so that every node in the network has the same notion of time as that of the sink node.

We use a modified version of Flooding Time Synchronization Protocol (FTSP) \cite{FTSP} in our application \footnote{Many other time synchronization approaches have been proposed in the research literature. Addressing 
the question of which approach is the best and the implementation of this approach are directions for future work.}. FTSP
is a time synchronization protocol which
uses periodic broadcast messages to synchronize the clocks of all the nodes in the network
with that of the sink node (node with smallest node id). Each broadcast message is timestamped at the transmitting node (considered as the global 
network time) and on reception of this message, each receiving node gets its local timestamp, providing it a global-local timestamp pair called as
reference point for time synchronization. The difference between global and local timestamp is known as the offset between the sender and 
the receiver node. This clock offset increases linearly with time due to the clock drift and to estimate the drift rate of the receiver clock with
respect to the sender's clock, a node needs to have enough reference points, which are spaced in time, provided by the periodic broadcast
messages. A node finds its offset and skew once it receives enough reference points and becomes a synchronized node. Then it starts to broadcast
periodic synchronization messages. Thus through flooding of synchronization messages by the synchronized nodes, all nodes in the network get
synchronized.

\begin{figure}
 \centering
\includegraphics[scale=0.47]{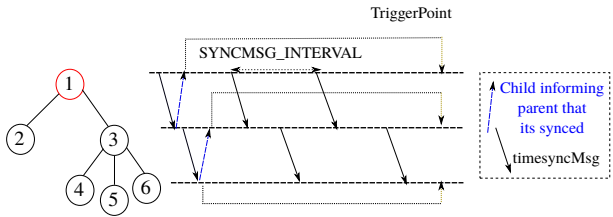}
 \caption{Time synchronization}
 \label{timesync_tree}
\end{figure}

In our approach a parent node synchronizes its child nodes using broadcast messages and 
this procedure starts from the sink node and thus all the nodes get synchronized to the sink node along the edges of the tree.
Every node (except sink node) maintains a time synchronization table which holds the reference points that are used for finding its offset
and clock skew. On reception of a synchronization message, each child node gets its local timestamp and thus gets a reference point 
for time synchronization which gets added to its synchronization table and the node's offset and clock skew get updated. A node simply drops
the synchronization messages received from non-parent nodes.
Now, we give an example to illustrate how the nodes of a tree get synchronized to the sink node.  For example consider a tree and its 
synchronization flow diagram shown in Fig. \ref{timesync_tree}.
In the considered example, the sink node (Node ID1) initiates the time synchronization phase by broadcasting 
a synchronization message once the data collection tree is built. The message is timestamped at 
the MAC layer, which is a feature supported by the tinyos platform \cite{TinyosTEP_packetleveltimesync}.
MAC layer timestamping helps to reduce the variations/ uncertainties 
associated with the time required to transmit a packet from one node to the other \cite{Elapsed_timeonarrival}.
As shown in Fig. \ref{timesync_tree}, once the child nodes (Nodes 2 and 3) receive the timesync message, they inform their parent that they are synchronized.
Multihop time synchronization can be considered as an extension of this procedure. That is, now Node 3 broadcasts timesync messages 
so that its child nodes get synchronized to it. Once a parent node comes to know that all its children are synchronized, 
it waits for a time instant in the future (after at most a few minutes) 
referred to as the \emph{trigger point}. The trigger point is a common point in time across all the nodes 
in the network and is used to initiate the data collection intervals nearly simultaneously at all the nodes.  

As a part of synchronization, a node needs to find out its offset and clock skew with respect to the reference node (sink node) in the network.
The initial transmission of a synchronization message is used to find out the clock offset between the nodes.
In order to find the clock skew, a node needs to have enough reference points which are spaced in time. For this purpose, each parent node
broadcasts time synchronization messages periodically (with large intervals-- example 30 seconds) till the data collection 
trigger point occurs. In Fig. \ref{timesync_tree}, the interval between consecutive time synchronization messages is denoted as ``SYNCMSG\_INTERVAL''. Also, during the data collection phase a node receives 
a sleep message from its parent in each data collection slot which adds a reference point entry in the time synchronization table (see Section~\ref{datacollection_section}).
We use simple linear regression for finding the clock skew as used
in FTSP. The performance of our synchronization scheme is evaluated in Section \ref{timesync_experiments}.

FTSP makes use of an ad hoc structure to transfer the global time of the sink node to all the other nodes-- any node which is in the transmission
range of the sender node, receives the broadcast message and gets synchronized. In contrast, our time synchronization happens along the data collection tree.
In FTSP, a node may receive synchronization messages from different senders since it employs broadcasting of the synchronization messages
from all the synchronized nodes and hence, it employs a redundant message handling mechanism; such a mechanism is not required in our scheme since a node only receives synchronization messages from its parent in the data collection tree. Also, FTSP employs a sink election process since it does not consider a dedicated sink node to address node failure, due to which there can be multiple sinks in the network at a point in time and hence, it includes techniques to handle this scenario. 
In our approach, these techniques are not required because of our assumption that the sink node is 
never going to fail (unless some hardware damage occurs which is a rare scenario)
since the node is placed in a room which is not a harsh environment and the node has an unlimited source of energy 
as electricity is available. Another advantage of our scheme is that during each time slot, the use of a sleep message 
(see Section \ref{datacollection_section}) as a  reference point for time synchronization eliminates the need for a 
separate packet transmission only for time synchronization.

\subsection{Data collection}
\label{datacollection_section}
Data collection begins after the formation of a synchronized data collection tree. Data collection phase consists of periodic sleep and 
wake up (active) stages (see Fig. \ref{periodic_operations_toplevel}). During the active state each node turns its radio on, collects data from all its children, 
aggregates them into a single packet along with its own sensed data and forwards it to the parent node.
If the packet size is above the maximum size limit, the data is forwarded as two separate packets with an extra byte 
indicating data message full. Each data packet transfer is made reliable with the help of an acknowledgment.
Thus the sink node collects the complete sensed data from all the nodes in the network. Now the sink node broadcasts a
sleep message which is propagated down the tree. This sleep message is meant for two purposes: (i) as sleep indication, and 
(ii) as a reference point for the child nodes which is added in their time synchronization table which is used for clock skew calculation.
Every parent node in the tree rebroadcasts the sleep message which is timestamped at the MAC layer for their children. On reception of this
sleep message, each node updates its clock skew, rebroadcasts the sleep message to its children and then goes to the sleep state. There is a timeout (large enough) associated with this so that
a node which could not receive the sleep message successfully does not remain active till the next data collection time slot, 
thereby wasting its energy. This helps to increase the lifetime of the network. Our sensor network is mainly designed for low duty cycle 
applications, for example in agricultural monitoring. In this application we collect data once in every three hours. Thus an active period is a 
few seconds long while the intermediate sleep states are of the order of a few hours.
\subsection{Complexity analysis of tree formation stage}
In this section, the various message transfers occurring in the network during different phases of the tree formation stage are analyzed.
In the neighbour discovery phase, each node broadcasts $k$ NDM messages at regular intervals and hence, for $n$ nodes, there are $k \times n$
message transmissions occurring in the network.

After the neighbour discovery phase, every node tries to convey its neighbour details to the sink node through flooding as detailed in
Section~\ref{flood_sink}. Consider a NBM originating from node $u$. All the neighbouring nodes which receive the NBM of $u$
rebroadcast the message and this process continues till the NBM reaches the sink node. In the worst case, the NBM that originated from $u$ 
will be rebroadcasted by all the other nodes in the network except the sink node. This results in a total of $(n-1)$ message transmissions.
Since each node in the network except the sink has a NBM to forward to the sink node, in the worst case there will be $(n-1)^2$ message transmissions.
Also, each node's NBM broadcast is accompanied by ACKs from the neighbouring nodes. Let $D$ be the maximum number of neighbours of any node
in the network. Thus there will be atmost $(D+1)\times (n-1)^2$ message transmissions in the network during the flooding stage.

Once all the information reaches the sink node it builds the data collection tree and informs each node about its parent and children
as detailed in Section~\ref{tree_construction}. That is, along each edge of the data collection tree, there is a CDM transfer accompanied
by its ACK resulting in a total of $2(n-1)$ message transfers. Similarly, the time synchronization mechanism detailed in 
Section~\ref{timesync} involves a worst case broadcast transmission of $m$ synchronization messages by each nonleaf node 
in the network.
This phase also consists of another $(n-1)$ message transfers (one along each edge) corresponding to the message transmission
by each node to its parent informing that it is synced. Thus the time synchronization phase consists of atmost a total of
$m(n-1)+(n-1) =$ $(m+1)(n-1)$ message transfers.

Considering all these, the tree formation mechanism involves a total of atmost $kn+(D+1)(n-1)^2+2(n-1)+(m+1)(n-1)$ message transfers
which can be considered as having a complexity of $\mathcal{O}(D(n-1)^2))$ or in the worst case $\mathcal{O}(n^3)$ since $D \leq n$.

\subsection{Handling node failures in the data collection tree}
\label{nodefailure_schmes}

A wireless sensor network consists of a large number of nodes deployed in harsh conditions. They are
supposed to function for a long time with minimum human intervention.
In this section, 
we discuss about how node failures are handled in an energy efficient manner in our implementation.

The sensor data is collected through a data collection tree which is rooted at the sink node. Any sensor node
can fail due to numerous reasons. We consider mainly the complete discharge of battery
or hardware failure as the main reasons for the node failure. A failed node cannot receive or transmit
any messages. The network architecture is such that the complete tree rebuilding happens very rarely and nodes are
in the data collection phase for most of the time.  If a node fails during the data collection phase, the parent node 
will not be able to receive data from the failed node in the upcoming data collection time slots. Consecutive data misses from a node is recognized
as a node failure and the handling mechanisms are detailed below. (If a node fails in the middle of a tree building phase, the corresponding next
stage in the tree building phase removes the failed node from the network-- the mechanisms used for achieving this are similar 
to those given in Section \ref{singlenode_failure_section} for handling node failures in a data collection phase and are omitted for brevity.)

\subsubsection{Single node failures}
\label{singlenode_failure_section}

\begin{figure}
\centering
\subfloat[Node 5 informing the sink node about the failure of node 11]{\label{singlenode_failure}\includegraphics[scale=0.4]{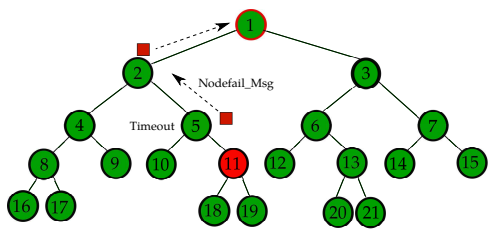}}
\hfill
\subfloat[The new data collection tree after the removal of node 11]{\label{singlenode_failure_corrected}\includegraphics[scale=0.4]{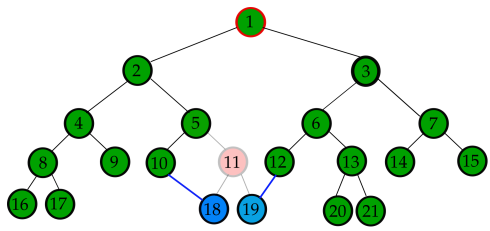}}
\caption{Single node failure}
\label{singlenodefailure_total}
\end{figure}

Fig. \ref{singlenode_failure} shows an instance of a single node failure. Once node 11 fails, node 5 will be waiting
for the data from node 11 during the next data collection time slot. When the data collection timeout occurs consecutively for a particular number 
of time slots (two in our implementation), node 5
will detect the failure of node 11 and will inform the sink node as shown in the figure by sending a NodefailMsg along the tree 
The data collection timeout is selected sufficiently large (few minutes) such that consecutive data 
collection timeout implies the failure of a node.
Since the sink node has complete information (all the vertices and edges) about the network, it will remove the failed node, 11 from the
graph and builds a new data collection tree as shown in Fig. \ref{singlenode_failure_corrected}.

\subsubsection{Multiple node failures}

 \begin{figure}
\centering
\includegraphics[scale=0.4]{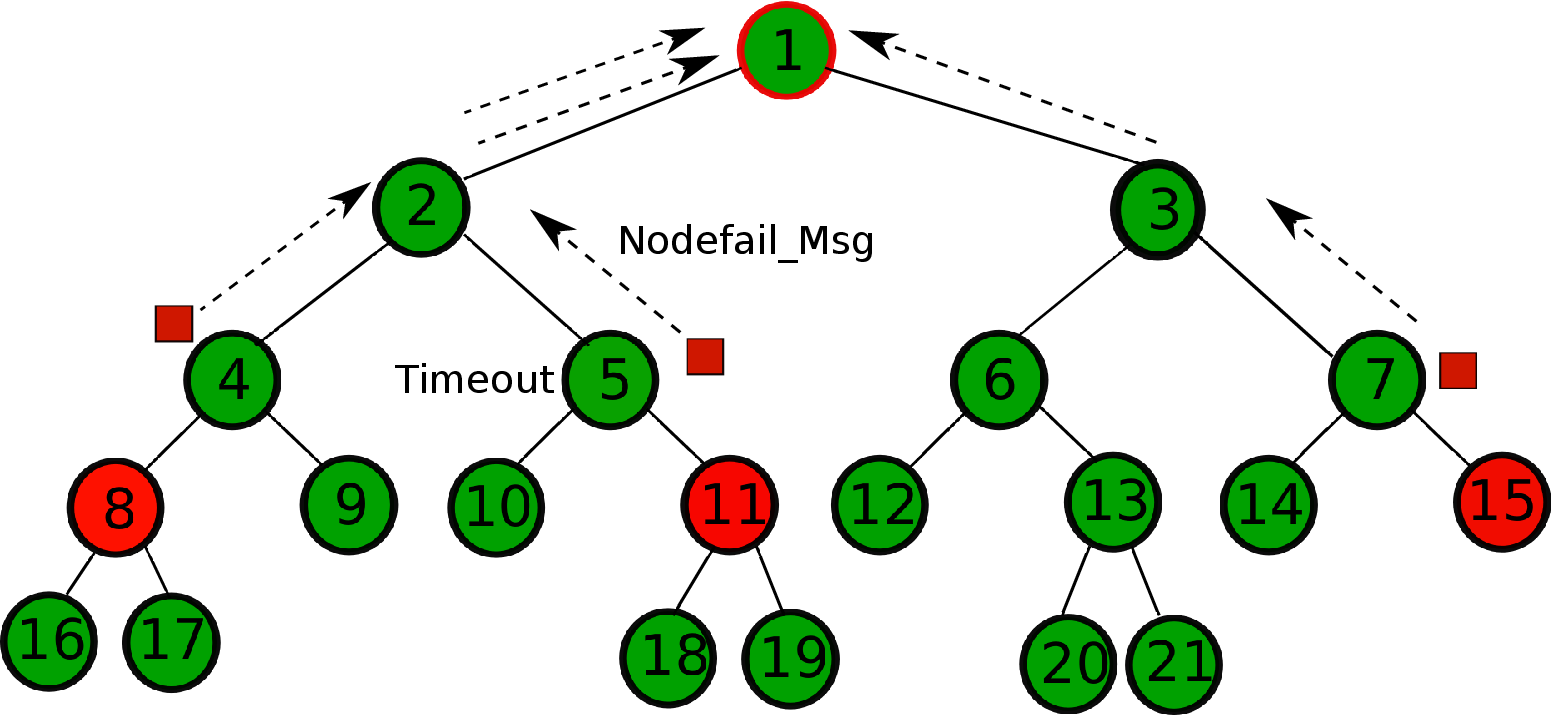}
 \caption{Multiple node failures (in different branches)}
 \label{mutiplefailure_differentbrances}
 \end{figure}

 Fig. \ref{mutiplefailure_differentbrances} shows a particular instance of multiple node failures in the network. This can be
considered as simultaneous single node failures. The respective parents will report all the node failures and the sink node
reconstructs the tree accordingly by removing the failed nodes 8, 11, 15 from the graph. 
 
 \begin{figure}[h]
\centering

\subfloat[Multiple node failures (in the same branch)]{\label{mutiplefailure_samebrances}\includegraphics[scale=0.4]{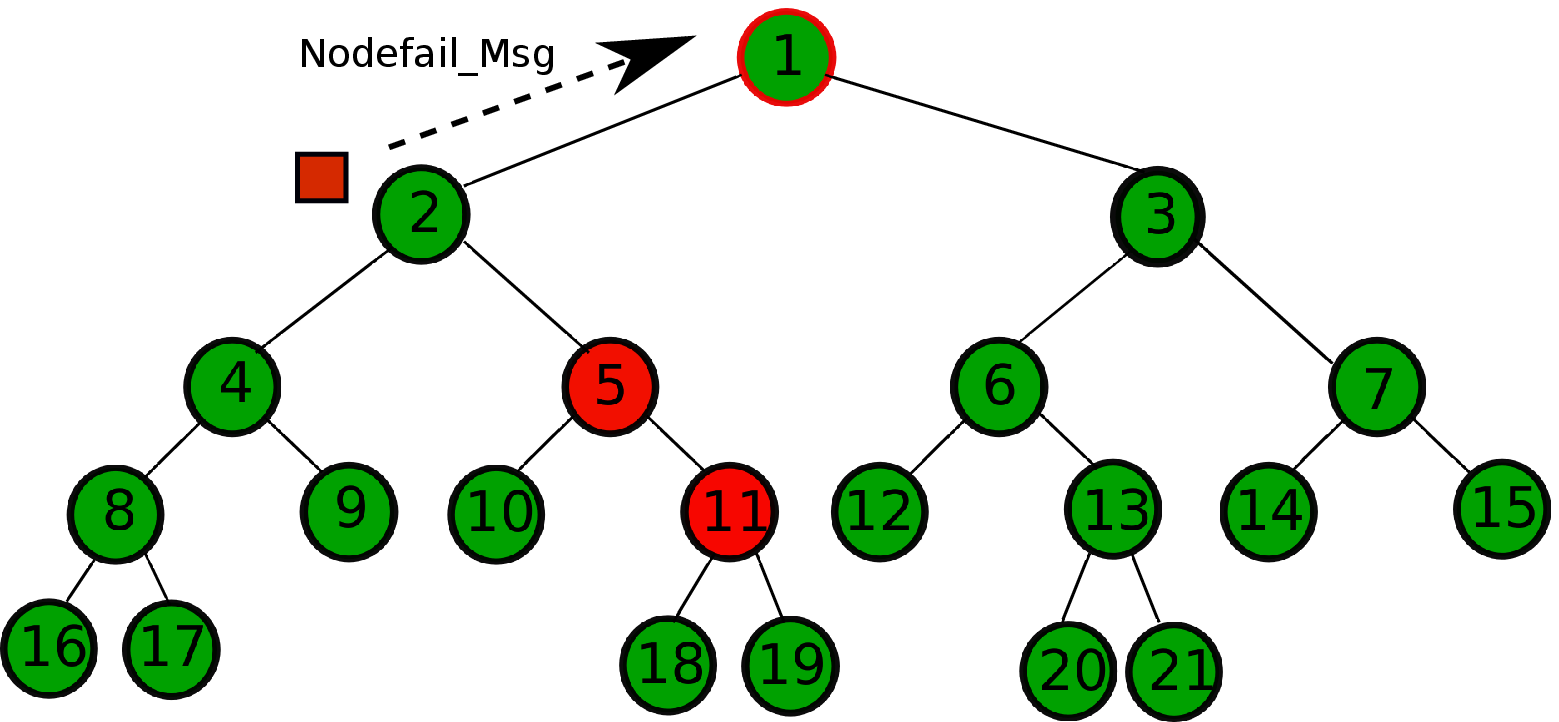}}
\hfill

\subfloat[Reconstructing the tree through a failed node]{\label{multiplenodefailure_correctingfirst}\includegraphics[scale=0.4]{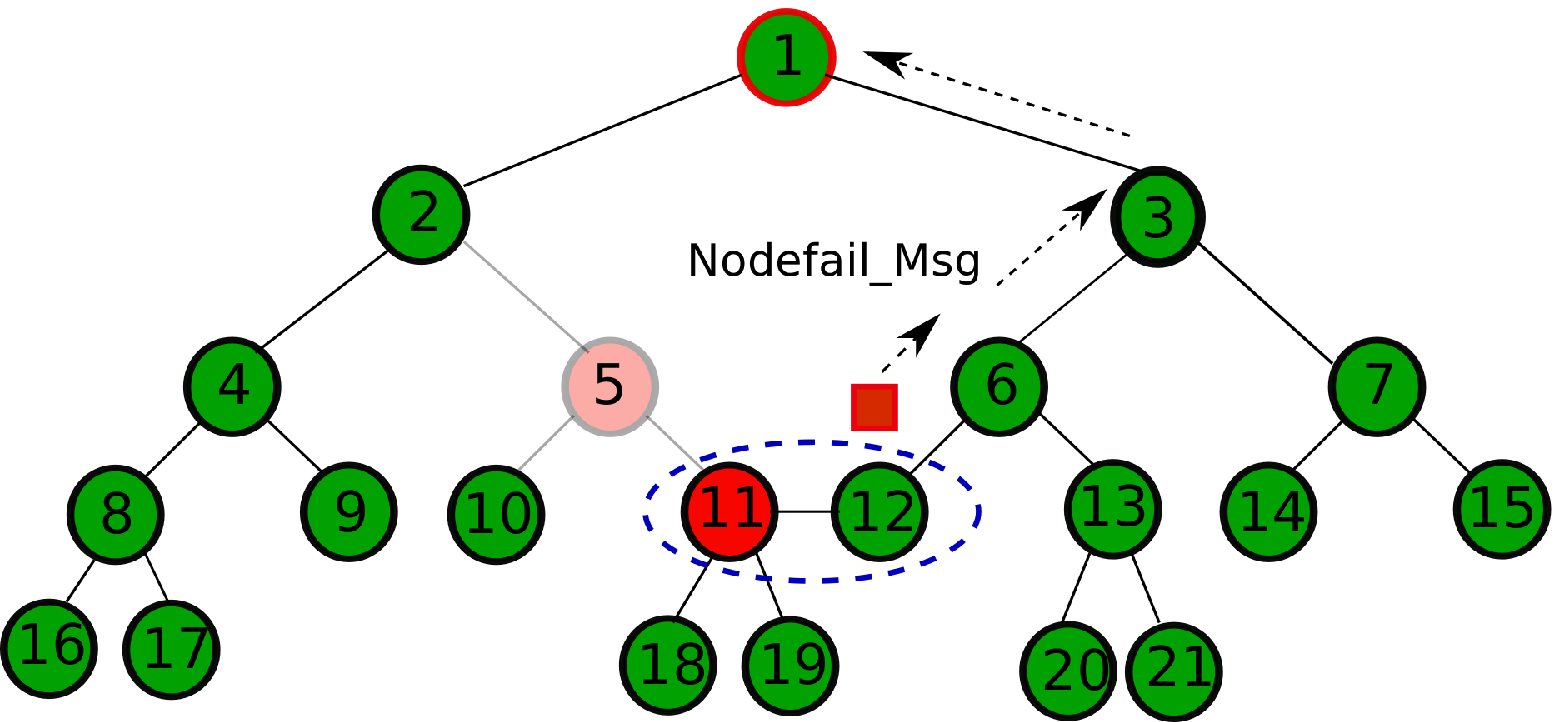}}
\hfill

\subfloat[Reconstructed tree through multiple stages]{\label{multiplenodefailuresamebranch_corrected}\includegraphics[scale=0.4]{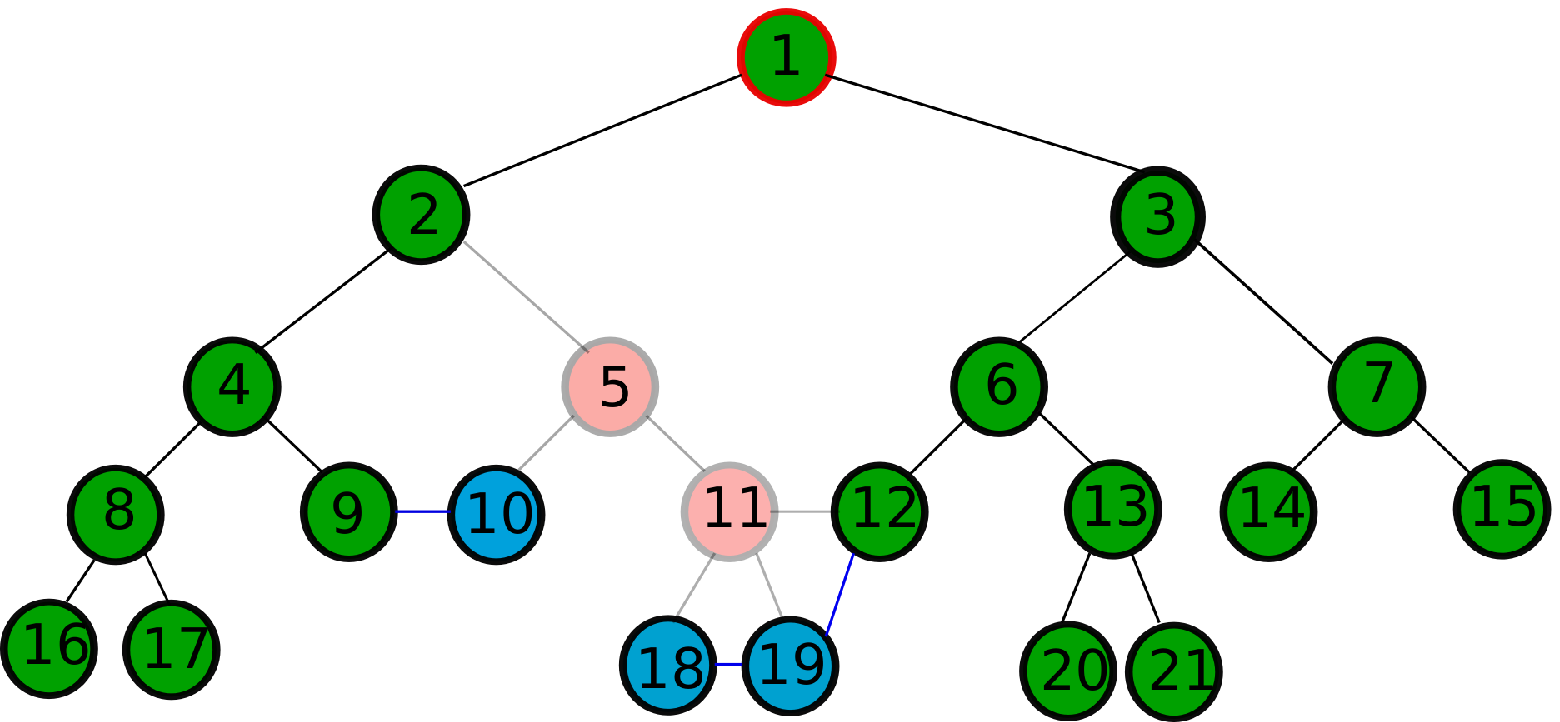}}
\hfill
\caption{Handling multi node failures}
\label{multiodefailure_total}
\end{figure}

Fig. \ref{mutiplefailure_samebrances} shows another instance in which the failed nodes are in the same branch. In this case node 2 will report 
only about the failure of node 5 and the sink node will try to reconstruct the tree through the failed node 11. While
reconstructing, the non responsiveness
of node 11 will be reported as another node failure by node 12 as shown in Fig. \ref{multiplenodefailure_correctingfirst}. Thus a new tree will be formed in two
different stages as shown in Fig. \ref{multiplenodefailuresamebranch_corrected}.

\section{Implemented Network and Experimental Methodology}
\label{experimental_methodology}
\subsection{Network structure and experimental methodology}
\begin{figure}[h]
 \centering
\includegraphics[scale=0.4]{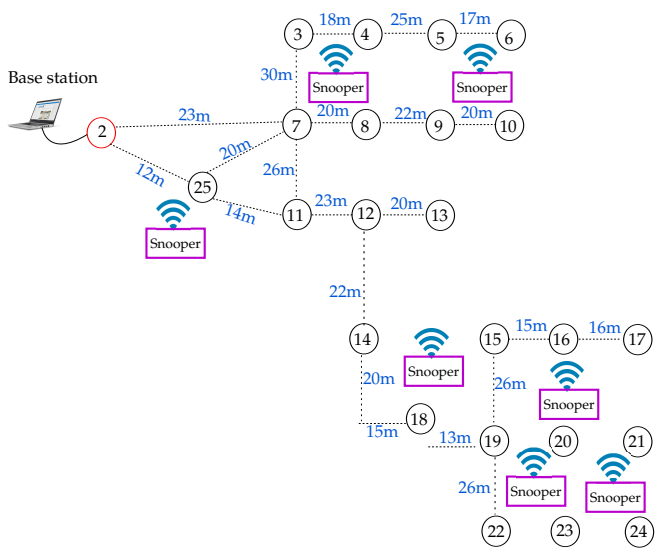}
 \caption{Field installation overview}
 \label{implementation_field}
\end{figure}

We have implemented a wireless sensor network consisting of 24 wireless sensor nodes which are equipped with various sensors; this network is deployed in a maize
farm of approximate 3 acre size. Fig. \ref{implementation_field} shows an overview of field installation. The sensor nodes are represented by 
circles with their respective node ids.
The sink node (Node Id - 2) is connected to a powerful device (we call it as base station, 
which is a laptop in our field installation) which logs the complete sensed parameters which may be accessed from a remote location through the 
Internet. We assume that the sink node has an unlimited source of energy as it is connected to a laptop which is kept in a small building
like a storage room or a pump-house where electricity is available for the motors to provide irrigation. We do not make any assumptions about the 
location of the sink node.

The major objective of the implementation was to capture the variations in the temperature and humidity both in the
atmosphere as well as in the soil which can be used as a reference for controlling the irrigation systems in an optimal manner which can result
in an improved yield as well as conservation of water resources. We aim to understand the effectiveness and performance of the proposed
simple data collection approach through synchronized data collection tree described in Section~\ref{Network_architecture}.
We are more interested in understanding the various activities and network behaviour of the sensor  nodes in the field. To capture 
the various actions occurring in the network, we have installed 7 snooper systems (a TelosB mote connected to a data logging laptop) where 
each sensor node's packet transmissions are captured in at least one snooper system. We use the data logged in the base station and snooper systems 
as the reference dataset for our evaluations and observations.

Here, we address some of the practices/ guidelines followed by us during the installation of the systems in the agricultural farm.
The system mainly consists of 4 parts: solar panel, an IP-65 enclosure box which contains all the electronic modules, 
different sensors and supporting rods. The solar panel is installed at a sufficiently large height above 
the plant canopy level (3.5m from the ground in this work) facing south direction. The enclosure box is also installed
above the plant canopy level so that the plants/ leaves do not hinder the wireless transmissions. Another factor which needs to be 
considered is the depth at which the soil moisture sensor is installed. It is preferred to install the sensor at the root zone 
of the plants which is different for different crops (for example up to 30 cm for maize \cite{rootzone_maize} and up to 
90 cm for grapes \cite{du2008water}).

\subsection{Sensor node -- internal structure}
\label{internal_structure}
\begin{figure}
\centering
\subfloat[Sensor node in the field]{\label{node_field}\includegraphics[scale=0.35]{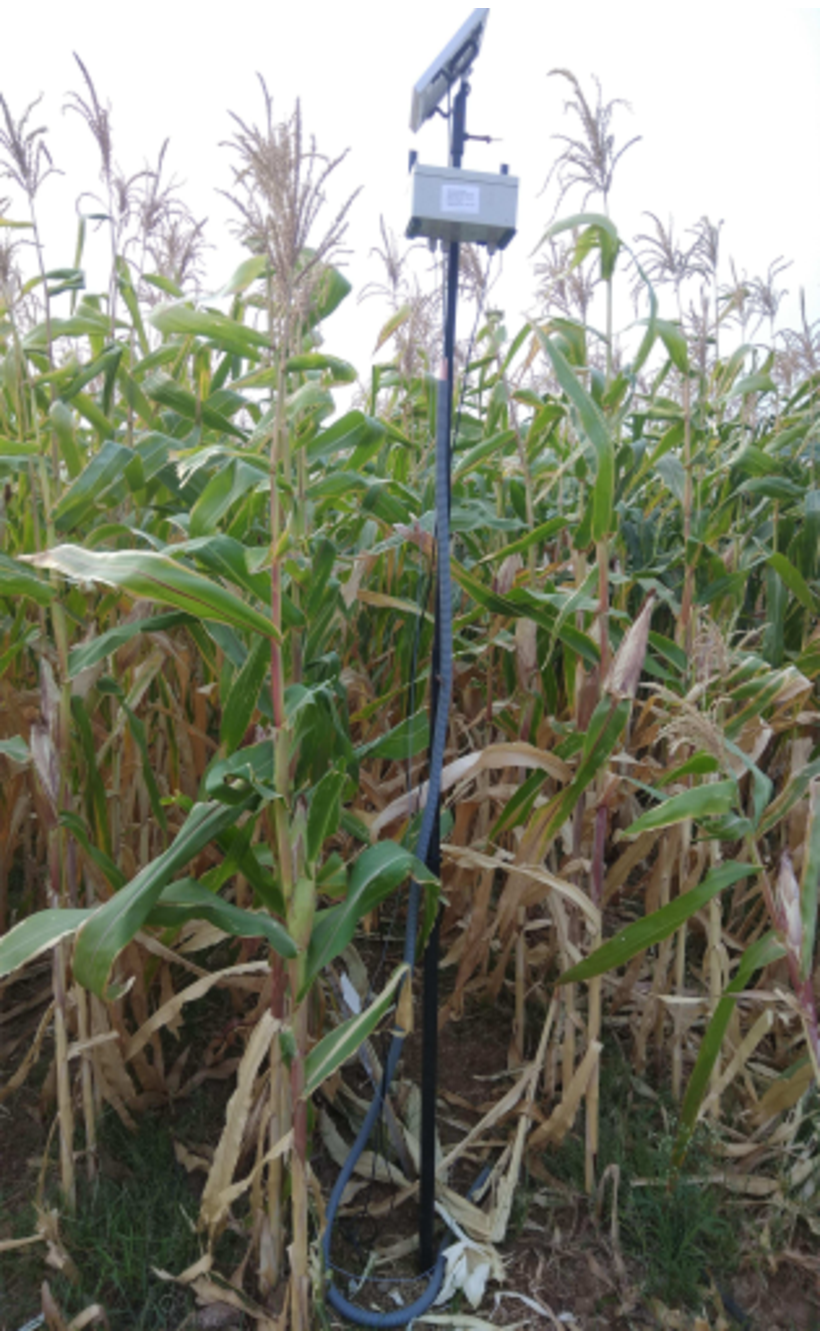}}
\hfill
\subfloat[Internal structure of the node]{\label{nodestructure}\includegraphics[scale=0.27]{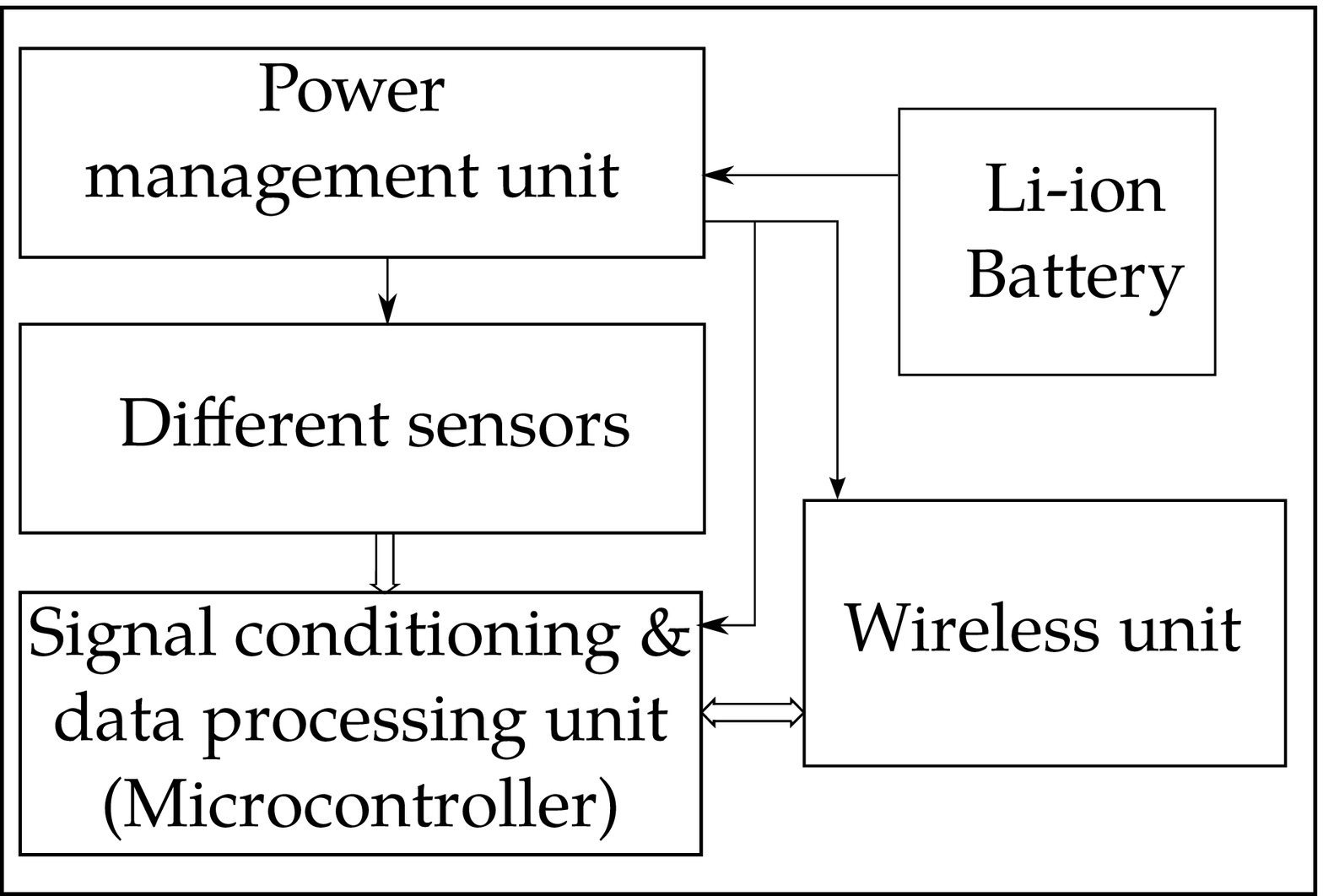}}
\hfill
\caption{Sensor node}
\label{field_installation_internal}
\end{figure}

\begin{figure}
 \centering
\includegraphics[scale=0.25]{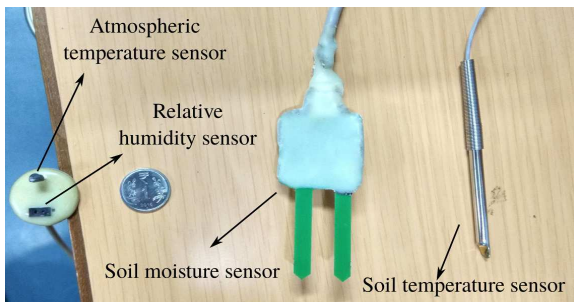}
 \caption{Different sensors in the system}
 \label{different_sensors}
\end{figure}

Fig. \ref{field_installation_internal} shows the filed installation and the internal structure of an individual sensor node. 
Each node is deployed with soil moisture, soil temperature, ambient temperature
and ambient humidity sensors as hown in Fig. \ref{different_sensors}. We are using a custom developed 
capacitive based sensor for soil moisture measurement which is 
detailed in \cite{capacitivework}. The soil moisture sensor uses two PCB probes and when inserted in soil, the soil acts as the dielectric 
medium between them. The sensor output is a square wave signal whose frequency varies with the moisture content present in the soil.
MCP9700A \cite{MCP9700A} is used for sensing atmospheric temperature and its packaged form is used for soil temperature sensing. 
HIH5035 \cite{HIH5031} is used for sensing the ambient humidity.  
The system is powered by a lithium ion battery of capacity 2200mAh which is recharged using solar energy. The power management unit has the circuitry for
providing the required power supply to the other modules. We use a TelosB mote for wireless communication.

\section{Motivations Behind the Proposed Network Architecture}
\label{motivations_architecture}

This section describes some of our experiments and the learnings which helped us to understand the deployment environment as well as
the hardware platform. The various factors which led us to the proposed network architecture for the data collection are detailed below.
\subsection{RSSI is a good indicator to capture link dynamics}
We use TelosB motes in our sensor nodes for wireless communication. TelosB is equipped with an IEEE 802.15.4 compliant
radio CC2420. Received Signal Strength Indicator (RSSI) and Link Quality Indicator (LQI) are the two measures provided by the radio which 
can be used as a measure to capture link dynamics.
During transmission, the radio chip CC2420 splits each byte into two symbols of 4 bits each and each symbol is mapped to one of the 
16 pseudo-random sequences of 32 chips each. RSSI can be considered as the power received at the RF pins and is an averaged value over 8 symbol periods.
LQI gives the strength/ quality of a received packet and can be considered as a measure of chip error rate. A value of 110 for LQI indicates maximum quality
and 50 indicates lowest quality frame for CC2420.

\begin{figure}
\centering
\subfloat[]{\label{rssi_indoor}\includegraphics[scale=0.34]{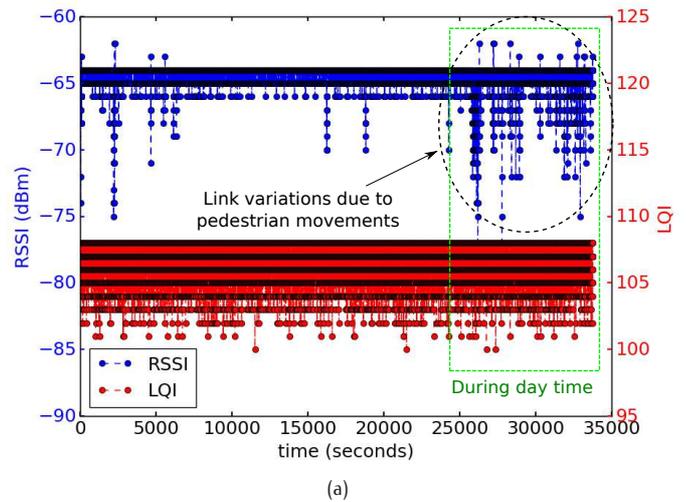}}
\caption{RSSI and LQI variations during night and day times} 
\label{RSSILQI}
\end{figure}
To understand the variations, we have programmed two motes, one as receiver and the other as transmitter which periodically transmits a message with 
maximum transmission power.
Fig. \ref{RSSILQI} shows the measured LQI and RSSI at the receiver mote which was kept at a distance of 10m
from the transmitter. Both motes were placed in a corridor of an indoor building and the transmission interval between two packets 
was one second. The interference/ disturbance in the link due to pedestrian movements in the daytime is captured significantly in the RSSI but not in
the LQI. This indicates that RSSI is a good indicator for capturing link dynamics.
Hence, we choose to use RSSI as a measure of link quality in our experiments.

\subsection{Wireless links in agricultural fields can be considered as almost static links}

To understand the link dynamics in our deployment environment, \emph{i.e.}, an agricultural farm, we have conducted an experiment which captured the 
variation of RSSI and LQI between different pairs of nodes continuously for a long time.
There are many works \cite{greenorbs}, \cite{woo2003taming} which report the variation of signal strength for different distances between the transmitter and the receiver node
and it is clear that the signal strength decreases with the distance between the transmitter and the receiver.
In addition to this fact, here we report the major observations from our experiments. We use three pairs of transmitter and receiver nodes for 
our experiment where each pair communicates through a different 802.15.4 wireless channel.

\begin{figure*}
\centering
\subfloat[Link dynamics (pair 1)]{\includegraphics[scale=0.18]{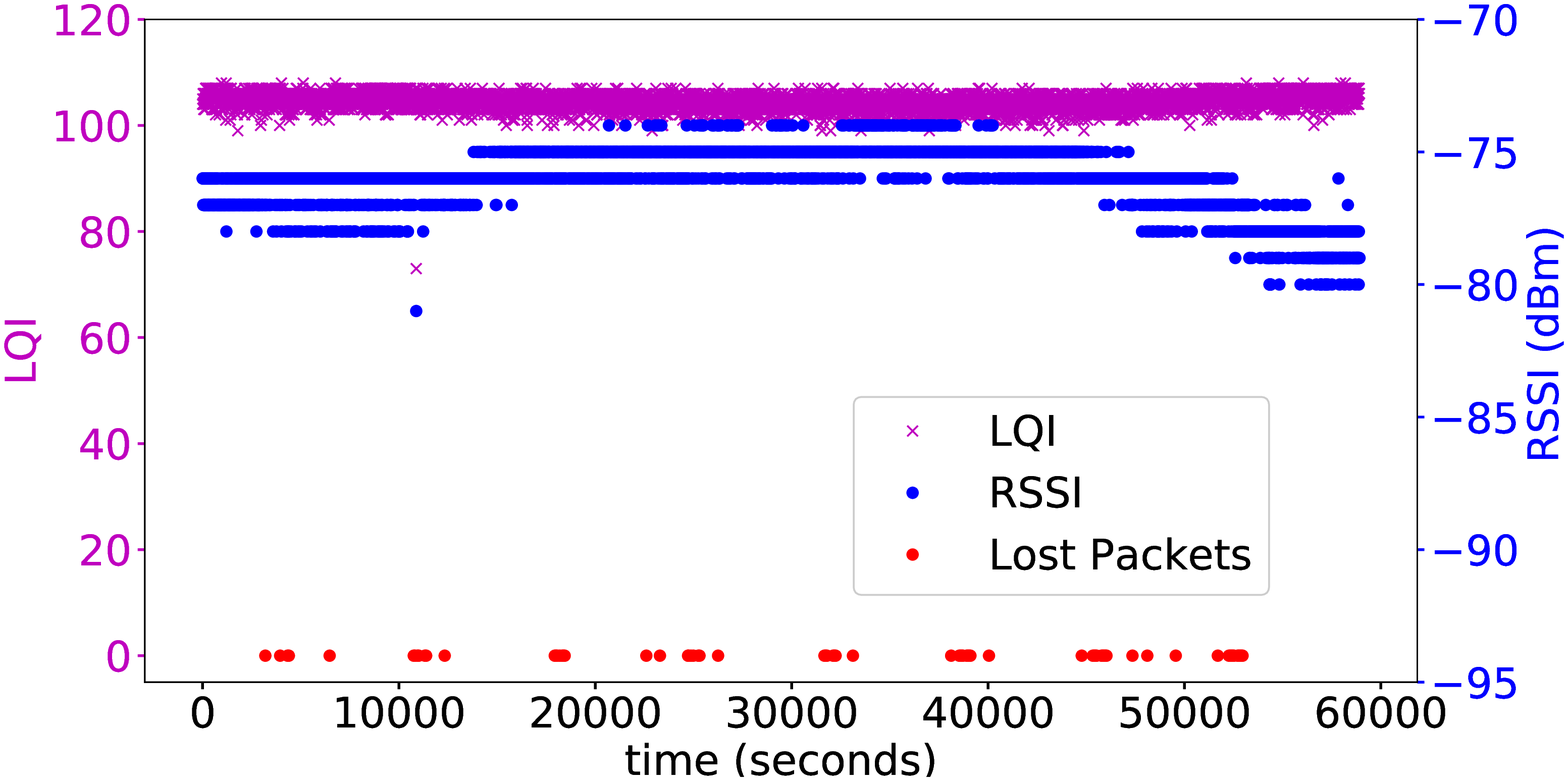}
\label{Link_dynamics_node15}}
\subfloat[Averaged RSSI variations (pair 1)]{\includegraphics[scale=0.185]{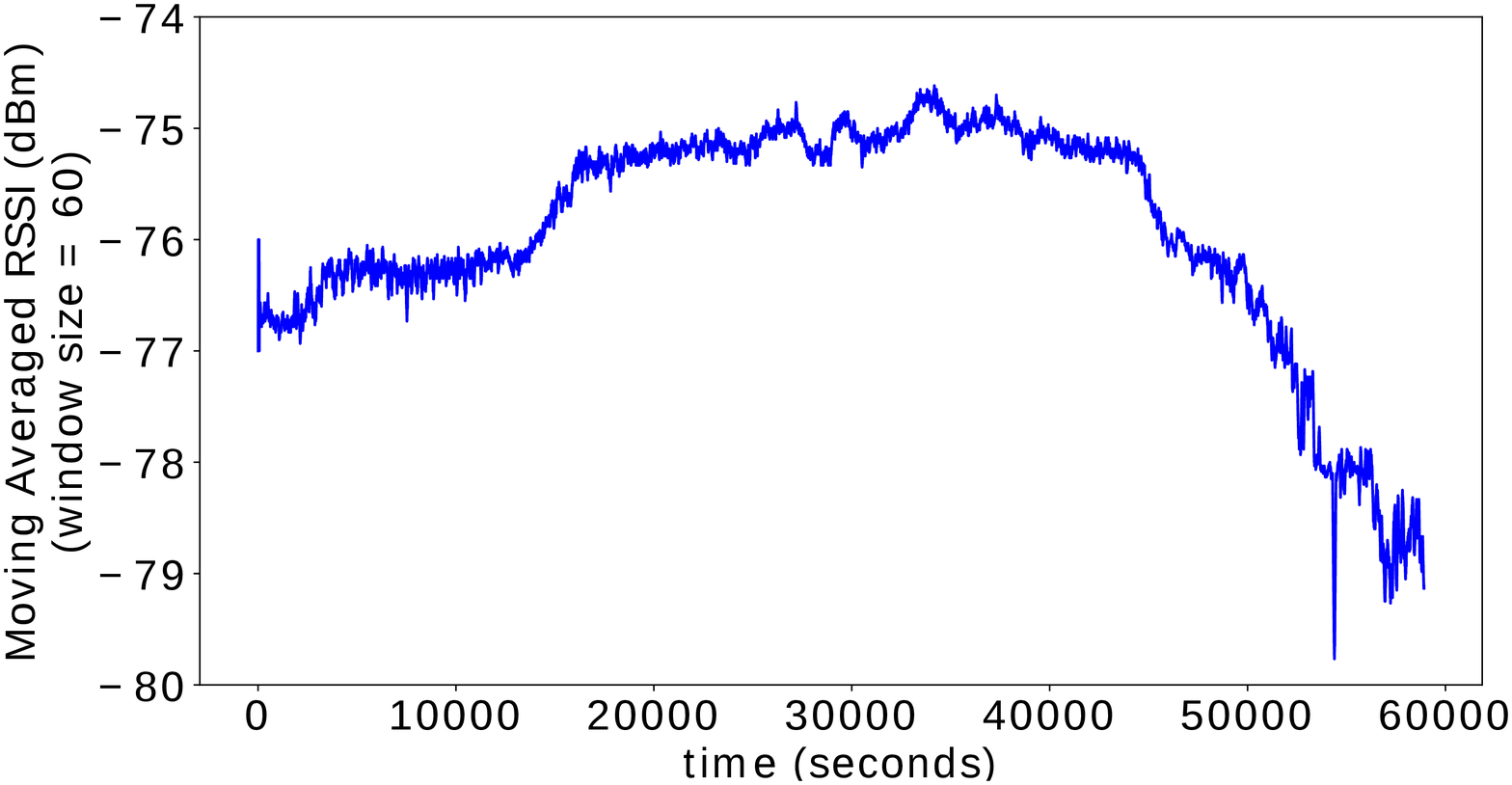}
\label{Linkrssi_node15}}
\hfill
\subfloat[Averaged LQI variations (pair 1)]{\includegraphics[scale=0.185]{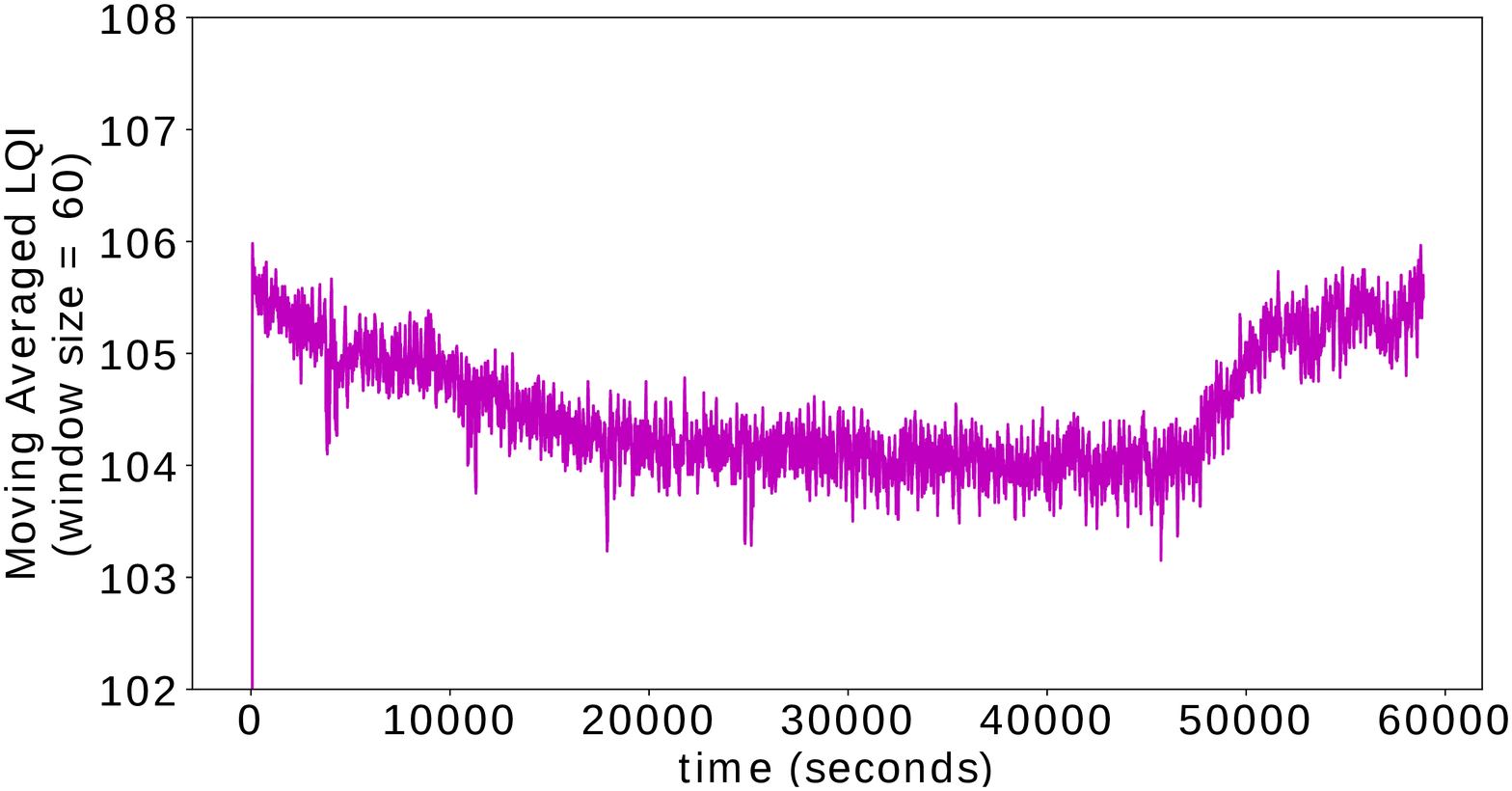}
\label{Linklqi_node15}}
\hfill
\subfloat[Link dynamics (pair 2)]{\includegraphics[scale=0.178]{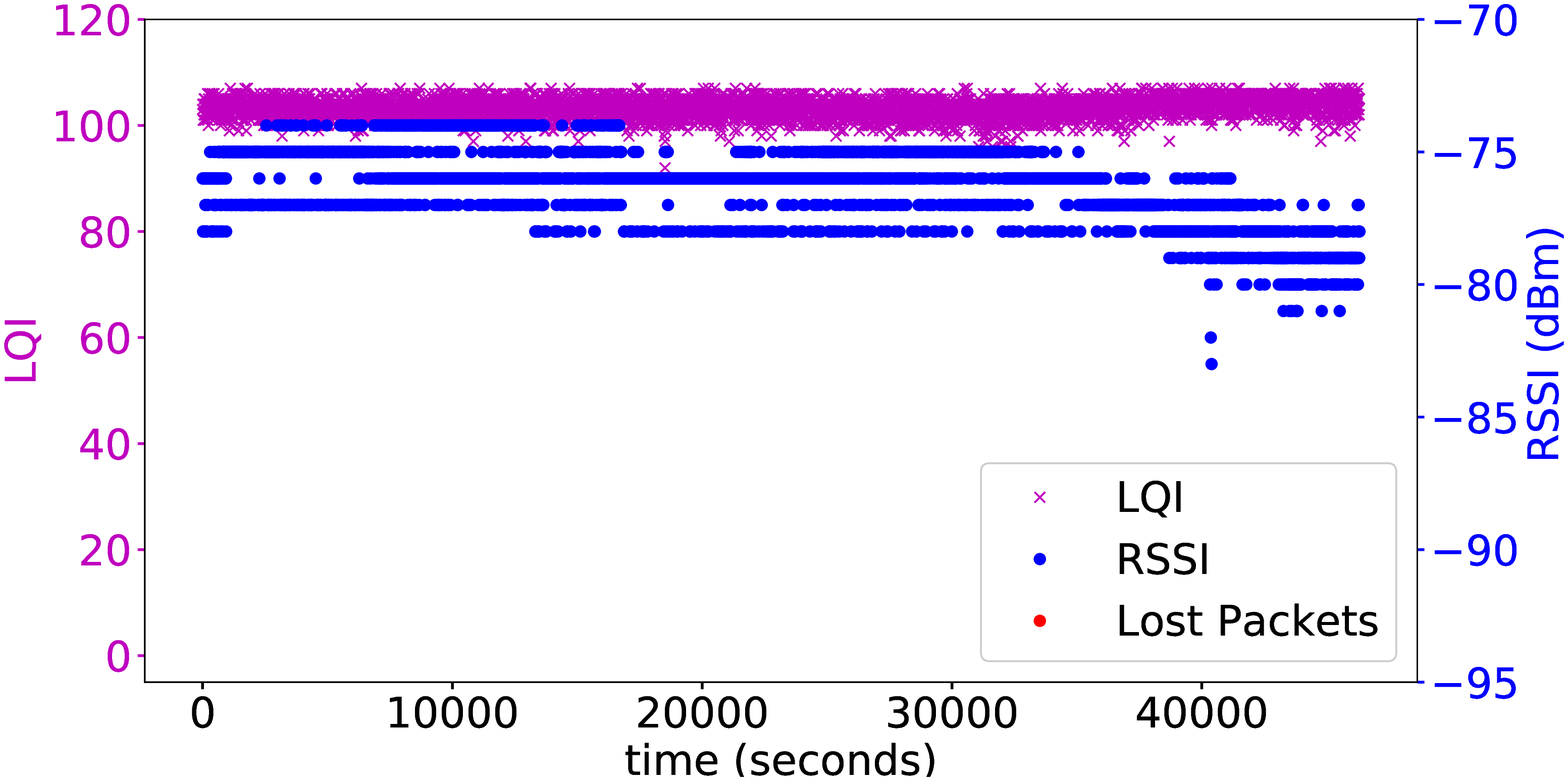}
\label{Link_dynamics_node22}}
\hfill
\subfloat[Averaged RSSI variations (pair 2)]{\includegraphics[scale=0.187]{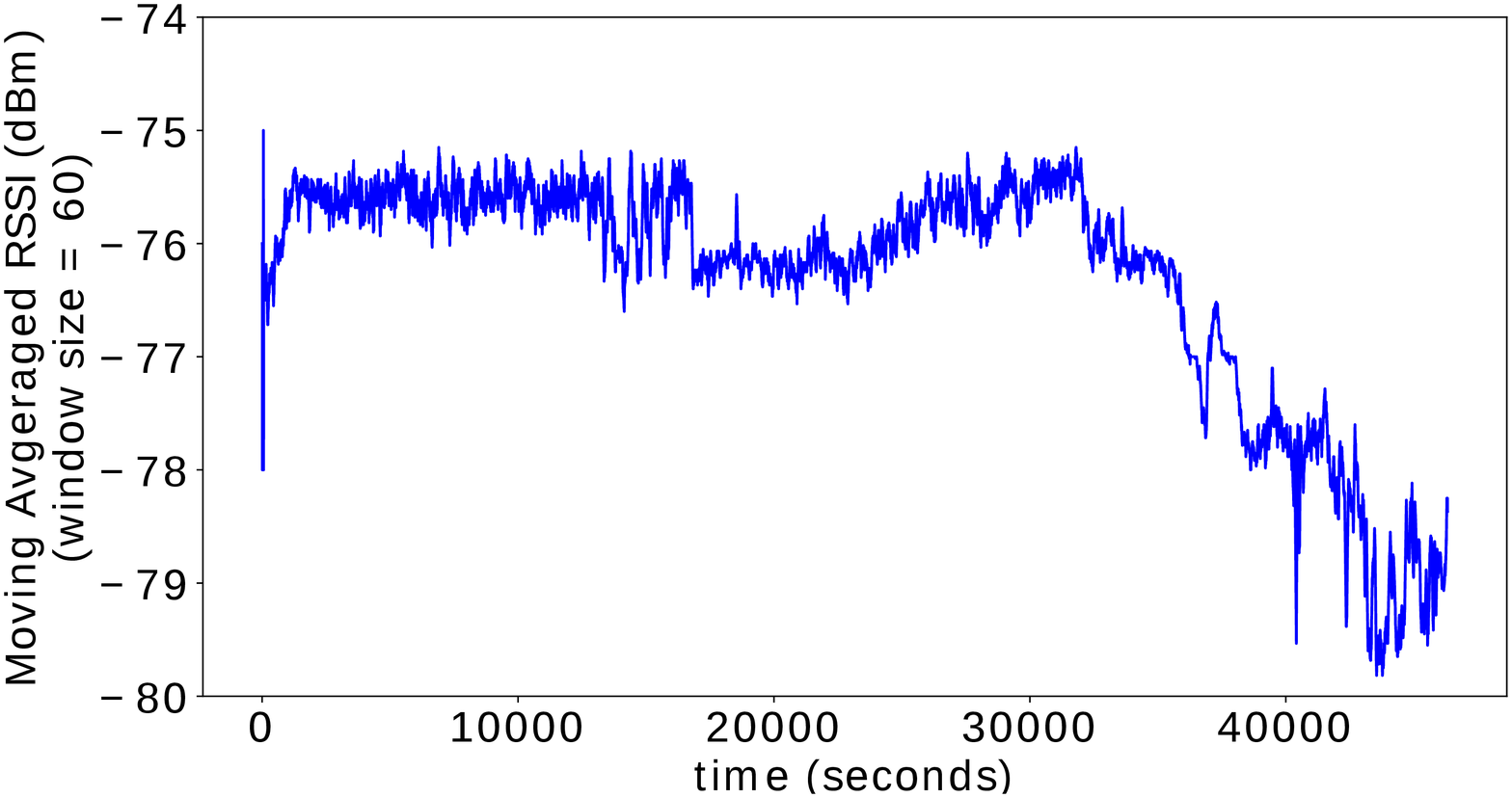}
\label{Linkrssi_node22}}
\hfill
\subfloat[Averaged LQI variations (pair 2)]{\includegraphics[scale=0.187]{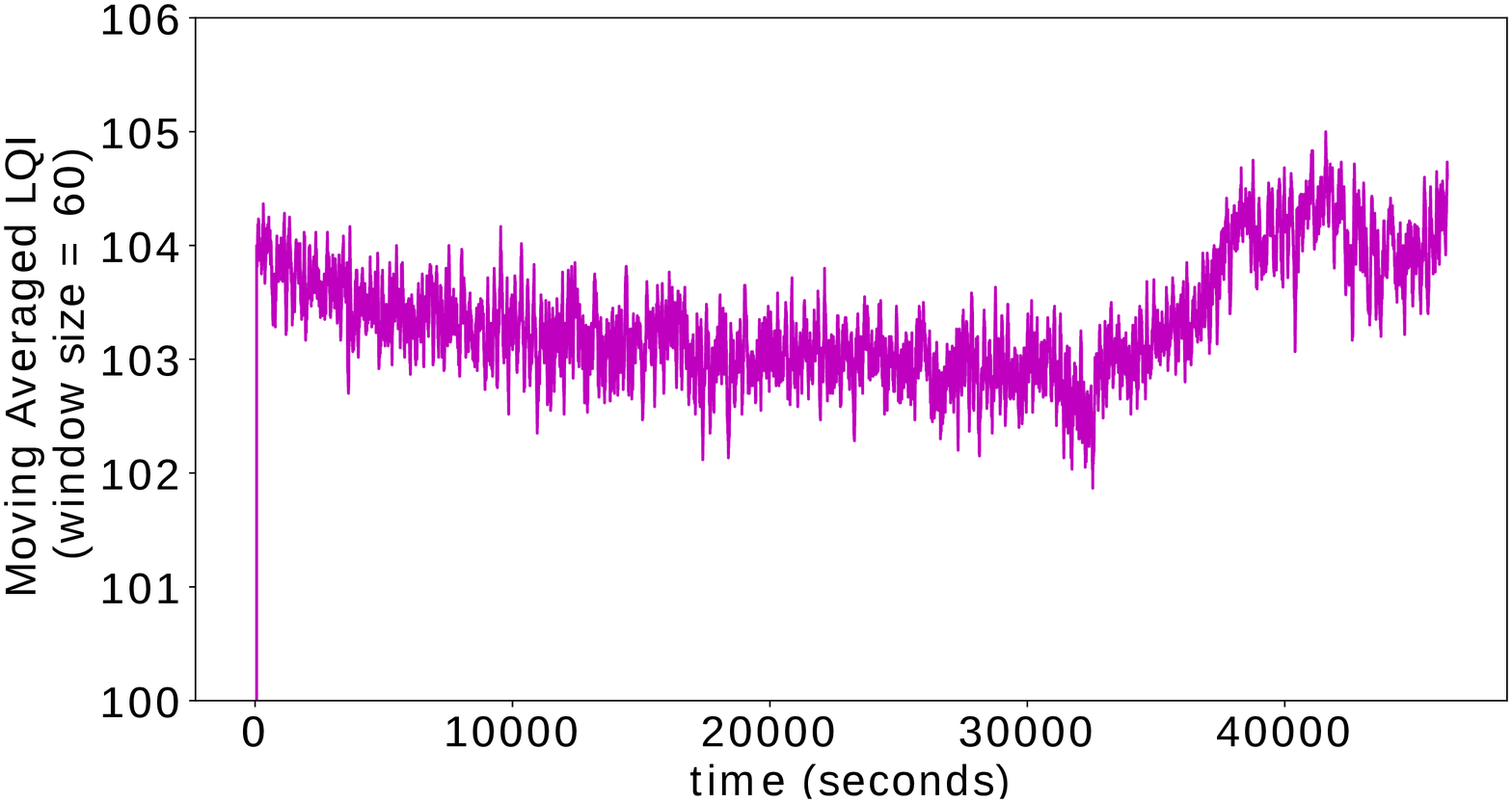}
\label{Linklqi_node22}}
\hfill

\subfloat[Link dynamics (pair 3)]{\includegraphics[scale=0.185]{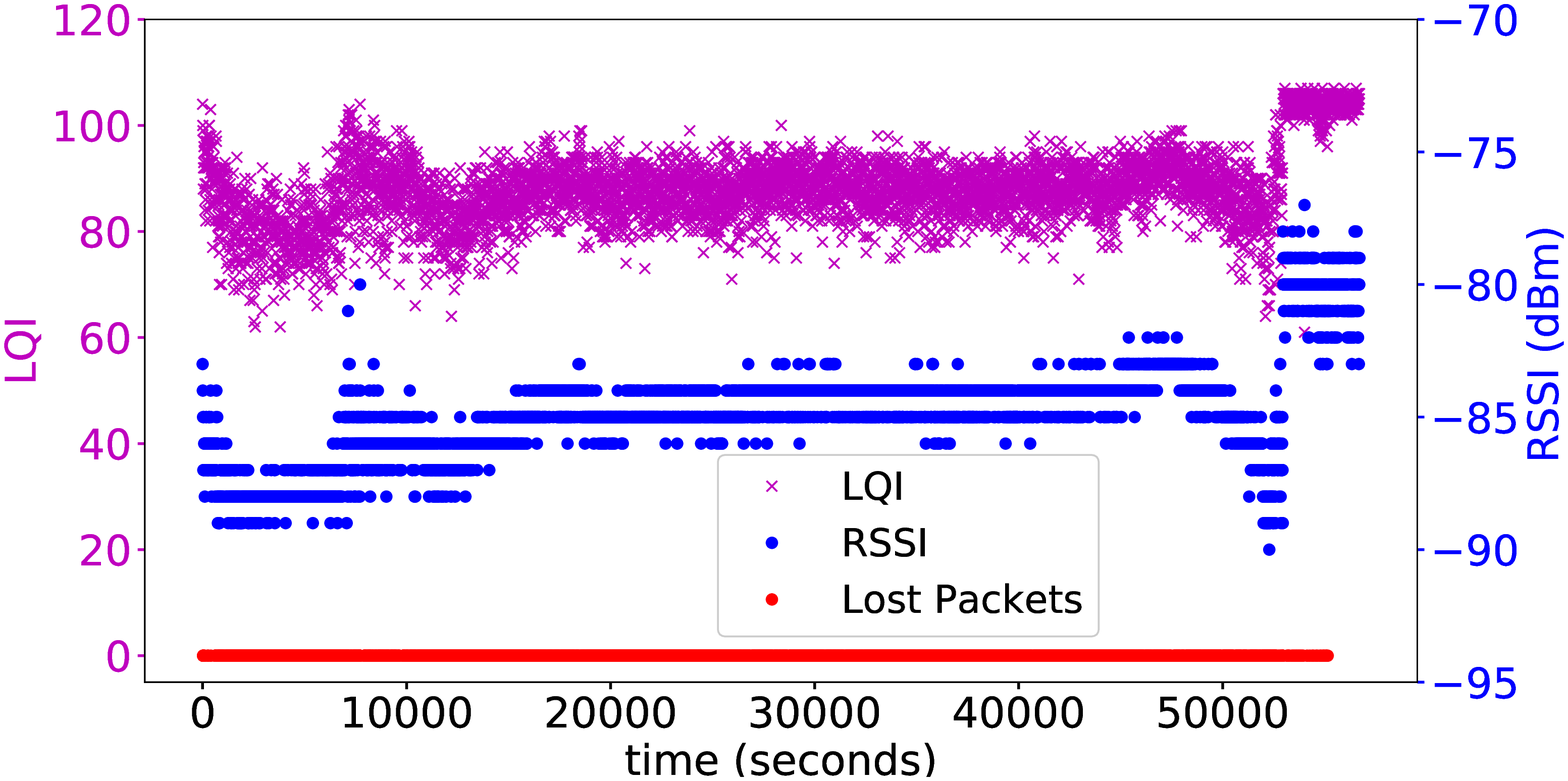}
\label{Link_dynamics_node6}}
\hfill
\subfloat[Averaged RSSI variations (pair 3)]{\includegraphics[scale=0.187]{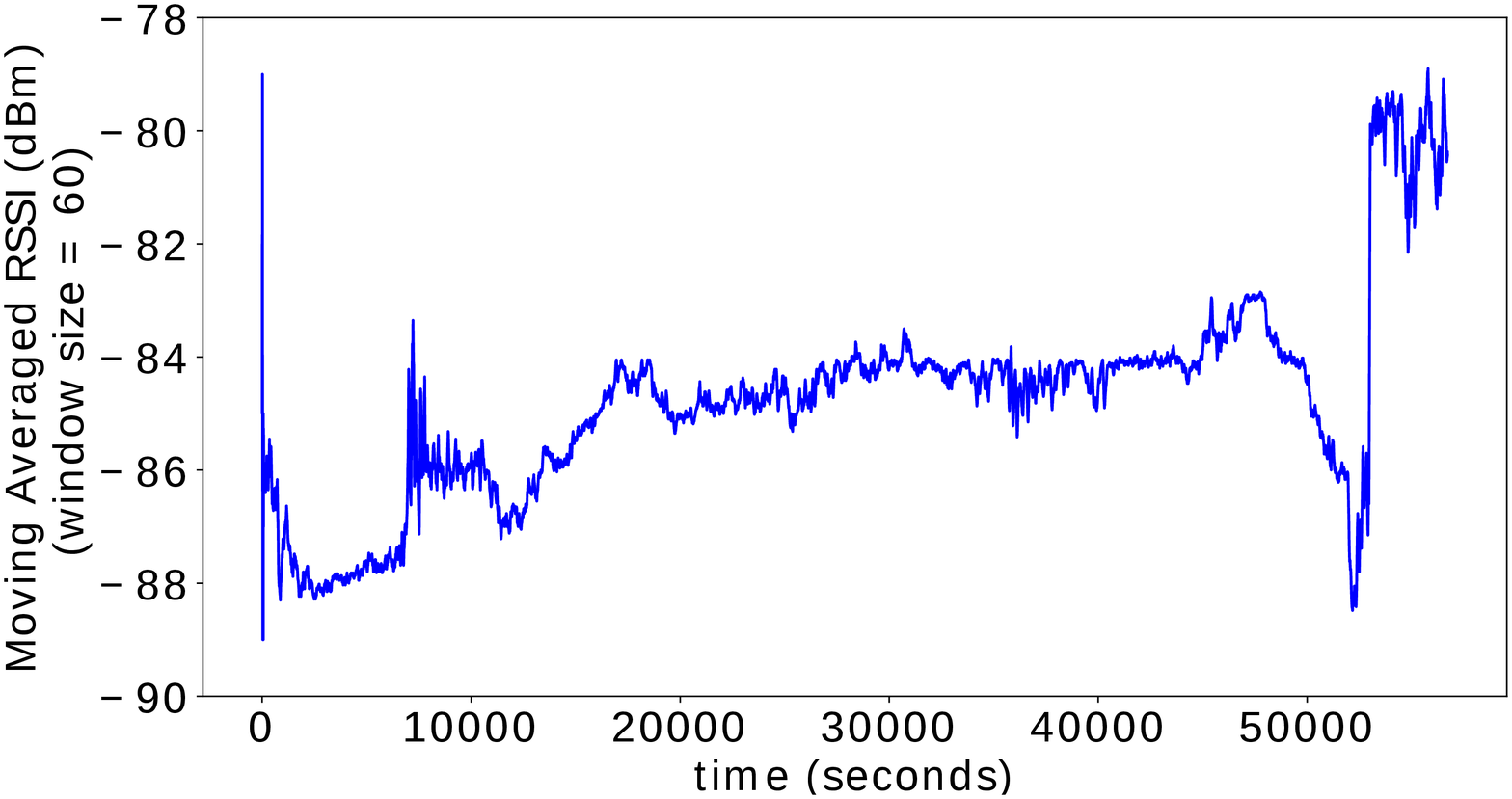}
\label{Linkrssi_node6}}
\hfill
\subfloat[Averaged LQI variations (pair 3)]{\includegraphics[scale=0.187]{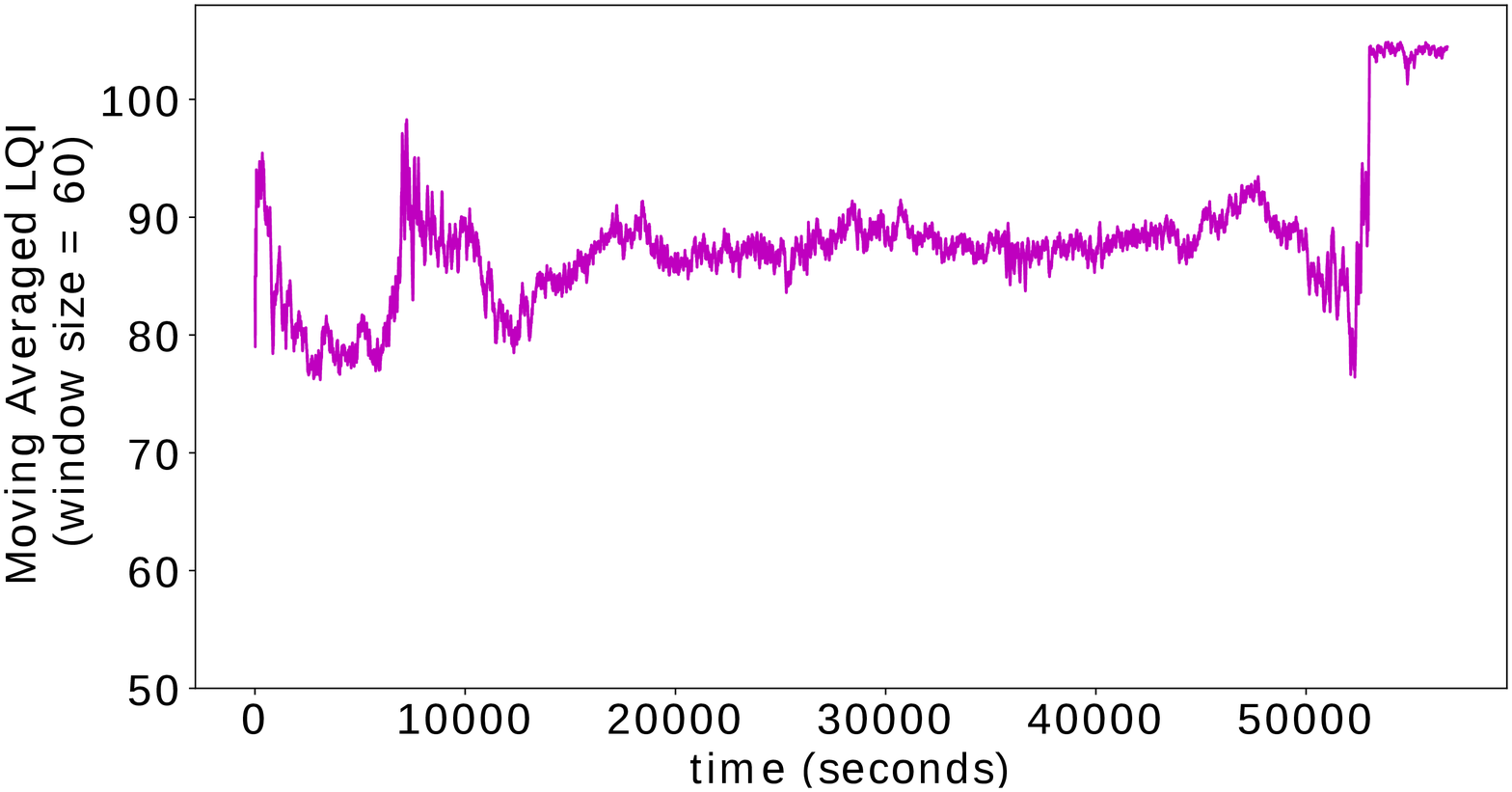}
\hfill
\label{Linklqi_node6}}
\caption{Wireless link dynamics in agricultural farm}
\label{linkdynamics_farm}
\end{figure*}

Each transmitter TelosB mote was programmed to send a packet once every one second and the receiving node calculated the RSSI and LQI 
for each packet it received. This experiment was continued for more than 10 hours 
and Fig. \ref{linkdynamics_farm} shows the dynamics of the three captured links. Table \ref{linkdynamicstable} details about the three links which were captured.

\begin{table}
\caption{Properties of the links analyzed}
\label{linkdynamicstable}
\centering
\begin{tabular}{|>{\centering\arraybackslash}p{3.5cm}|>{\centering\arraybackslash}p{1cm}|>{\centering\arraybackslash}p{1cm}|>{\centering\arraybackslash}p{1cm}|}
\hline
Parameter&Pair-1&Pair-2&Pair-3\\
\hline
802.15.4 wireless channel used for wireless communication &15&26&20\\
\hline
Total duration for which the link is captured &16hrs& 13hrs & 15.5hrs\\

\hline
Distance between the transmitter and receiver node &20m&97m&18.5m\\
\hline
Number of packet losses &69&0&5315\\
\hline
Number of packets sent &58842&46318&56715\\
\hline
\end{tabular}
\end{table}

\begin{figure*}
\centering
\subfloat[Graph 1]{\includegraphics[scale=0.5]{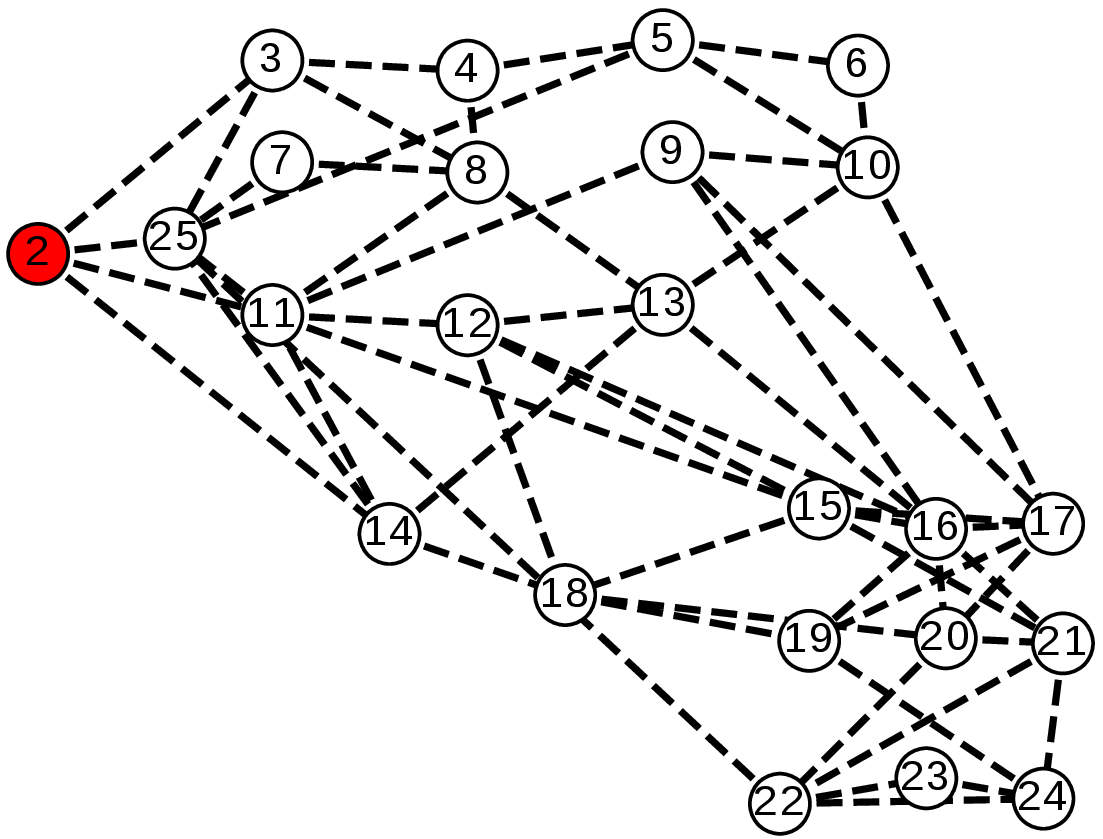}}
\hfil
\subfloat[Graph 2]{\includegraphics[scale=0.5]{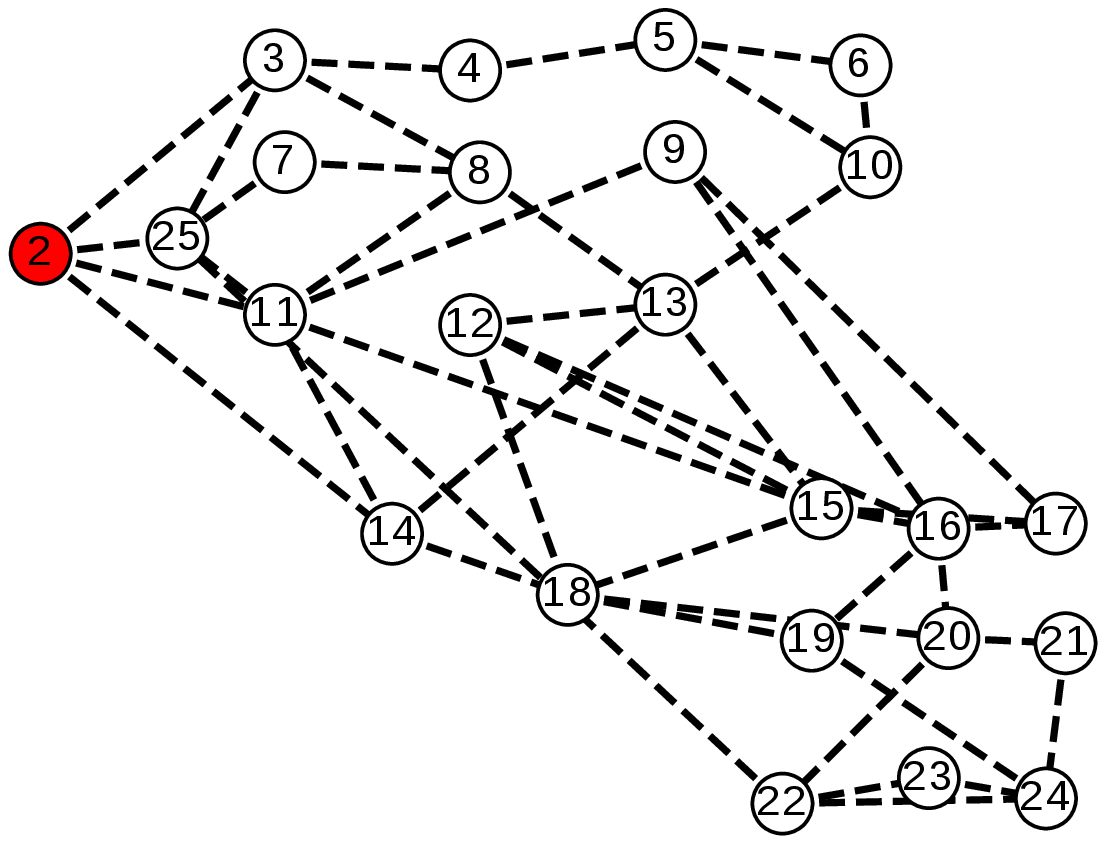}}
\hfil
\subfloat[Graph 3]{\includegraphics[scale=0.5]{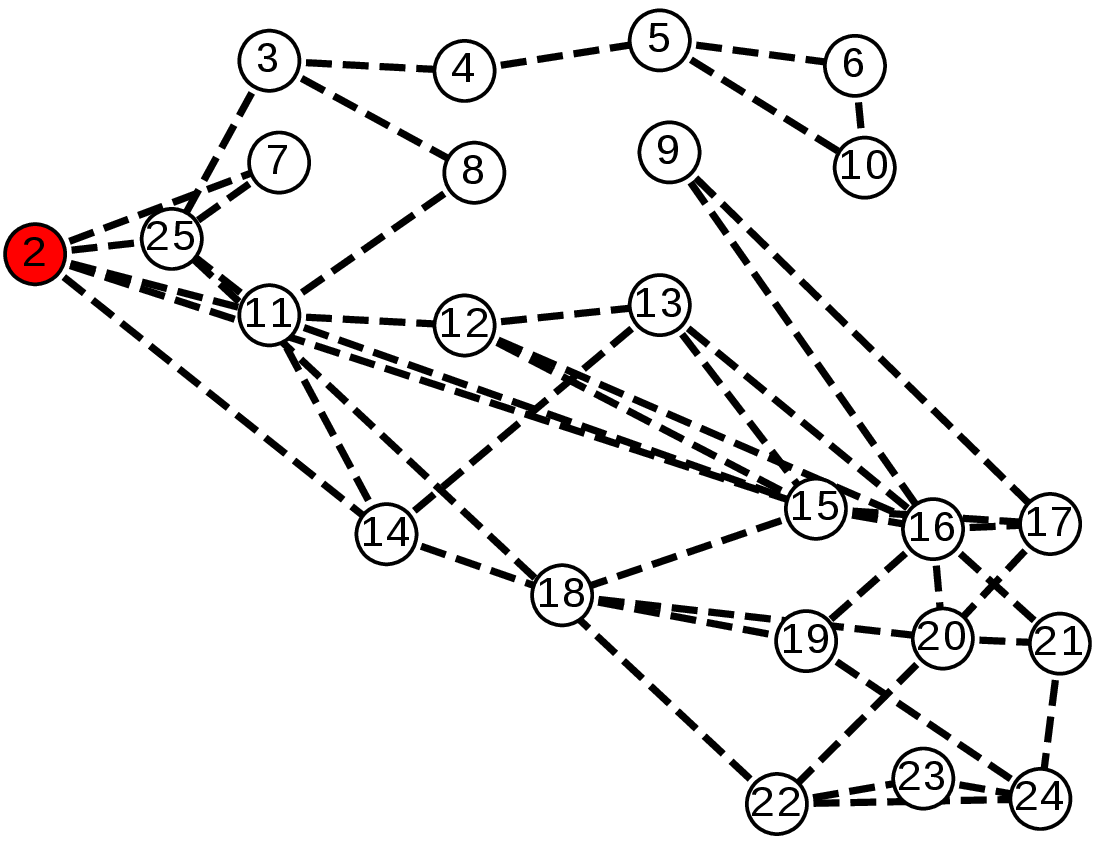}}
\hfil
\subfloat[Graph 4]{\includegraphics[scale=0.5]{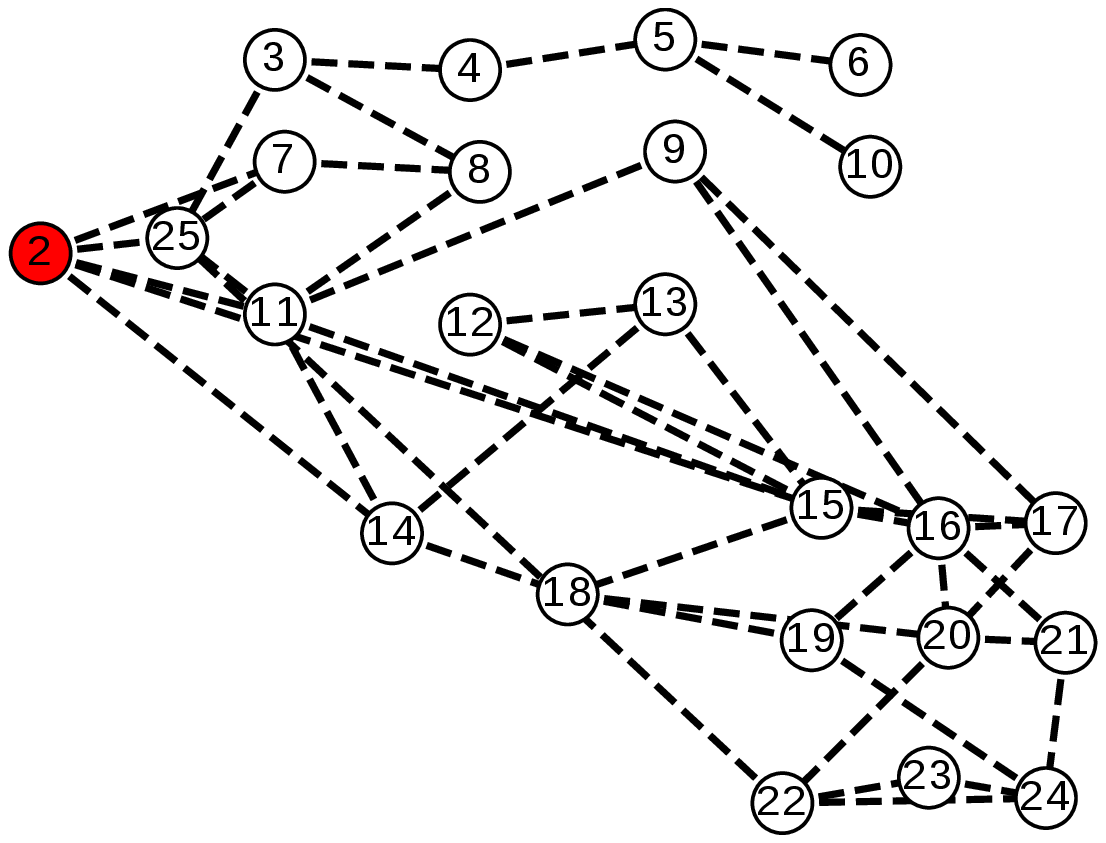}}
\hfil
\subfloat[Graph 5]{\includegraphics[scale=0.5]{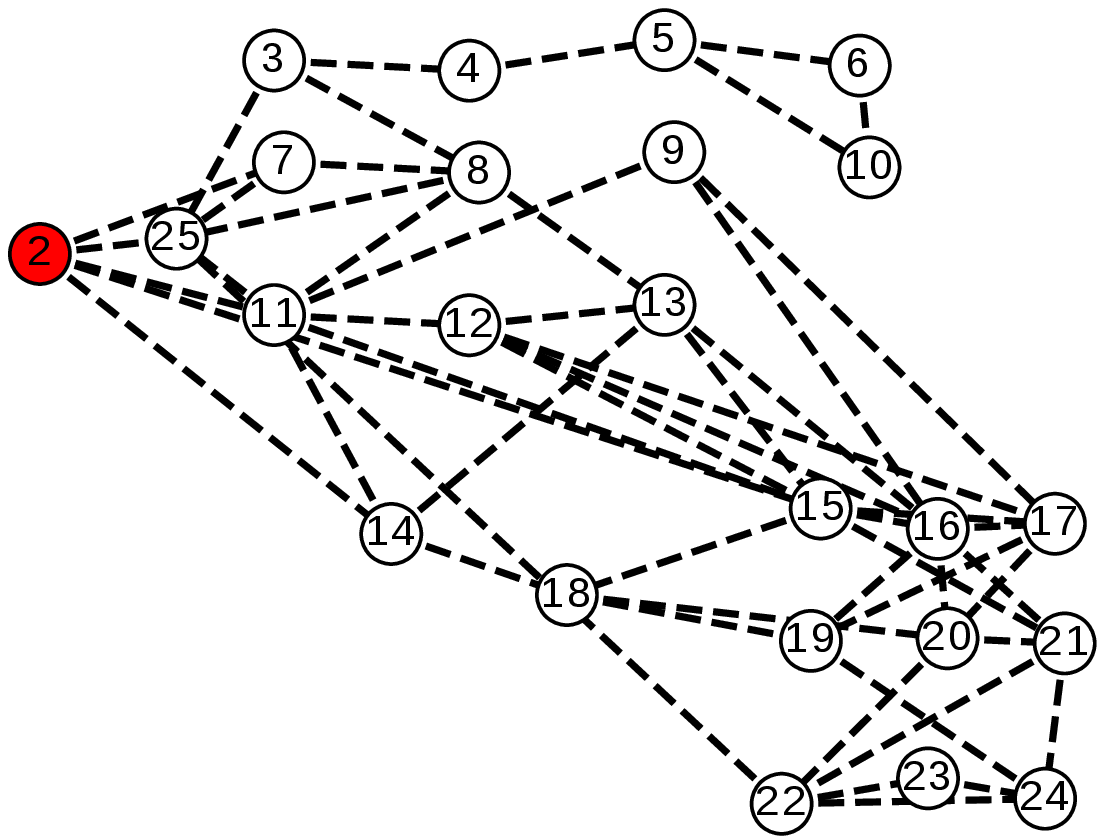}}
\caption{Consecutive graphs show that most of the edges are common}
\label{consecutive_graphs}
\end{figure*}

The major observations from the analysis of link dynamics are:
\begin{enumerate}
 \item {Links in agricultural fields are almost static with small variations, which can happen due to various environmental factors like
 temperature, humidity and mobility in the environment (e.g., movement of leaves/ plants due to wind).}
 \item{We know that RSSI decreases with increase in distance between the transmitter and the receiver. This does not imply that 
 all nearby nodes in an agricultural farm will have good wireless links.
 The radiation pattern of the antenna is not omni-directional, which has an impact on the 
 received signal strength in a particular direction \cite{sky2007ultra}. 
 It also depends upon the obstacles between the nodes.
 Fig. \ref{linkdynamics_farm} and Table \ref{linkdynamicstable} capture various packet transmissions that occurred in three different pairs of links in the farm.
 Pair-2 (Fig. \ref{Link_dynamics_node22}) does not have any packet losses, pair-1 (Fig. \ref{Link_dynamics_node15}) had intermittent packet 
 losses while pair-3 (Fig. \ref{Link_dynamics_node6}) had more packet losses. We now explain these trends. There were no obstacles between the transmitter
 and the receiver of the pair-2 link as they were located along two adjacent corners of the field and hence, had a good line of sight between them.
 There were a few maize plants whose canopy level was almost blocking the line of sight between the pair-1 devices. Pair-3 devices 
 were installed in a thick cultivation area and hence, many plants caused greater hindrance to wireless communication.
 Thus, in our experiment, we observed less packet losses for the farthest pair of nodes because of the uneven variations in the obstacles
 between the nodes (reflected in Table \ref{linkdynamicstable}).
 Assuming almost same obstacles between every pair of nodes is not a fair assumption even though the land has uniform cultivation.
 There can be cases where long distance nodes have a better link than nearby ones. 
  Even such links are static over a long period of time.}
 
 \item {We can set a threshold on the RSSI below which a link will have many packet losses even though communication can happen through that link.
 We set this limit as $-85$ dBm for our implementation.}
 \item {Packet losses may not always be reflected as variations in RSSI or LQI.}
 \item {Even though the enclosure boxes containing the TelosB modules were kept 1 feet above the plant canopy level, the wireless transmission range was 
 getting reduced. Using an external antenna mounted at a higher height would have helped to increase the effective received 
 signal strength at the receiver nodes.}
\end{enumerate}

\textit{The above experimental results suggest that wireless links in  agricultural farms are quite stable and hence, we can use a static tree for 
data collection over a long period of time. This eliminates the need for frequent rebuilding of the data collection tree and thus 
reduces the wastage of node's energy used for tree building.} RSSI thresholding and packet losses are incorporated in the network
architecture design in the neighbour discovery phase. A node $u$ identifies node $v$ as its neighbour only if it receives a fraction of the total
neighbour discovery messages sent by node $v$ above a certain threshold.

\subsubsection*{How different are the graphs over consecutive tree builds?}

To confirm the above findings that the links are almost static, we have conducted
some experiments where data collection happens once every 10 minutes and tree
building occurs once every 2 hrs. Fig. \ref{consecutive_graphs} shows five consecutive graphs which
were collected at the sink node as part of the tree building phase. Fig. \ref{consecutive_graphs} along with Table \ref{consectivegraph_comparison}
shows that most of the edges in the graphs are common to all the five graphs.

\begin{table}[h]
\caption{Analysis of consecutive graphs}
\label{consectivegraph_comparison}
\centering
\scalebox{0.85}{
\begin{tabular}{|>{\centering\arraybackslash}p{2.5cm}|>{\centering\arraybackslash}p{0.95cm}|>{\centering\arraybackslash}p{0.95cm}|>{\centering\arraybackslash}p{0.95cm}|>{\centering\arraybackslash}p{0.95cm}|>{\centering\arraybackslash}p{0.95cm}|}
\hline
Parameter&Graph1&Graph2&Graph3&Graph4&Graph5\\
\hline
Total no. of edges& 56 & 45 & 45 & 44 & 53\\
\hline
No of edges which are newly added compared to the previous graph & - & 1 & 6 & 2 & 9\\
\hline
No of edges which got removed compared to the previous graph & - & 12 & 6 & 3 & 0\\
\hline
No of  edges that remained same as compared to the previous graph & - & 44 & 39 & 42 & 44\\
\hline
No of  edges common to all the graphs & \multicolumn{5}{c|}{37}\\
\hline
\end{tabular}}
\end{table}

\subsection{Our time synchronization strategies are accurate enough for this application}
\label{timesync_experiments}
We are using the simple time synchronization method which is detailed in Section \ref{timesync} to synchronize all the nodes in the data collection 
tree to the sink node. This section discusses the results of our synchronization mechanisms for different data collection intervals
for both single hop and multihop networks. 

\begin{figure}
\centering
\subfloat[]{\label{timesync_singlehop}\includegraphics[scale=0.6]{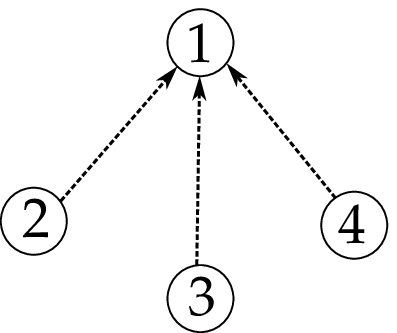}}
\qquad
\subfloat[]{\label{timesync_singlehop_errors}\includegraphics[scale=0.28]{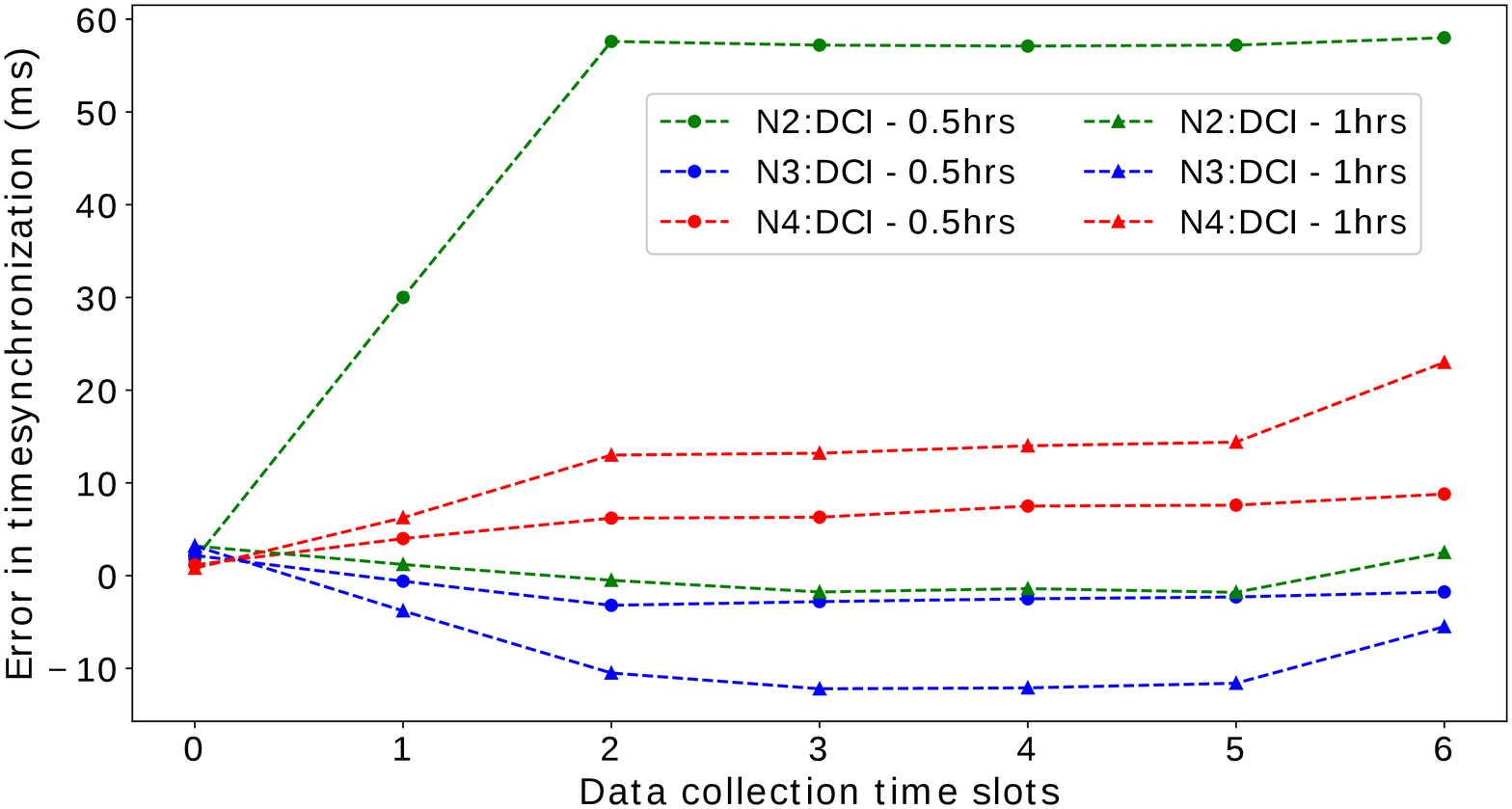}}
\caption{A singlehop network and its synchronization errors} 
\label{synchronizationerrors_singlehop}
\end{figure}

Fig. \ref{synchronizationerrors_singlehop} shows  a single hop network
and the associated synchronization errors where nodes 2, 3, 4 are the children of the sink node 1. We capture the synchronization
error for two different data collection intervals (DCI)-- half an hour and one hour intervals. Data collection time slot zero refers to the 
trigger point (detailed in Section \ref{timesync}), data collection time slot 1 refers to the first data collection time slot and so on.
\footnote{When the data collection timer fires, an I/O pin of the TelosB mote in the respective node
is enabled and the time difference of each node with respect to the sink's test I/O pin is calculated by using a DPO to which all these test 
I/O pins are connected.}

\begin{figure}
\centering
\subfloat[]{\label{timesync_multihop}\includegraphics[scale=0.6]{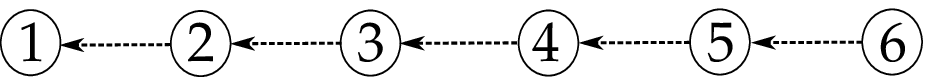}}
\qquad
\subfloat[]{\label{timesync_multihop_errors}\includegraphics[scale=0.28]{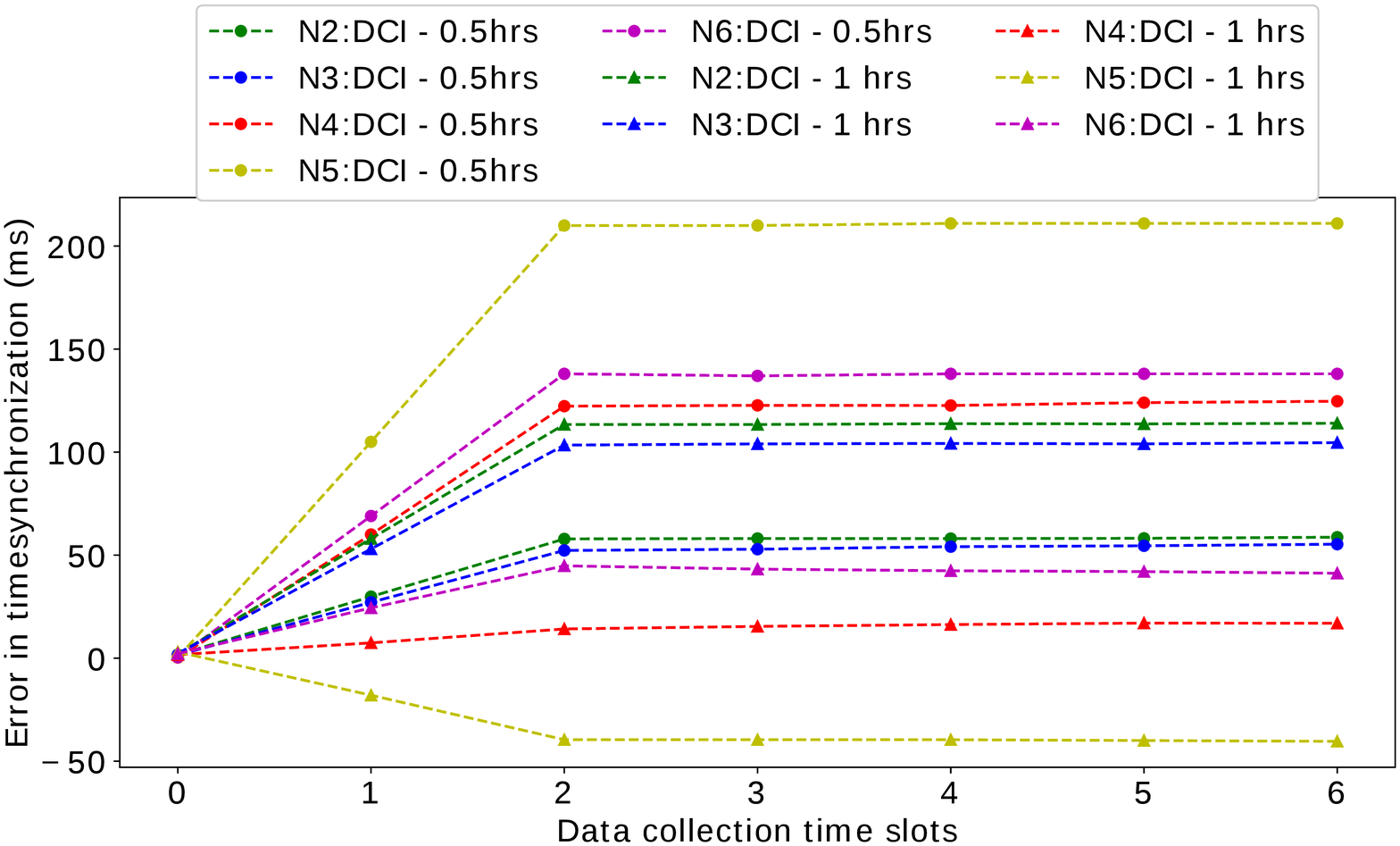}}
\caption{A multihop network and its synchronization errors} 
\label{synchronizationerrors_multihop}
\end{figure}

Fig. \ref{synchronizationerrors_multihop} shows the synchronization errors for a multihop scenario.
For this test network, we have manually programmed the parent of each node as shown in the figure.

The nodes can have positive or negative errors in time synchronization resulting in an early wake-up or late wake-up with
respect to the expected wake-up time. The synchronization error increases in the initial time slots and becomes almost constant 
after a few time slots . This is because each node needs to have enough reference points which are well spaced in time to find
its clock skew correctly.
\footnote{The time synchronization errors introduced because of the clock skew calculated from a small number of reference points can result 
in random errors and hence, we cannot guarantee that the nodes farther from the sink node will have larger errors. The same applies to 
time synchronization errors with different data collection intervals.}
During a data collection time slot each node gets one reference point from its parent in the form of a sleep message
as detailed in Section \ref{timesync}. Each node could find its skew within the initial two or three time slots after which the 
error remains almost constant in the upcoming time slots. \textit{These errors are in the range of milliseconds and for our monitoring application, 
we believe that we can tolerate these error levels instead of employing a complicated time synchronization protocol.} 

\subsection{Relation between battery voltage and remaining capacity}
\label{BVtoRC}
We use lithium ion batteries in our sensor nodes as the energy source and they are charged using solar energy. Improving the lifetime of the nodes
is one of the major design criteria in sensor network applications. This can be achieved by utilizing the nodes with higher energy to relay 
more data packets. In some prior works, \emph{e.g.}, \cite {barrenetxea2008sensorscope}, battery voltage is used as a measure to identify the nodes
with higher energy. In this paper, we use remaining capacity of the battery as a measure to identify the nodes with higher remaining energy 
(see Section \ref{edgewtcalculation}) since the remaining battery capacity more directly corresponds to the remaining energy than the battery voltage. 

Current integration and voltage based measurements are the two general methods used to find the remaining capacity of a lithium-ion 
battery \cite{barsukov2004challenges}. In current integration technique, the charging and discharging currents are continuously integrated
over time to calculate the remaining capacity of the battery. Voltage based remaining capacity calculation is valid only if the load is very low
during the measurement. Our sensor nodes use  Li-ion batteries of capacity 2200mAh and the load is quite low ($<$ 10mA) during the battery 
voltage measurement, since all the major current consuming modules (sensors and the wireless radio) are in off state. 
Therefore, we use the dependency of the state of charge (SOC) on the open circuit voltage (OCV) (``SOC - OCV relation'')
to find the remaining capacity of the node based on its current battery voltage \cite{barsukov2004challenges}. There are several fuel gas-gauge ICs available from various semiconductor 
manufacturers which can be used for this; however, we do not prefer them since they add to the node's cost as well as complexity. Instead, we use the scheme described below. 

\subsubsection{SOC-OCV relation}

The relation between the battery voltage and remaining capacity is obtained with the help of discharge profiles of lithium ion cells.
Generally the datasheet of Li-ion cells do not provide the discharge profile for very light loads \cite{lithiumion_datasheet}. Hence, we 
found out the light load discharge profiles experimentally.

\begin{figure}
\centering
\subfloat[Discharge profile of lithium ion cells of 2200mAh at 100mA load]{\label{discharge_profiles}\includegraphics[scale=0.29]{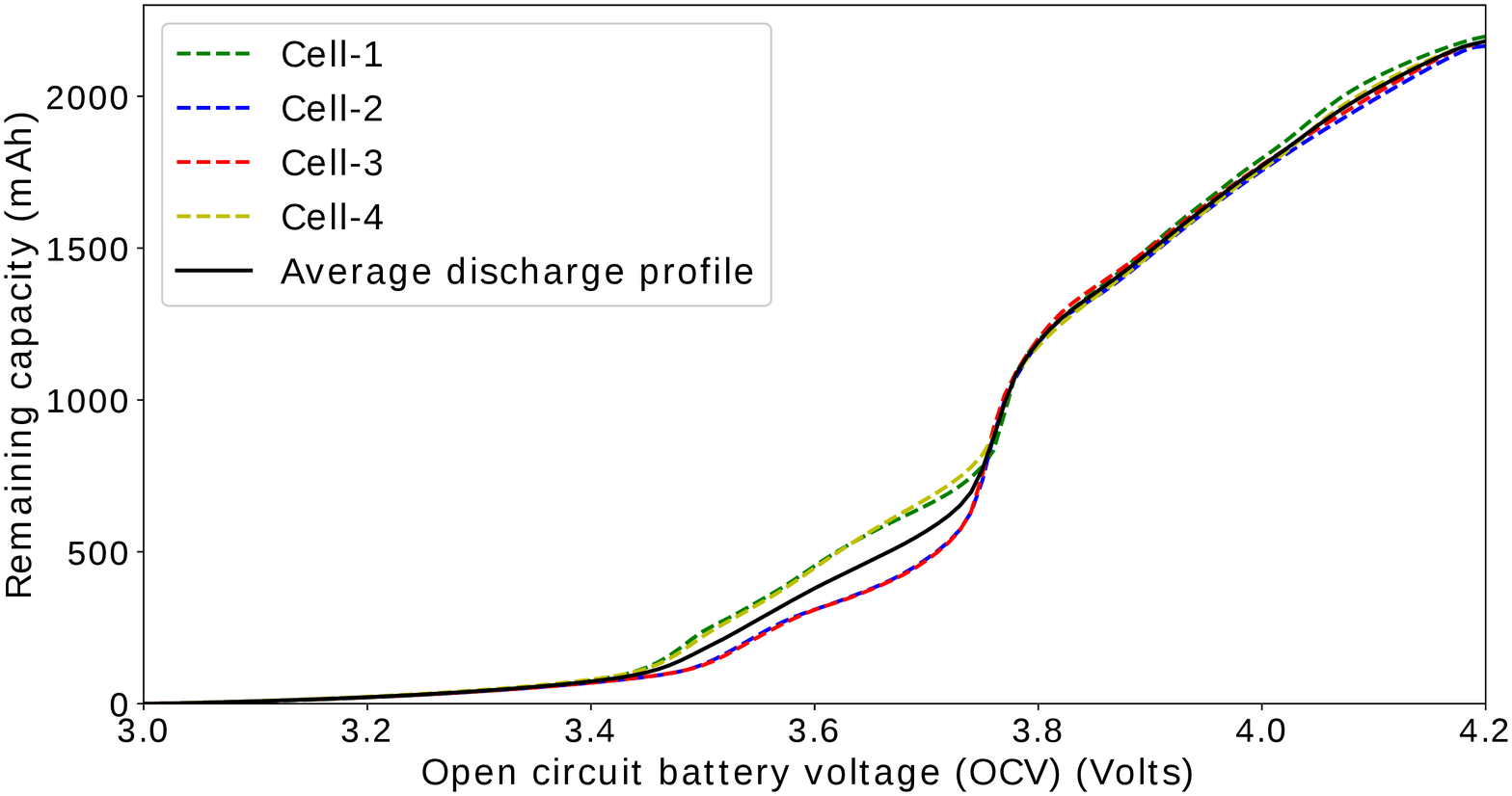}}
\qquad
\subfloat[Verification of the relation between SOC and OCV]{\label{dischargepoints}\includegraphics[scale=0.29]{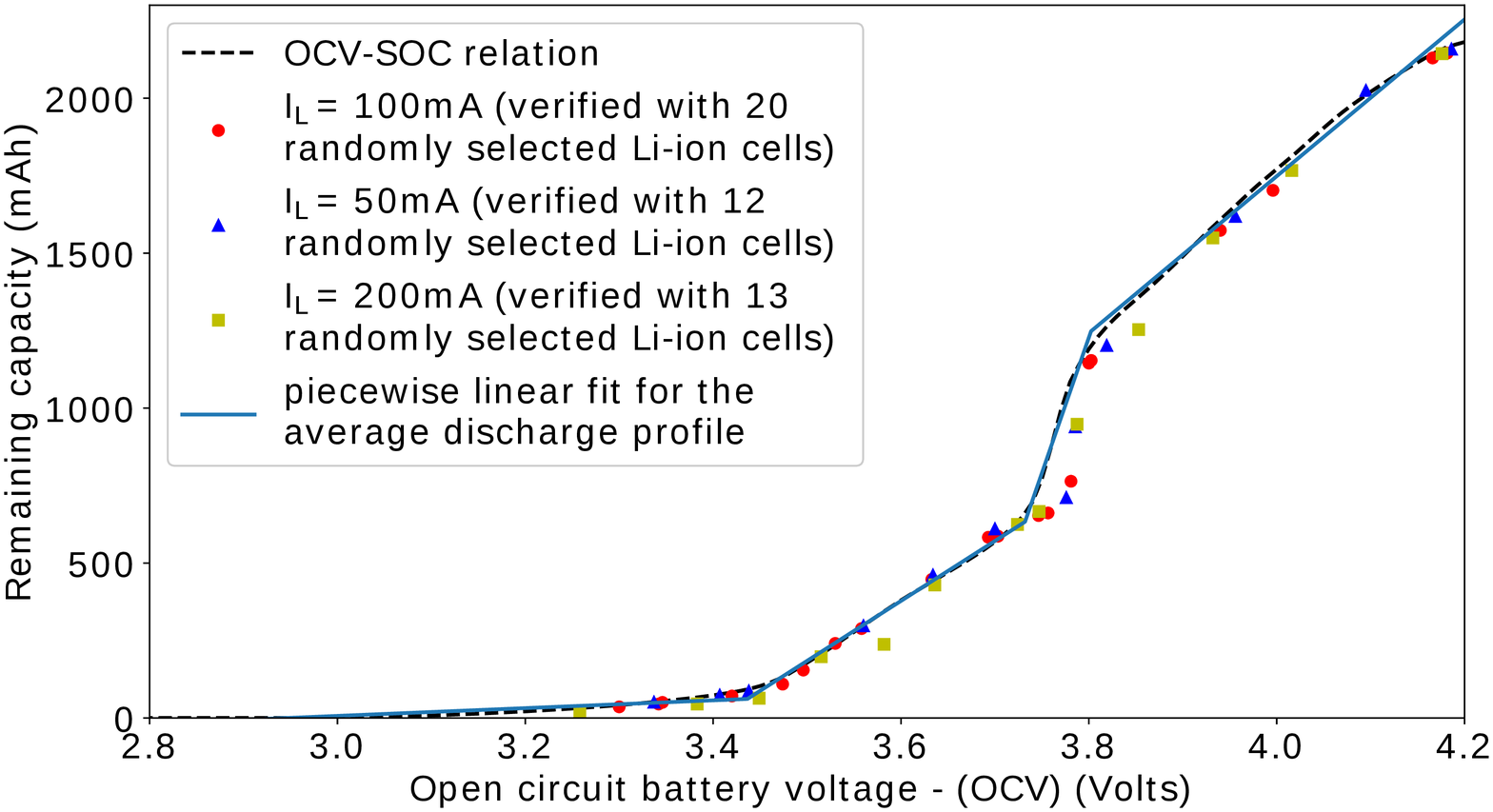}}
\caption{Relation between battery voltage and remaining capacity } 
\label{discharge_profiles_li_ion}
\end{figure}

Fig. \ref{discharge_profiles} shows the discharge profile of four Li-ion cells and their averaged profile which can be used as the 
SOC-OCV relation to find a node's remaining capacity from its current battery voltage.
The cells are discharged at constant 100mA load, which can be considered as a light load and the open circuit voltage of the battery 
is calculated by using the relation: 
\begin{equation}
\centering
 V_{oc} = V + I_L \times R_0
\end{equation}
where V is the measured voltage across the battery while loading a current of 100mA ($I_L$) and $R_0$ is the battery resistance at 100mA load.
We have used 0.15 ohm as $R_0$ for generating the SOC-OCV relation which is found out experimentally
for the lithium ion cells used for the above profiling. 

The generated  SOC-OCV relation is verified by discharging several Li-ion cells (which are not used for the SOC-OCV relation generation)
which have different SOC. Fig. \ref{dischargepoints} shows how closely we could estimate the remaining capacity of a node just by measuring its battery voltage.
As shown in the figure, we have discharged different Li-ion cells at different loads to see how good the generated SOC-OCV relation is. 
It is observed that the SOC-OCV relation can be considered as a combination of four piecewise-linear fits. This helps us to embed this
relation in an efficient manner in the microcontroller instead of representing it as a look-up table.
\section{Field Experiments for Data Collection}
\label {field_experiments}
Here, we detail the set of experiments conducted in the maize field to collect the sensor data from different sensor nodes
and report the various activities occurring in the network by analyzing the packet transmissions captured in the snooper nodes.
Each node reports
soil moisture, soil temperature, atmospheric temperature, relative humidity and its battery voltage to the sink node, along
the synchronized data collection tree as explained in Section \ref{Network_architecture}. We report three different test scenarios
of data collection with time interval between two consecutive data collection time slots as 10 minutes, 1 hour and 3 hours.
Each test case is evaluated for three continuous data collection rounds where one round involves data collection tree formation followed by
periodic data collection through it.

\begin{figure*}
\centering

\subfloat[Graph 1]{\includegraphics[scale=0.47]{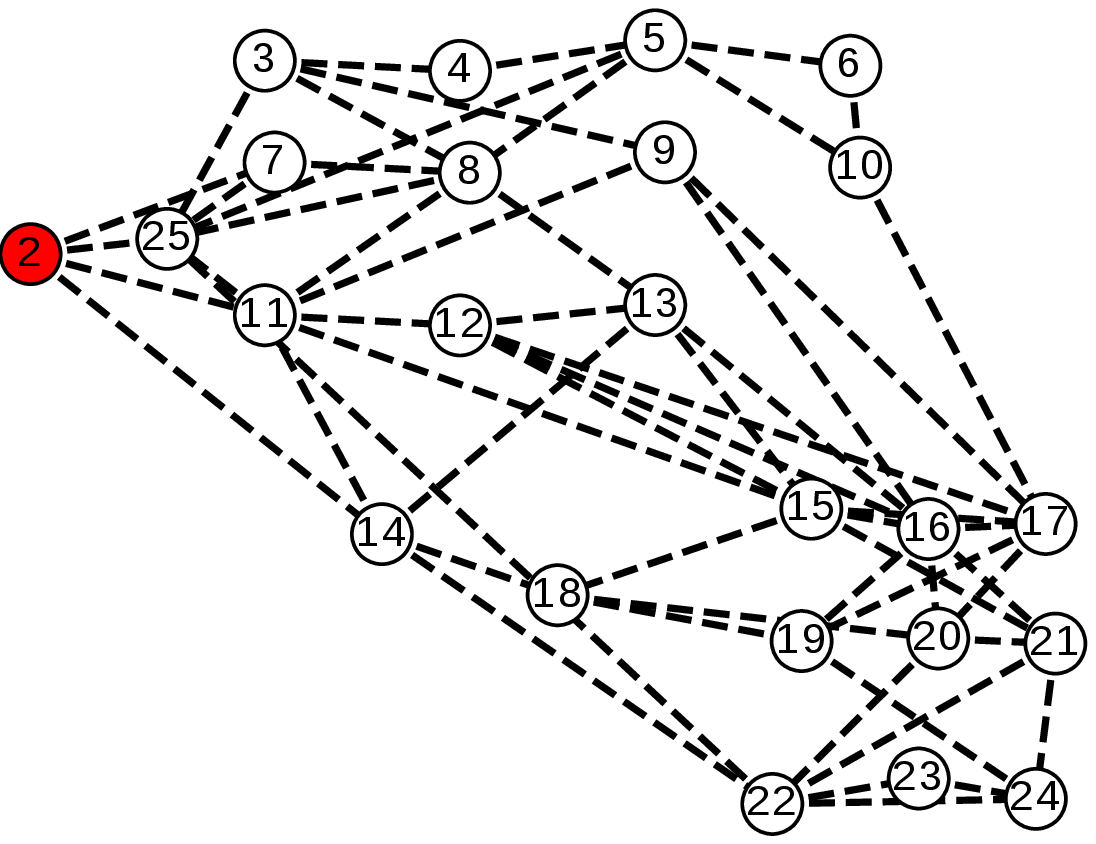}}
\hfil
\subfloat[Graph 2]{\includegraphics[scale=0.47]{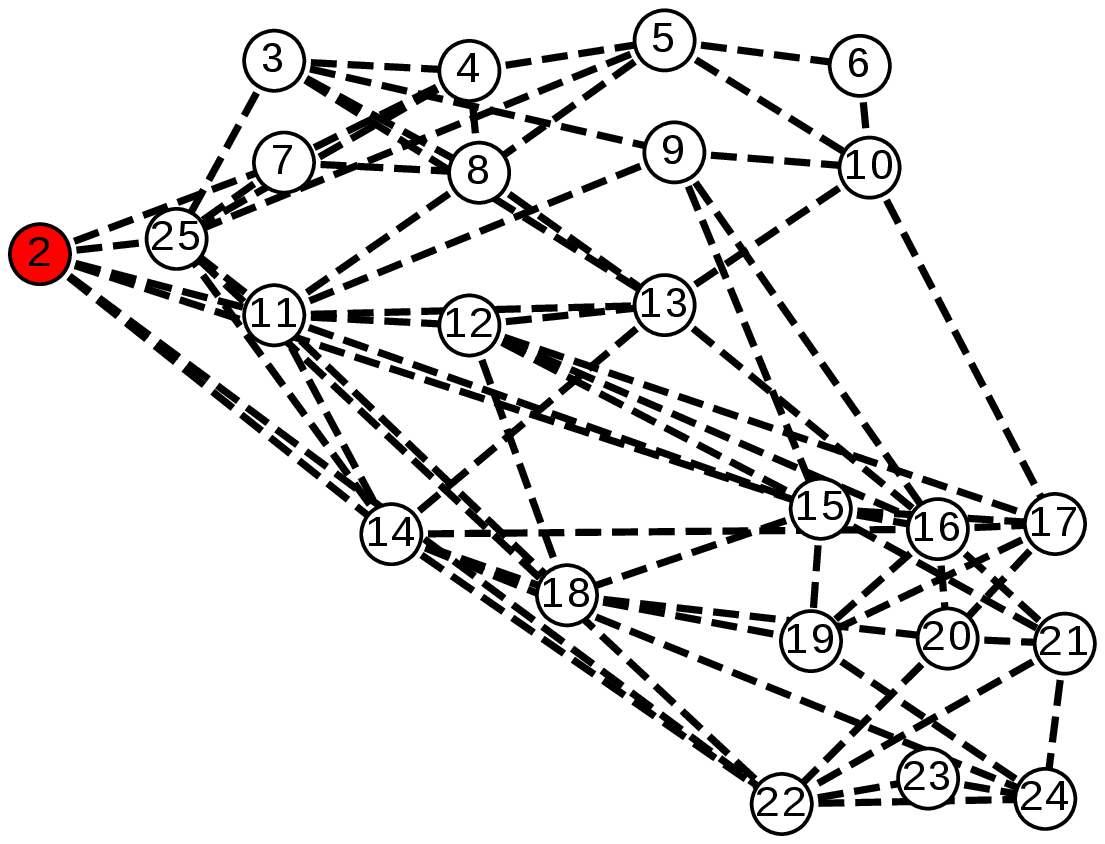}}
\hfil
\subfloat[Graph 3]{\includegraphics[scale=0.47]{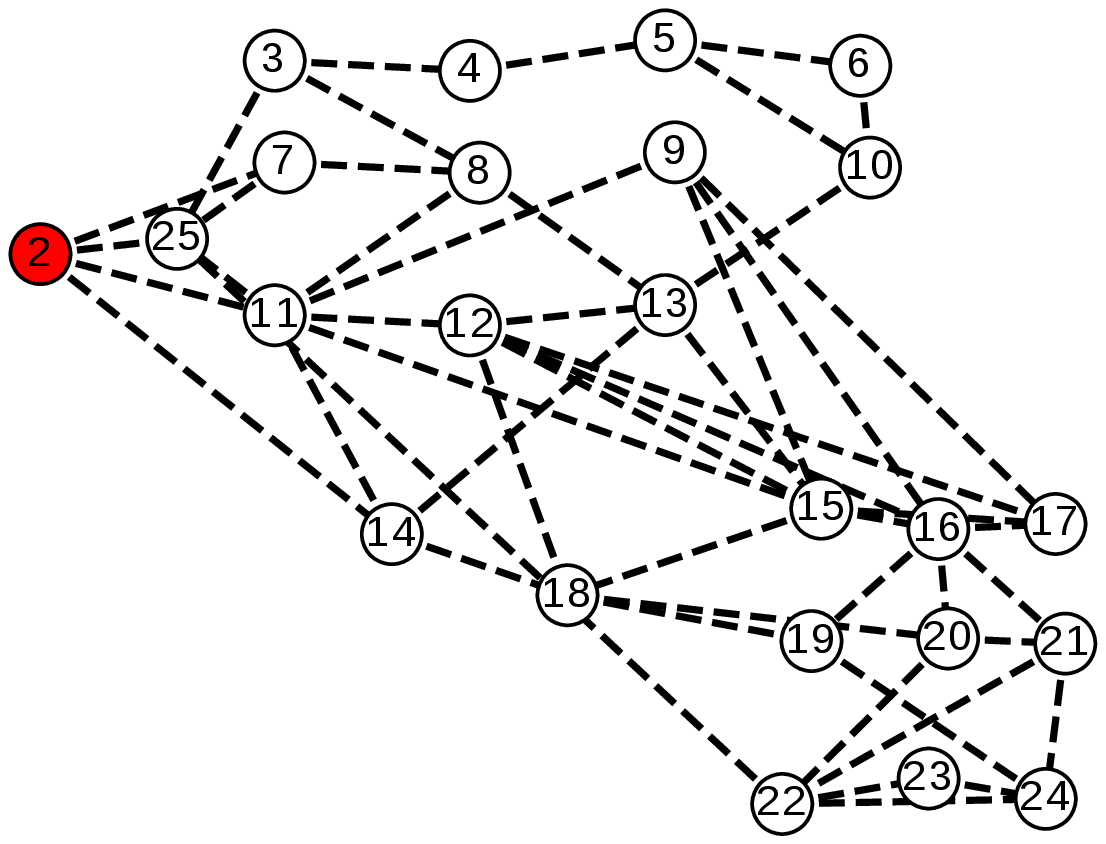}}
\hfil
\subfloat[Data collection tree formed from graph 1]{\label{dct_10mnt_tree1}\includegraphics[scale=0.26]{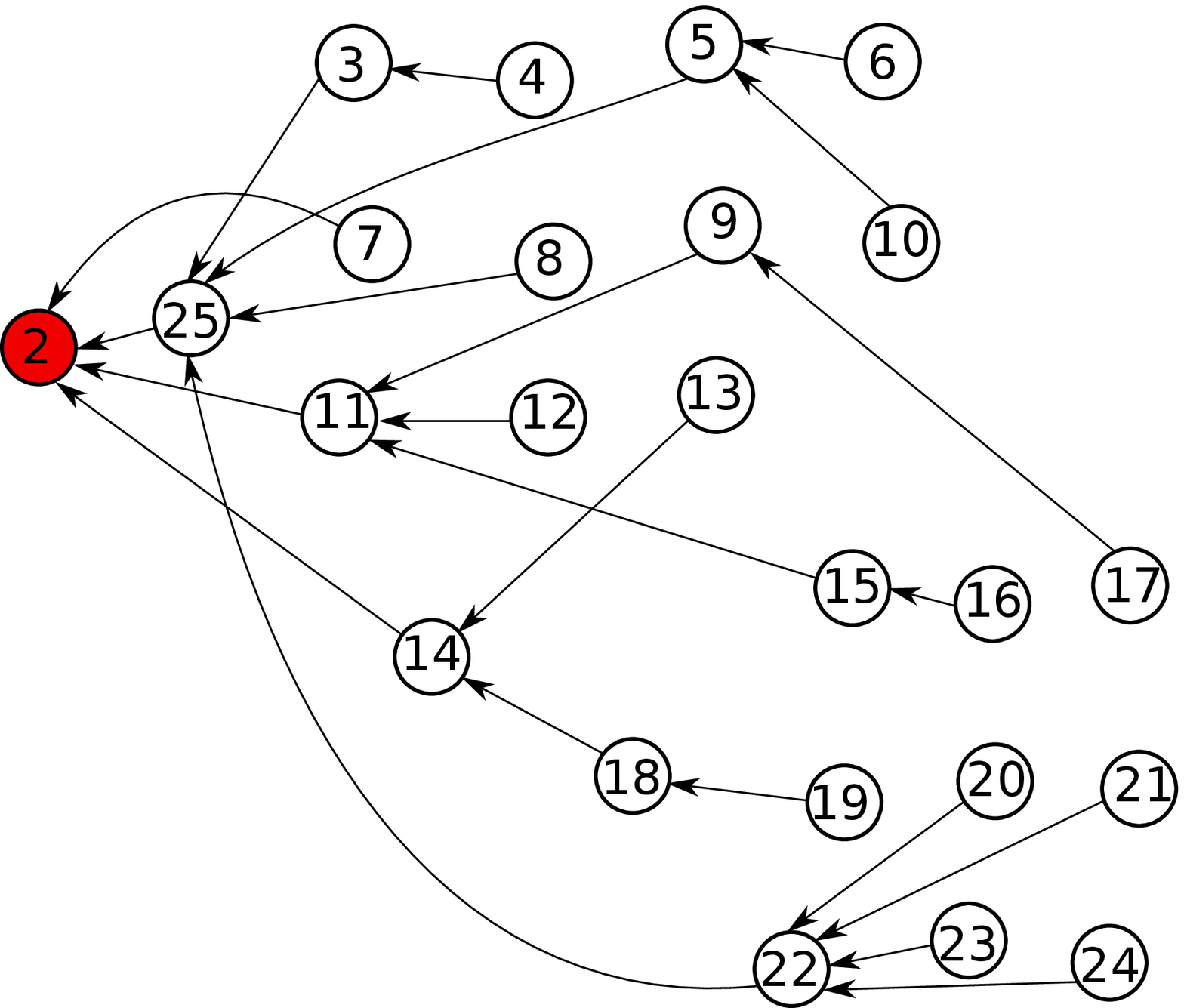}}
\hfil
\subfloat[Data collection tree formed from graph 2]{\includegraphics[scale=0.26]{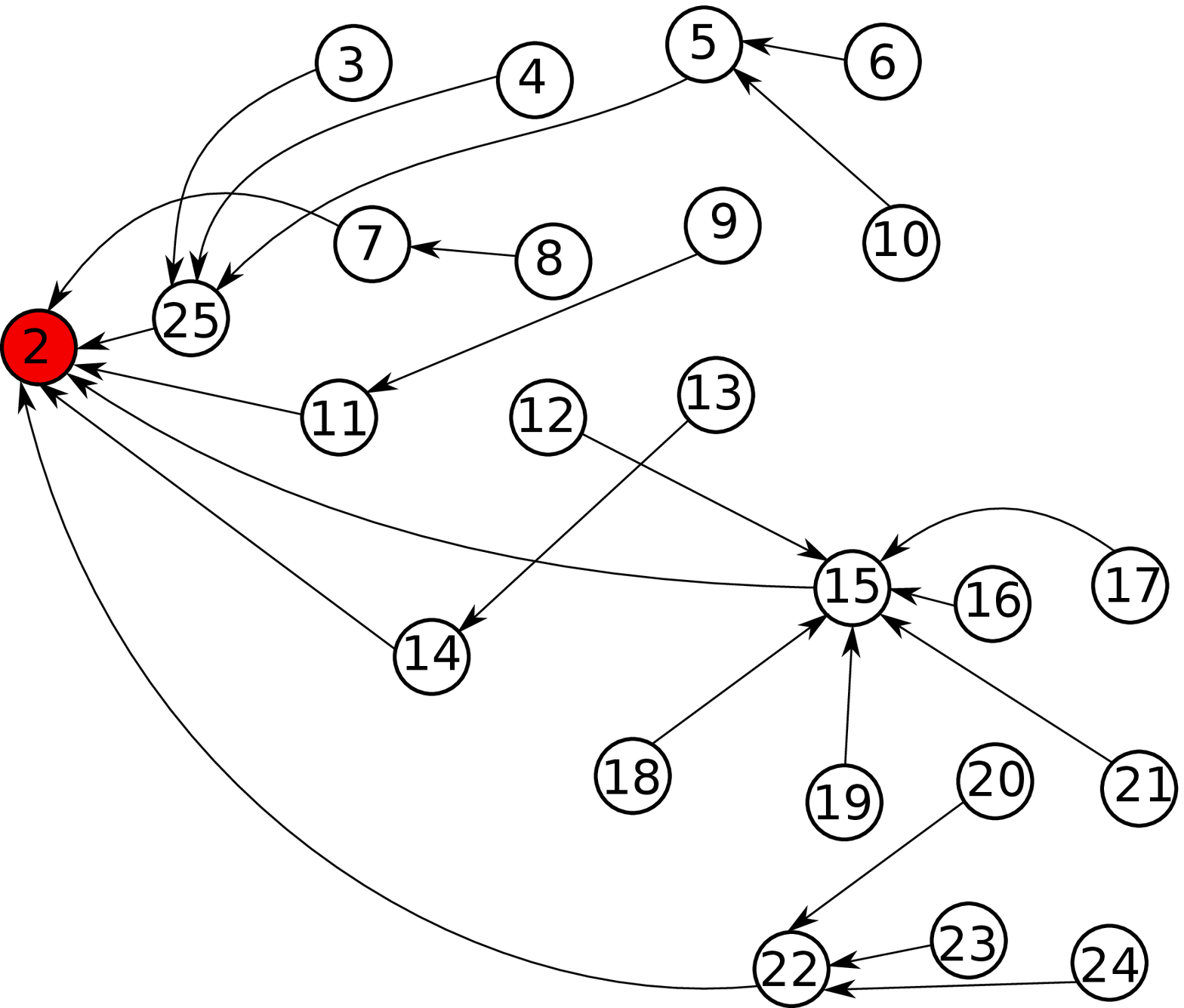}}
\hfil
\subfloat[Data collection tree formed from graph 3]{\includegraphics[scale=0.26]{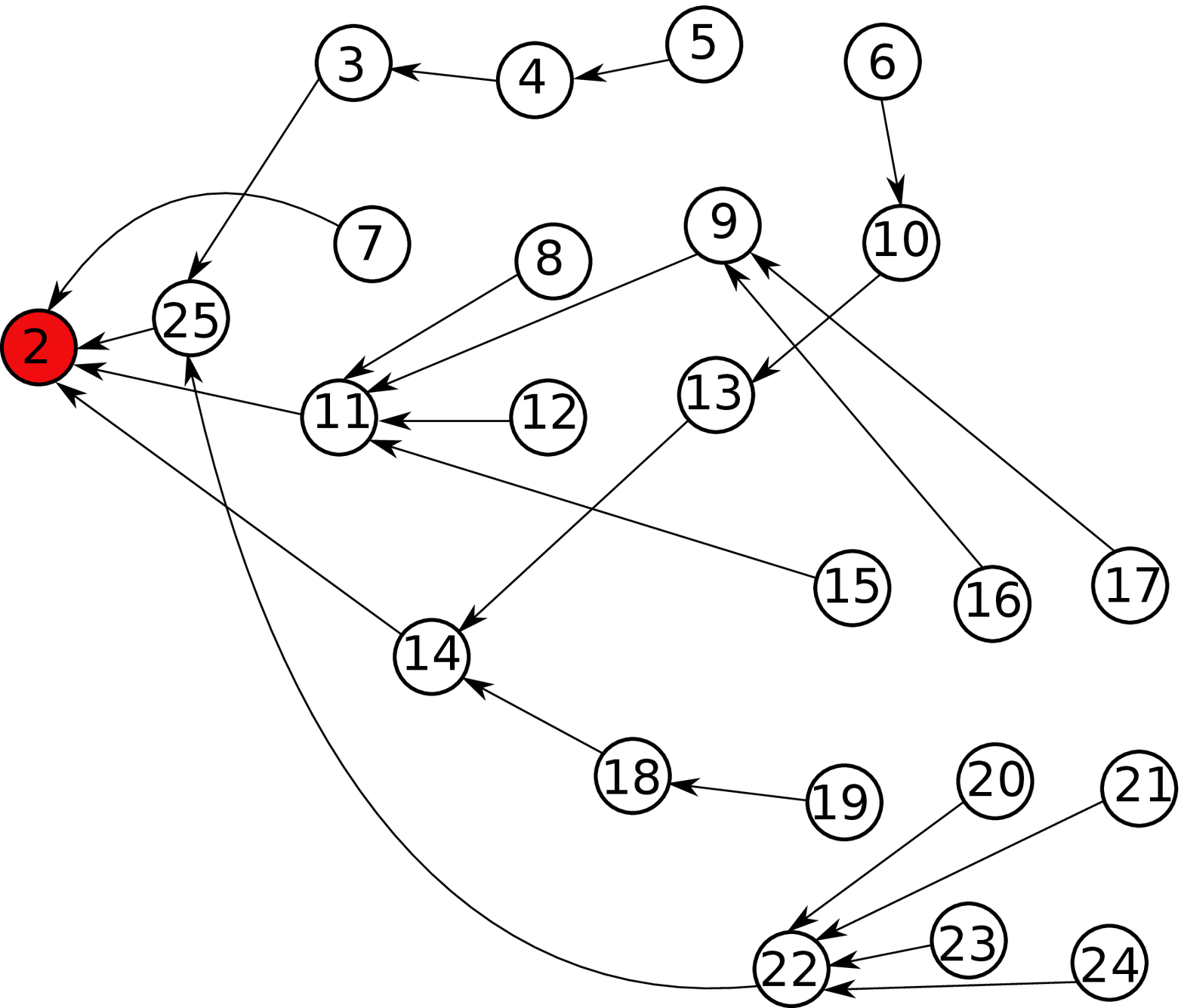}}

\hfil
\subfloat[Data collection round 1]{\includegraphics[scale=0.2]{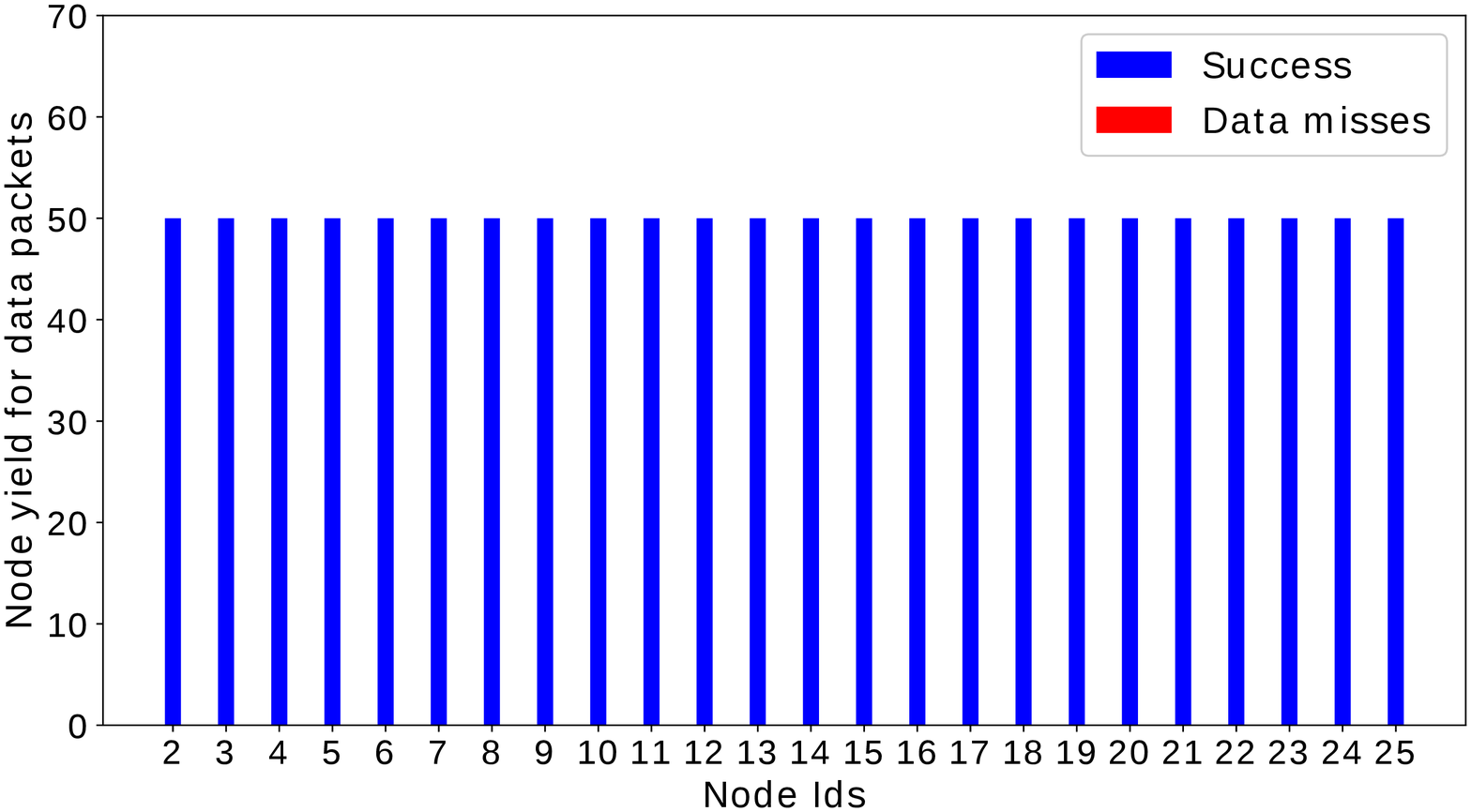}}
\hfil
\subfloat[Data collection round 2]{\label{nodeyield_10mnt_tree2}\includegraphics[scale=0.2]{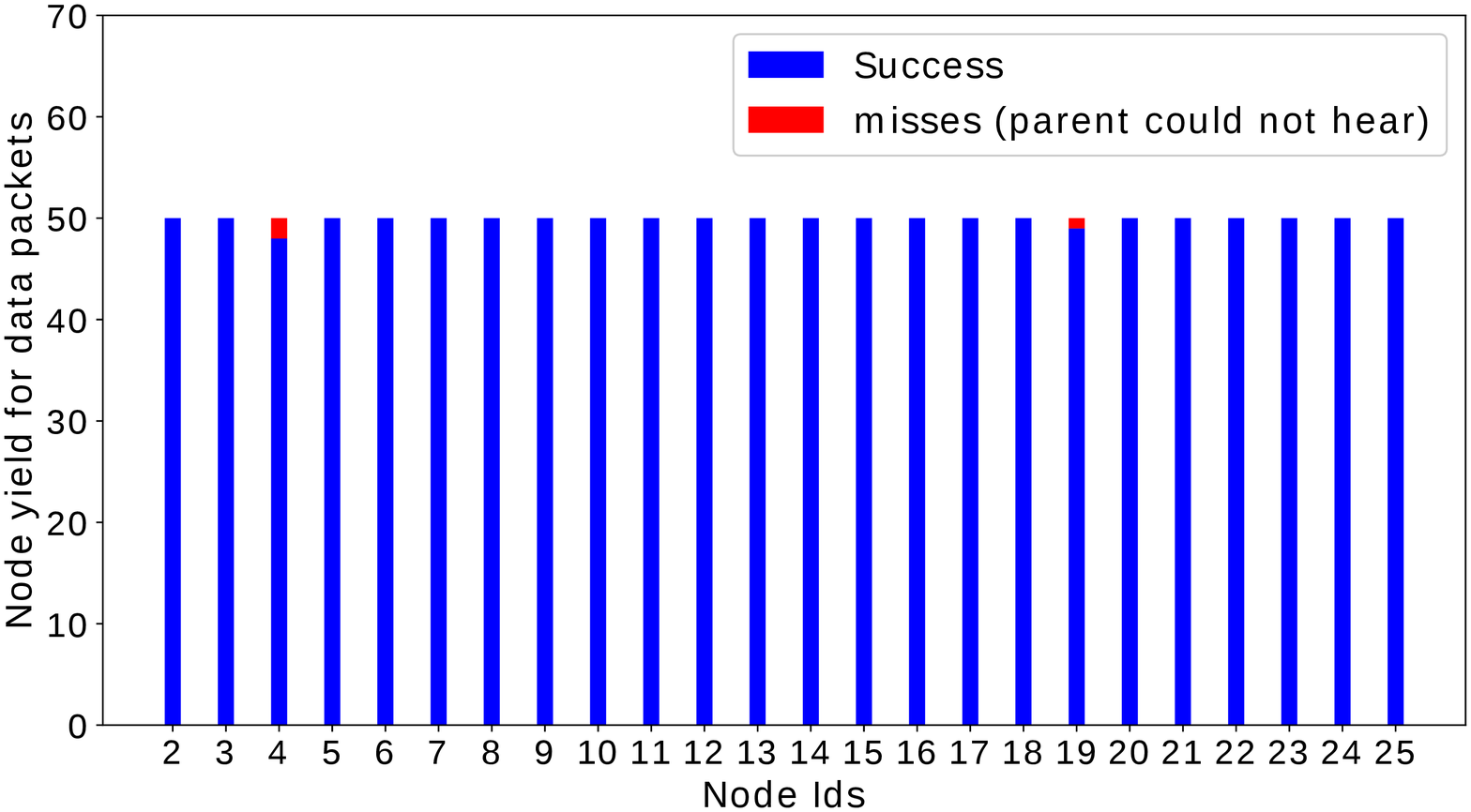}}
\hfil
\subfloat[Data collection round 3]{\label{nodeyield_10mnt_tree3}\includegraphics[scale=0.2]{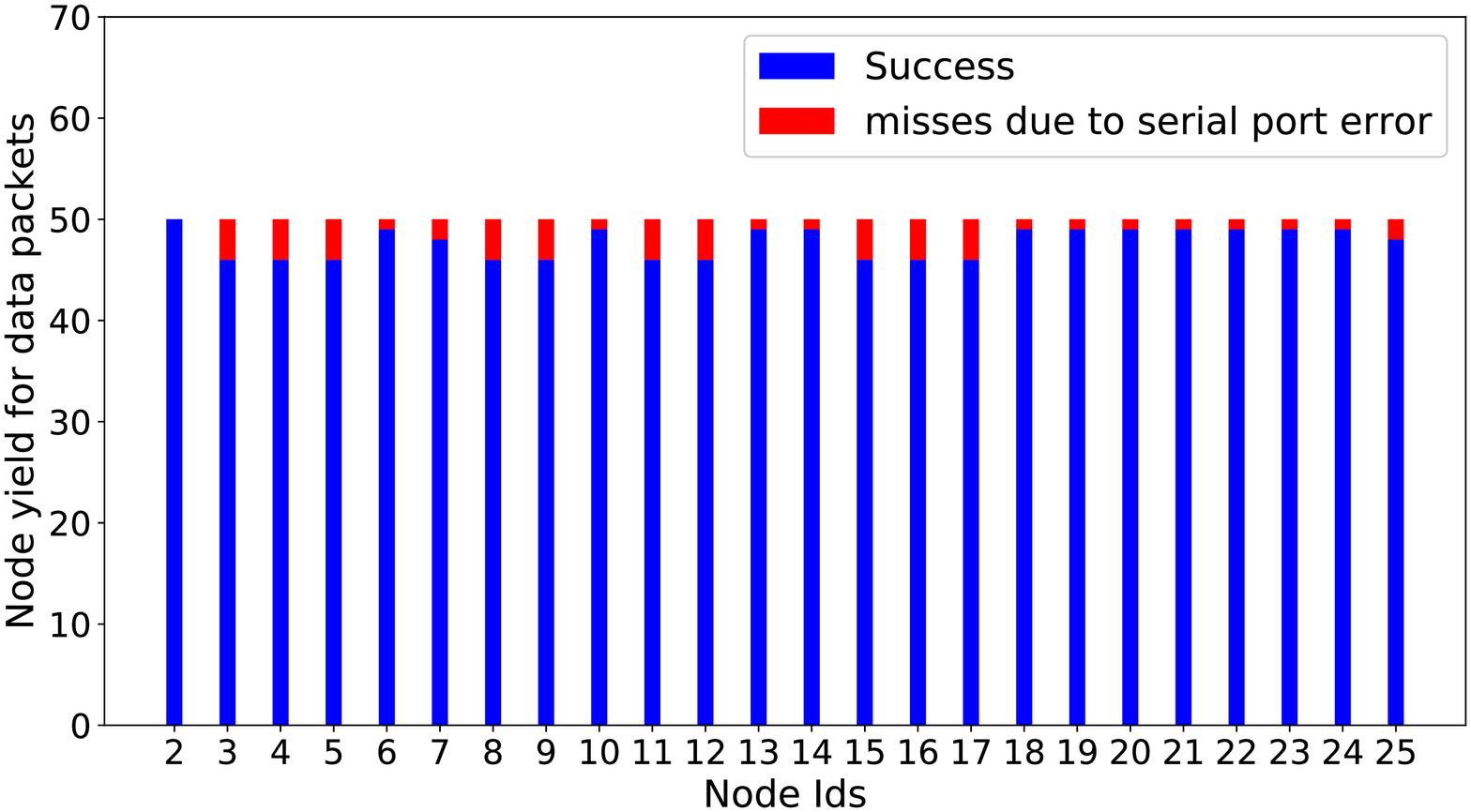}}

\caption{Consecutive graphs and respective data collection trees and the node yields for test case 1. Note that there was a temporary loose connection in the USB cable which is used to connect the sink node to the laptop for data logging 
and this is reflected as ``serial port error'' in Fig. \ref{nodeyield_10mnt_tree3}.}
\label{consecutive_graphs_10mnt}
\end{figure*}

\subsection{Test case 1}

Once all the nodes in the network are turned ON, with sink node being the last one to be turned ON, a synchronized data collection tree is formed
as explained in Section \ref{Network_architecture}.
In this test case, once the data collection tree is constructed, each node reports its data to its parent node once in every ten minutes.
The data collection is continued for consecutive 50 time slots, followed by the tree reconstruction procedure, initiated by the sink node.
Thus a new data collection tree is constructed once in every 8 hours ($50 \times 10 = 500$ minutes $= 8$ hours approx).

Fig. \ref{consecutive_graphs_10mnt} shows the network graphs collected at the sink node during the consecutive tree construction
phases and the respective data collection tree in each round, through which the data is collected. It also shows
the node yields of each node in the network. Node yield of a node represents the number of time slots for which its sensor information
is reported successfully at the sink node through multi-hop communication along the data collection tree. Ideally each node's data should 
reach at the sink node in each time slot resulting in no net data loss.

\begin{figure*}
\centering{
\subfloat[packet transmissions in round 1]{\label{combinedtree1_10mnt}\includegraphics[scale = 0.55]{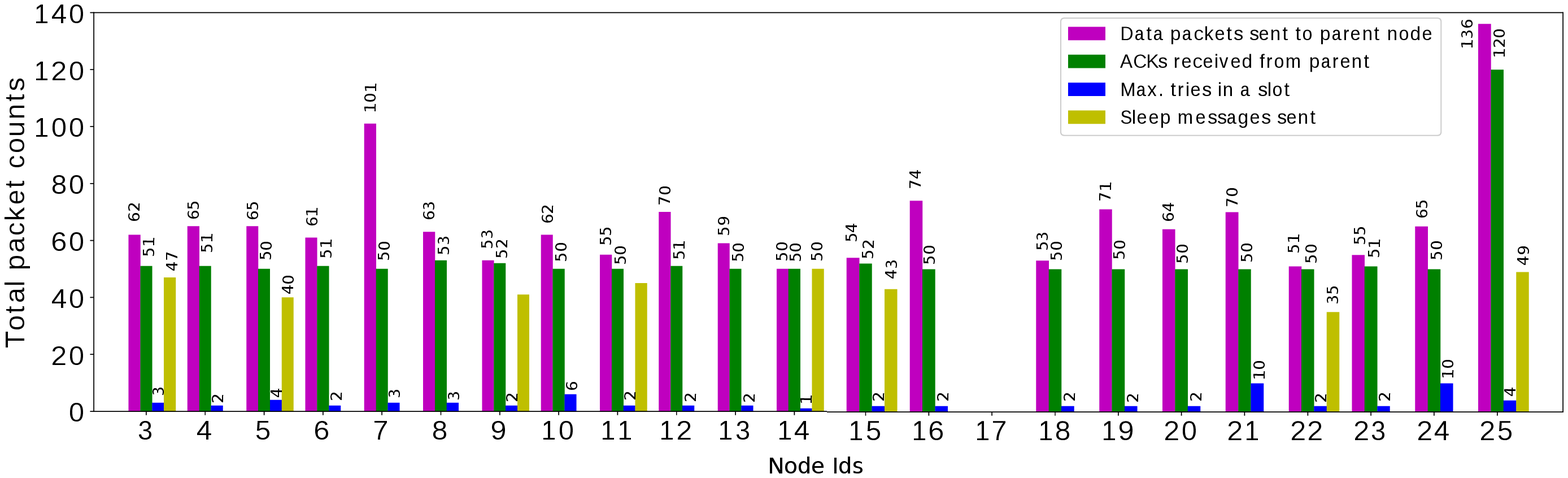}}
\qquad
\subfloat[packet transmissions in round 2]{\label{combinedtree2_10mnt}\includegraphics[scale = 0.53]{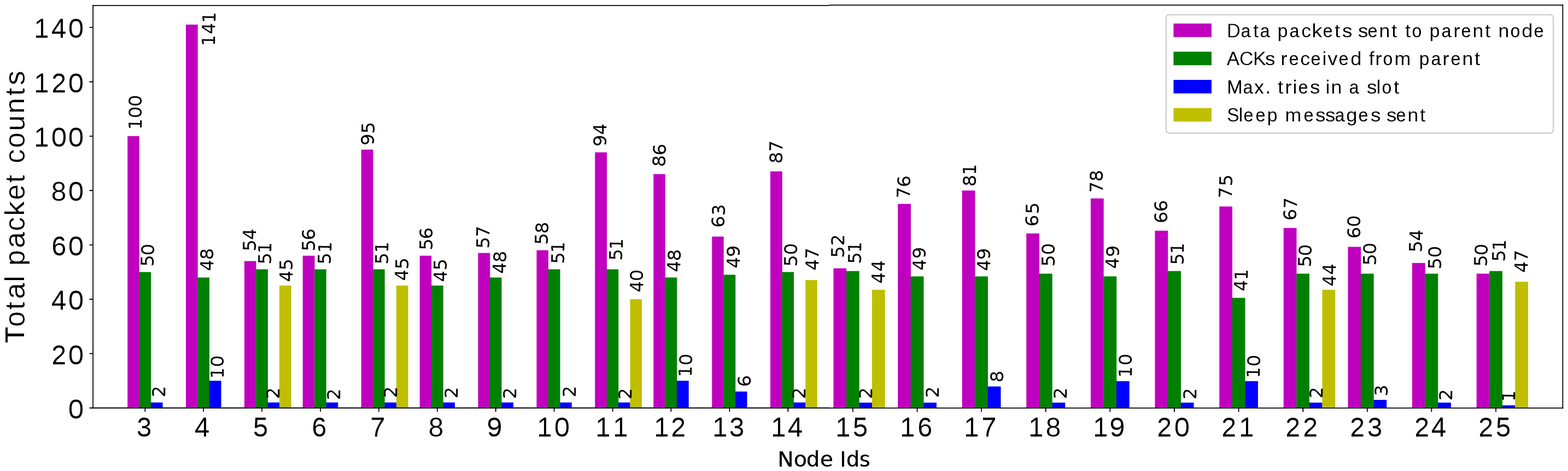}}
\hfill
\subfloat[packet transmissions in round 3]{\label{combinedtree3_10mnt}\includegraphics[scale = 0.49]{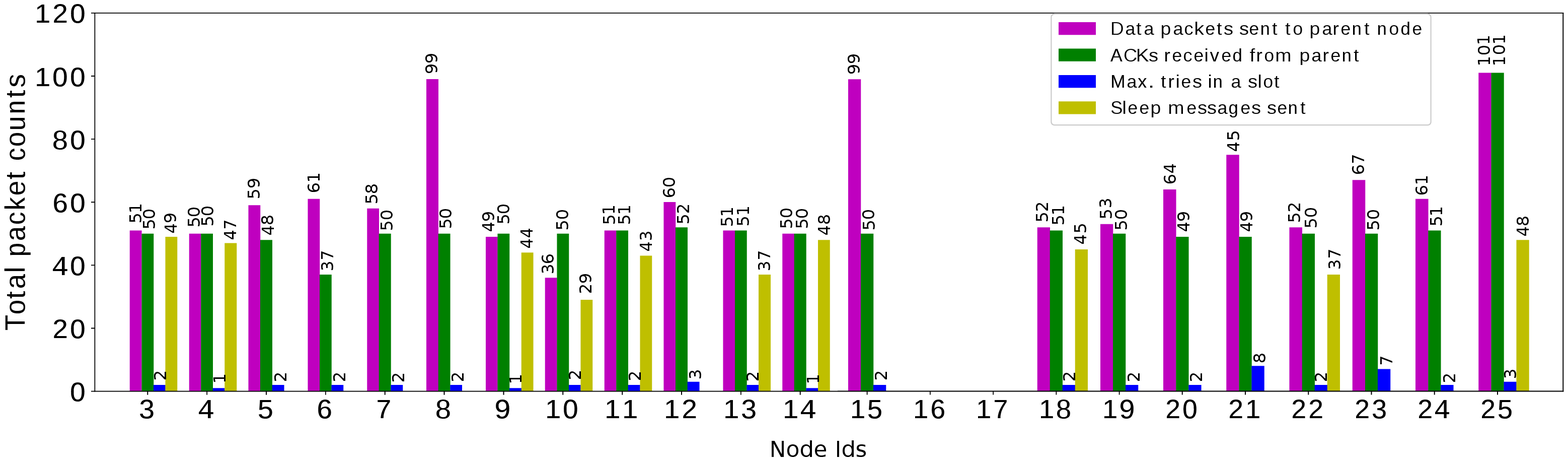}}
\caption{Comprehensive packet analysis from the snooper nodes for test case 1}
 \label{snooperanalysis_10mnt}
 }
\end{figure*}

Detailed analysis of all the packet transmissions captured in different snooper nodes helps us to understand the network performance and the different 
activities that occurred in the network. Fig. \ref{snooperanalysis_10mnt} shows 
a comprehensive analysis of different packet transmissions that occurred in the network during
the data collection time slots.

We should keep in mind that the snooper nodes need not capture all the packet transmissions occurring in the network. 
The analysis is based on the captured packet transmissions and here, we look for a qualitative analysis from the captured data rather
than an exact quantitative analysis. The major observations from Fig. \ref{snooperanalysis_10mnt} are:

\subsubsection*{Extra packet transmissions due to errors in time synchronization}
Ideally in each data collection time slot, a sensor node is expected to forward one data packet to its parent node
which contains the sensor data from its descendant nodes as well as its own sensor data.
It may forward more than one data
packet, if the node could not fit its sensed data and the data received from its descendant nodes into a single data packet due 
to the limitation on the maximum data payload length by the wireless radio chip.
Each forwarded data packet is expected to result in an ACK message receipt from the parent node.
After forwarding the data packet(s), the node is expected to broadcast a sleep message
on reception of the sleep message from its parent node (this is valid only for non-leafnodes). Thus for this test case having 50 
data collection time slots, ideally each node expects
50 data forwards to parent, 50 acks from parent and 50 sleep broadcasts (if it is a non-leaf node), like Node Id 14 in data collection round 1
(see Fig \ref{combinedtree1_10mnt}).
Maximum tries in a slot refers to the maximum of the numbers of times the same data packet is forwarded by the node to its parent
and has a value 1 in ideal conditions.

The number of data packets sent by some of the leaf nodes are almost double of the expected number of packet transmissions.
The timestamps of the packets captured in the snooper nodes reveal that this happened because of the errors in time-synchronization.
That is, in this test case implementation, each node was programmed in tinyos in such a way that whenever it enters into the active
state and the radio is turned ON, the node sends its data packet to its parent node immediately (if its a leaf node).
But the parent's radio may be in the off stage due to time synchronization error.
In such cases, the parent node will not be able to receive the data packet and no ACK will be send to the child node. This leads to 
an ack timeout in the child node and it retries to deliver its data packet to the parent. 

This happened in many of the nodes like node Ids 7, 12, 16, 19, 21 in data collection tree 1 (Fig \ref{combinedtree1_10mnt}),
node Ids 3, 16, 18,20 in data collection tree 2 (Fig \ref{combinedtree2_10mnt}), Node Ids 8, 15 in data collection tree 3 
(Fig. \ref{combinedtree3_10mnt}).
Note that the max tries in a time slot remains almost 2 for these nodes and it is 2 for almost all the time slots for many of these nodes.
At the same time, several other nodes could successfully deliver their data packet to their parent node in the first attempt itself.
One of the possible solutions to reduce the number of extra transmissions due to this time synchronization error is to
provide a small delay at the beginning of each active time slot. The radio of each node remains ON during this period.
This will also help to save energy in scenarios where the packet transmission power is greater than the reception power.

\subsubsection*{Packet losses due to link loss or concurrent transmissions}

Some of the nodes in the network had sent the same data packet to its parent node, multiple times in some data collection times slots (very few). 
This is because either the parent node could not receive the data packet or the child node could not receive the corresponding ACK from 
the parent. Both these cases will lead to an ack timeout at the child node and it will resend the data packet.
Most of the nodes, in all the three data collection rounds or trees, 
has this Max tries in a data collection time slot value as 2 or 3 which implies that either during the second or third try, the 
node could successfully deliver the data packet to the parent node and receive the ACK for the respective data packet.

\subsubsection*{Higher value for the max tries in a data collection time slot}

Some nodes have a higher value for the max tries in a data collection time slot - 10 or close to 10 (some 
packet transmissions may not be captured in the snooper).
This can be because of two reasons:

1) The parent node is not able to hear the child node even though it repeatedly sends for 10 times (maximum number of retries). 
This happens very rarely and it results in  loss of the
corresponding node data at the sink node. Such scenarios are observed in data collection tree 2 (Fig \ref{combinedtree2_10mnt}) for 
node Ids 4 and 19. The same is reflected in 
node yield plots (Fig \ref{nodeyield_10mnt_tree2}) as well.

2) ACK-SLEEP dualloss, which is defined as: assume node $v$ is the parent of $u$ and the following incidents occur in sequence.

\begin{itemize}
 \item {Node $u$ could not receive ACK for the $n^{th}$ data packet try in a 
	slot ($0<n\le10$) while $v$ received the data packet and forwarded it to the sink.}
 \item {Node $v$ broadcasts the sleep message of that data collection slot and enters into the sleep state.}
 \item {Node $u$ could not receive that sleep message also.}
 \item {Node $u$ tries to resend the data packet again to $v$ when the ack timeout occurs. Since $v$ is in sleep stage, it cannot send ACK and 
   thus $u$ repeatedly tries to resend the data packet for the maximum number of tries (10).}
\end{itemize}

The ACK-SLEEP dualloss will not result in any loss of data at the sink node and hence, does not affect the node yield but leads to a few
extra data packet transmissions at the respective node. 
The number of occurrences of ACK-SLEEP dualloss is also relatively small in number.
Since each node enters into sleep stage based on a sleep message timeout (which is quite large), 
the nodes where ACK-SLEEP dualloss occurs remain in ON stage for a longer time compared to the other nodes and spend 
more energy in a data collection time slot.

Propagation of a sleep message from the sink to all the nodes is done with the propagation of a broadcast message
along the edges of the data collection tree. Currently, this broadcast is not implemented
as a reliable data transfer. That is, a node broadcasts the sleep message just once and if its child receives it, the latter goes into sleep
stage (after rebroadcast if it is a non-leaf node). 
Some of the nodes do not receive this sleep message in some time slots and hence, they do not rebroadcast.
Thus the total number of sleep messages sent by the nodes decreases as the hop count increases from 
the sink node.
   
This results in more sleep timeouts for the nodes farther from the sink and hence, these nodes may have more energy loss due to this.
This effect can be taken care by converting the propagation of the sleep message into a reliable transfer. But it has its own disadvantages
(like every node will have to send ACK to its parent for the sleep message and then go to sleep stage. Now if a parent could not receive this ACK 
and the child went to sleep, the problem of being awake and spending more energy just shifts from the child node to the parent node.)
   
So making the sleep message a reliable transfer, will be considered after looking at the energy expenditures of all the nodes in detail.

\subsubsection{Sensor data variations}
\begin{figure}
\centering
\subfloat[Battery voltage of node-Id4]{\includegraphics[scale=0.21]{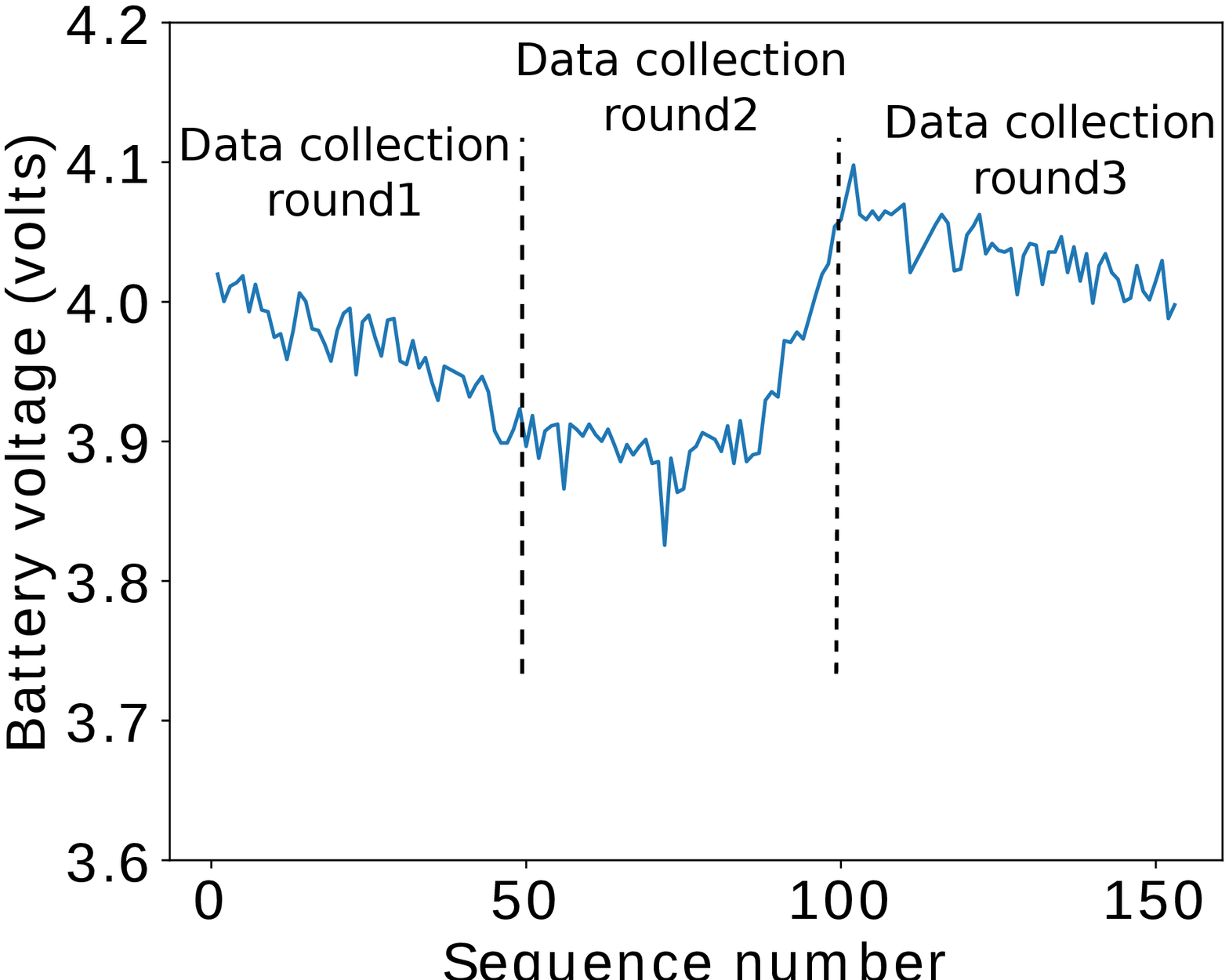}}
\hfil
\subfloat[All parameters of node-Id4]{\includegraphics[scale=0.21]{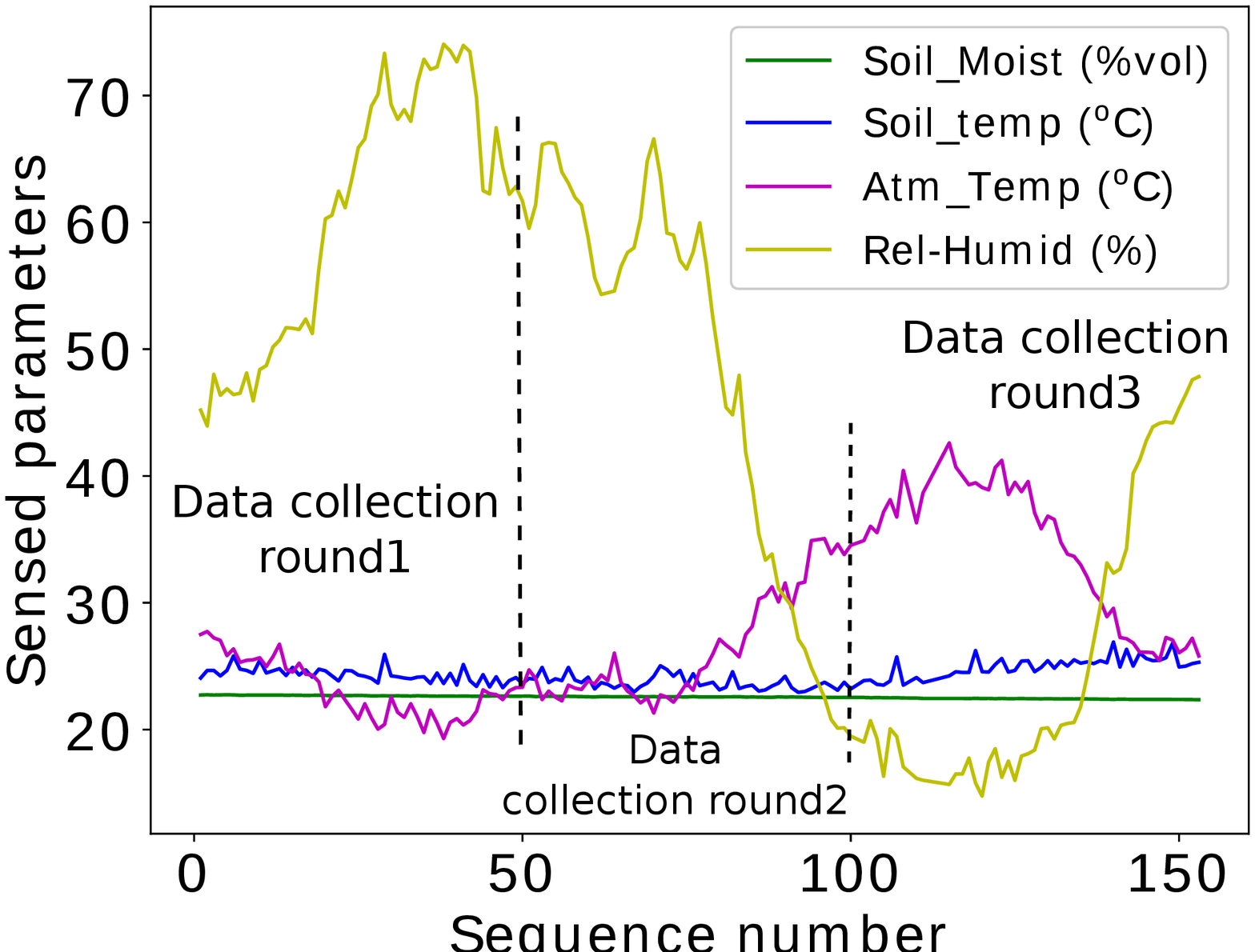}}
\hfil
\subfloat[Battery voltage of node-Id23]{\includegraphics[scale=0.2]{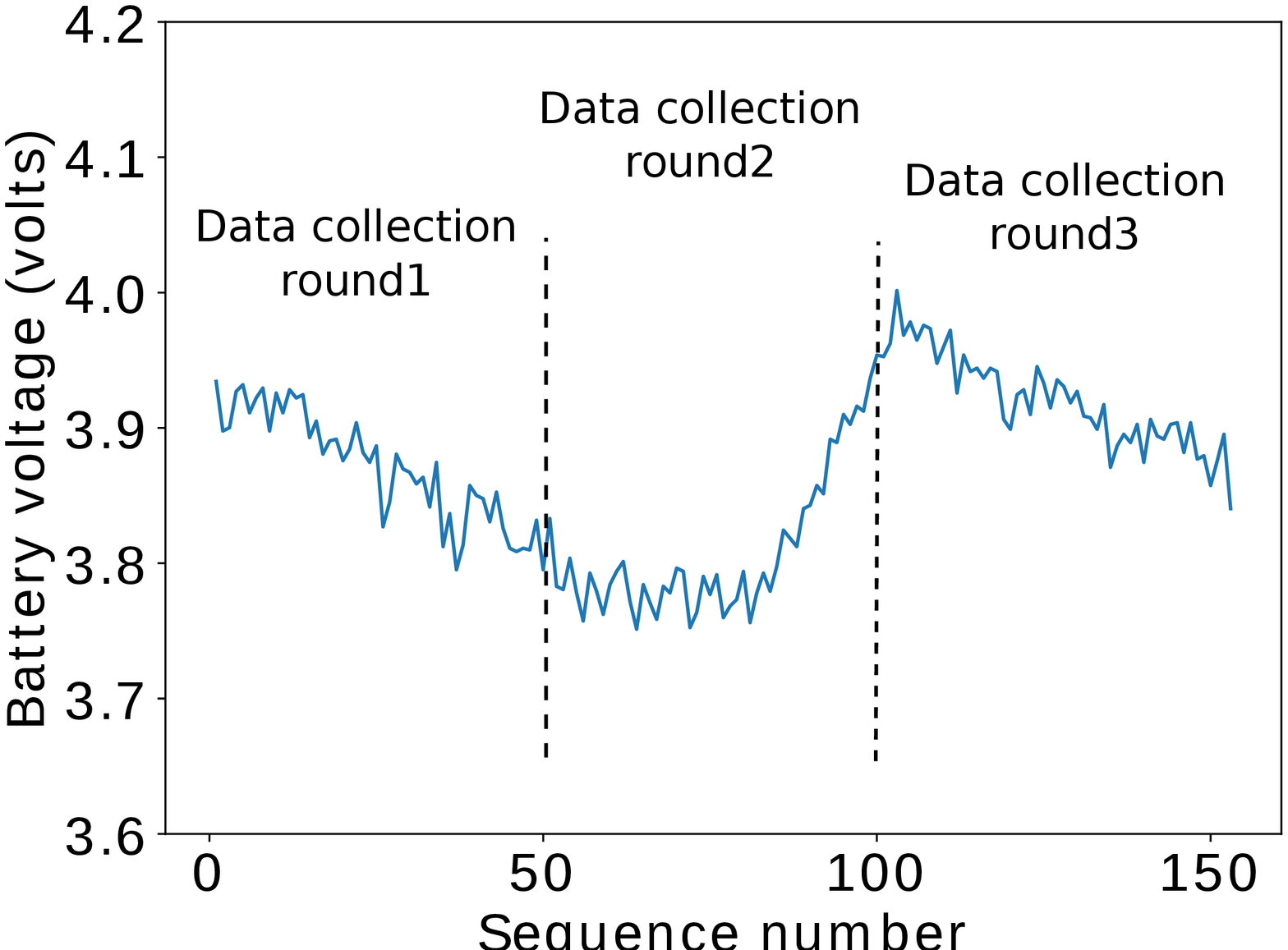}}
\hfil
\subfloat[All parameters  of node-Id23]{\includegraphics[scale=0.2]{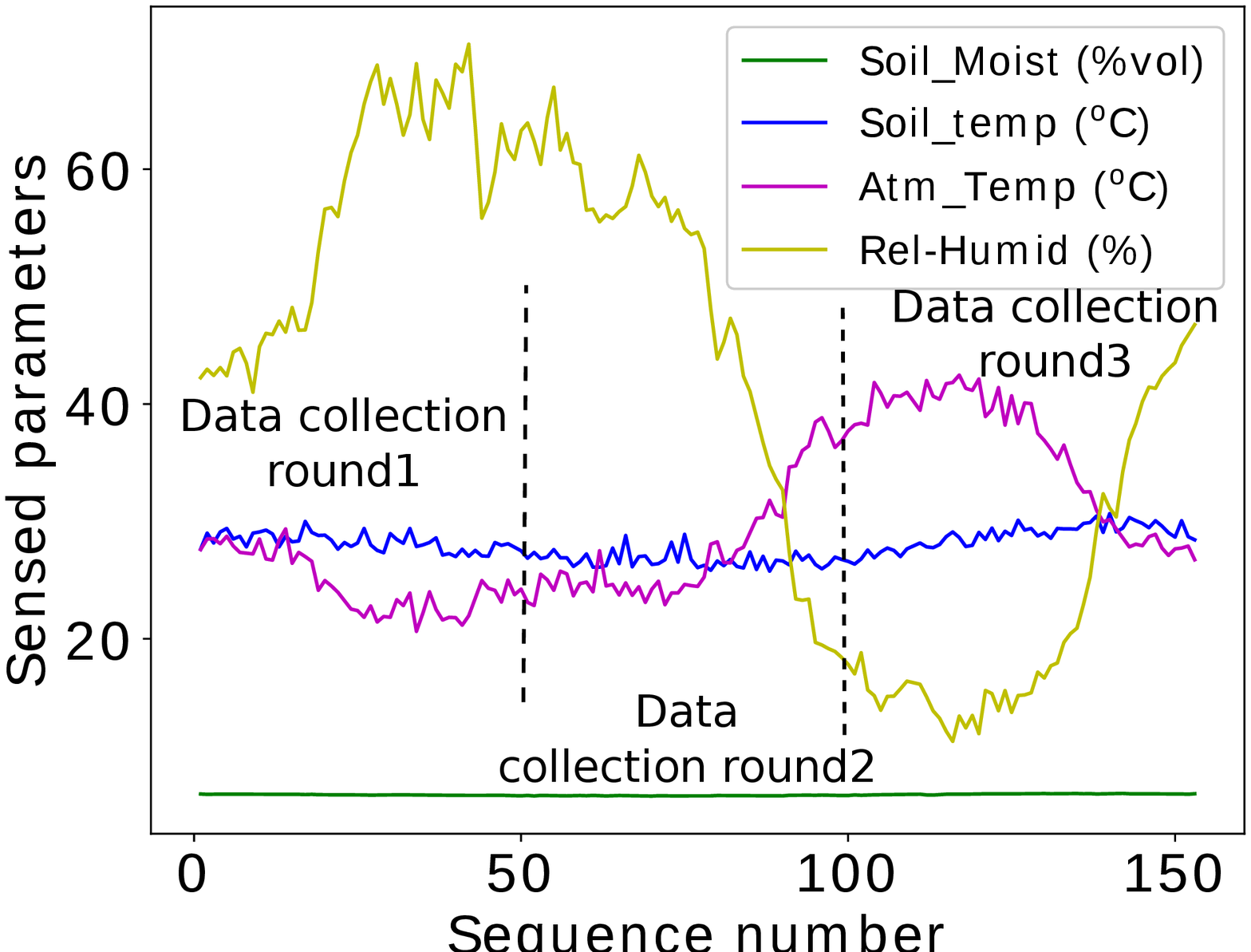}}
\caption{Sensed parameters - test case 1. Time interval between two consecutive sequence numbers in a datacollection timeslot is 10 minutes. }
\label{sennsordata_10mnt}
\end{figure}

Fig. \ref {sennsordata_10mnt} shows the variation in the sensed parameters for two different nodes in this test case.
As shown in Fig. \ref {sennsordata_10mnt}, the battery of each node gets charged during the day using solar energy which
gets discharged as the node consumes energy during its operation. The figure also captures the variation of soil moisture, soil temperature,
atmospheric temperature and relative humidity for the three rounds of data collection spanning approximately 1 complete day.
The reduction in soil moisture is very small in a day which clearly indicates that we do not need to measure it frequently.  

\subsubsection{Energy expenditure of the nodes}
\begin{figure}
 \includegraphics[scale = 0.27]{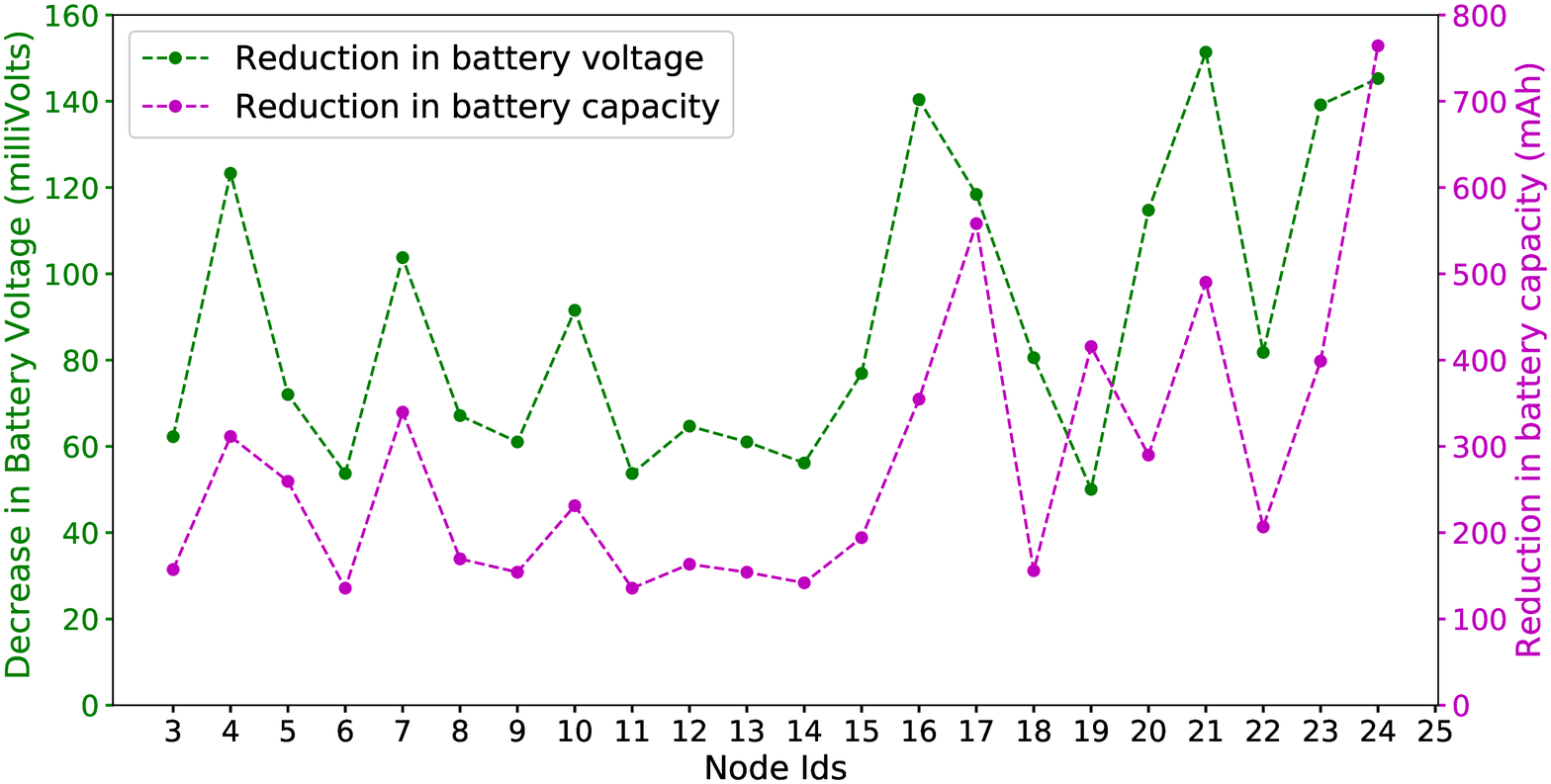}
 \caption{Energy expenditure for data collection round 1 in test case 1}
 \label{energyexpenditure_10mnt}
\end{figure}

In one of the data collection rounds (round 1 or tree 1), the first data collection time slot occurred at 6.30PM and the last 
time slot in the same round occurred on the  next day at 2.40AM. 
The battery was not getting charged via the solar panel during this time interval because of very low light intensity/ darkness.
Each node reports its battery voltage in addition to the sensor data to the sink node in each time slot.
The difference in the battery voltage (last slot battery voltage - first slot battery voltage) gives us the reduction in battery voltage 
and the corresponding energy difference calculated from the OCV-RC (Open circuit voltage - remaining capacity) curve gives us the
energy consumed for data collection in that round. This is shown in Fig. \ref{energyexpenditure_10mnt}
and Table \ref {energyexpenditure_10mnt_table} presents
the nodes in the decreasing order of energy expenditure. 

\begin{table}[!t]
\caption{Decreasing order of energy expenditure}
\label{energyexpenditure_10mnt_table}
\centering
\begin{tabular}{|>{\centering\arraybackslash}p{1cm}|>{\centering\arraybackslash}p{2.5cm}|>{\centering\arraybackslash}p{2.5cm}|}
\hline
Node ID & Reduction in Battery Voltage (mV) & Reduction in Battery Capacity (mAh)\\
\hline
24 & 145.26 & 764.26 \\
\hline
17 & 118.41 & 558.15 \\
\hline
21 &	151.37	& 490.22 \\
\hline
19 &	50.05 & 415.68 \\
\hline
23 &	139.16	& 398.89 \\
\hline
16 &	140.38	& 354.69 \\
\hline
7 &	103.76	& 339.69 \\
\hline
4 &	123.29	& 311.51 \\
\hline
20 &	114.75	& 289.92 \\
\hline
5 &	72.02	& 259.50 \\
\hline
10 &	91.55	& 231.32 \\
\hline
22 &	81.79	& 206.65 \\
\hline
15 &	76.90	& 194.31 \\
\hline
8 &	67.14	& 169.64 \\
\hline
12 &	64.70	& 163.47 \\
\hline
3 &	62.26	& 157.30 \\
\hline
18 &	80.57	& 155.98 \\
\hline
9 &	61.04	& 154.21 \\
\hline
13 &	61.04	& 154.21 \\
\hline
14 &	56.15	& 141.88 \\
\hline
6 &	53.71	& 135.71 \\
\hline
11 &	53.71	& 135.71 \\
\hline

\end{tabular}
\end{table}

\subsubsection*{Observations from the energy discharge profile of nodes}

\begin{figure}
 \centering
\includegraphics[scale=0.6]{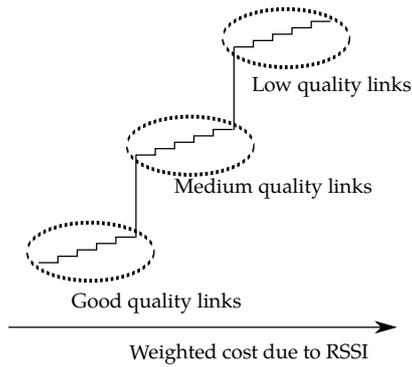}
\caption {Weighted cost due to RSSI Range}
 \label{rangeweightedrssi}
\end{figure}

\begin{figure*}[!b]
\centering
\subfloat[Graph 1]{\includegraphics[scale=0.47]{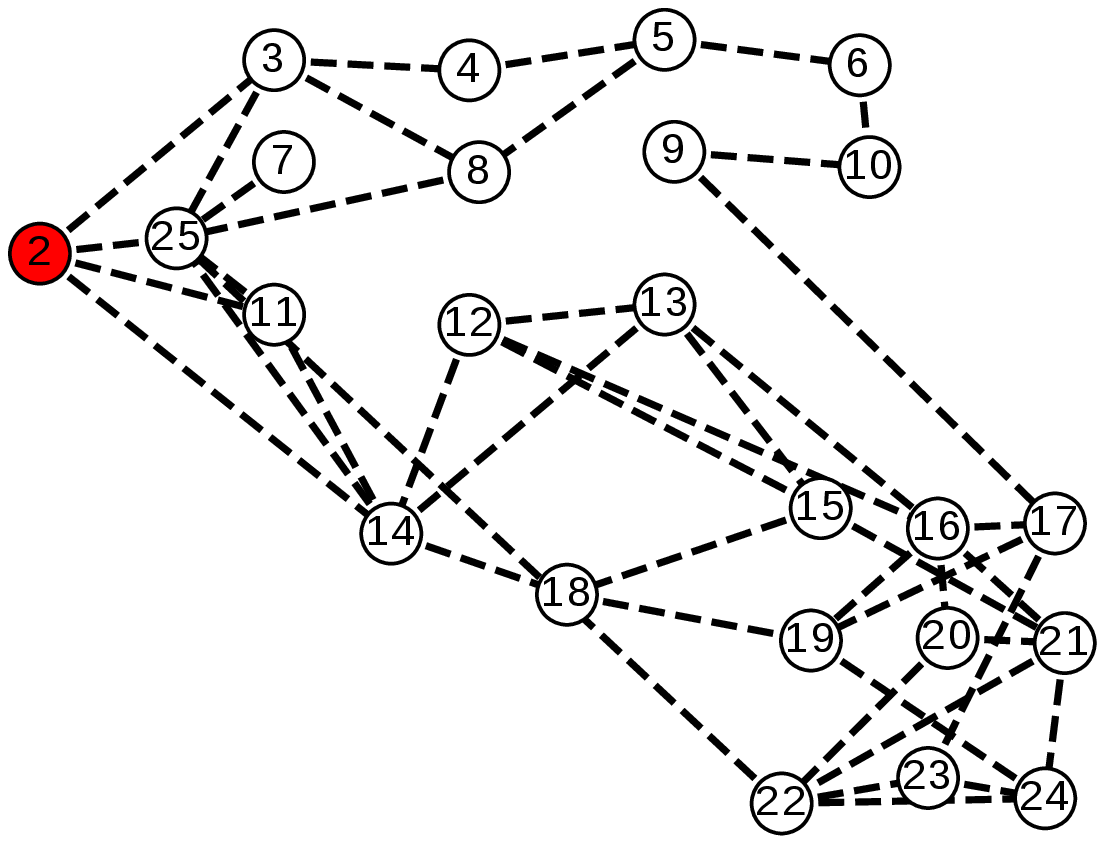}}
\hfil
\subfloat[Graph 2]{\includegraphics[scale=0.47]{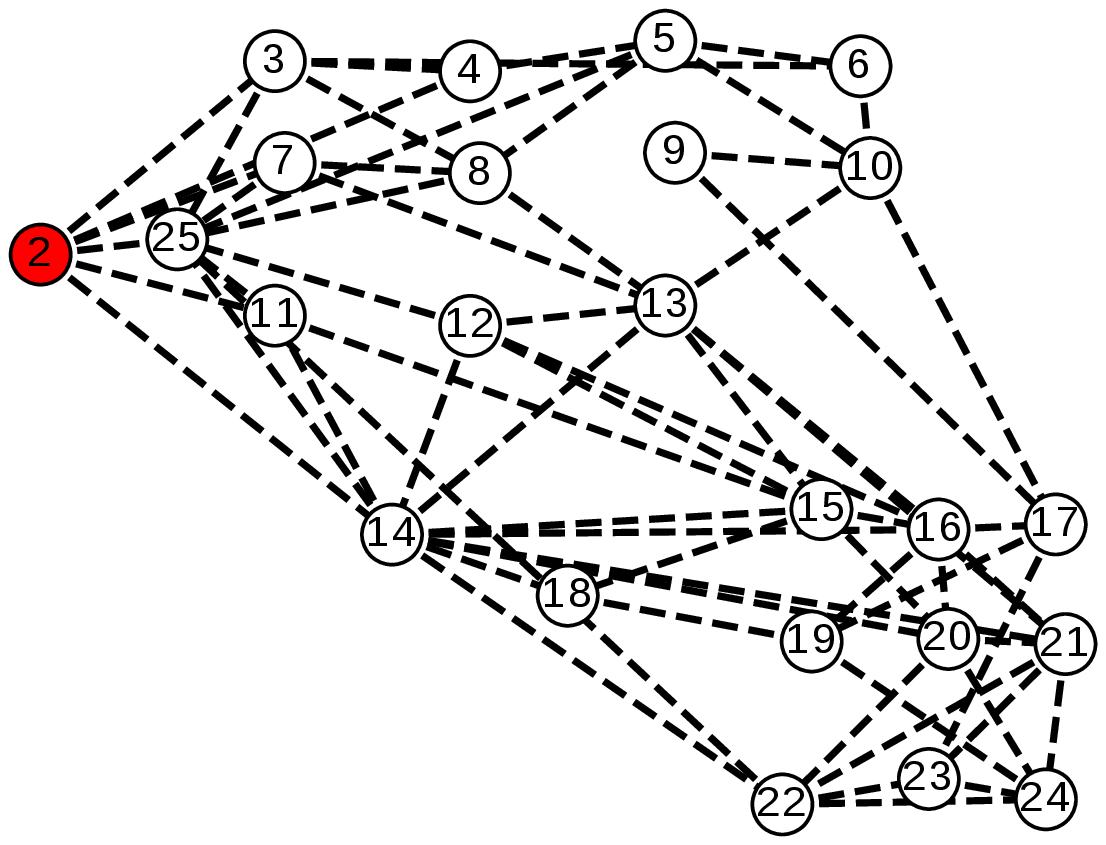}}
\hfil
\subfloat[Graph 3]{\includegraphics[scale=0.47]{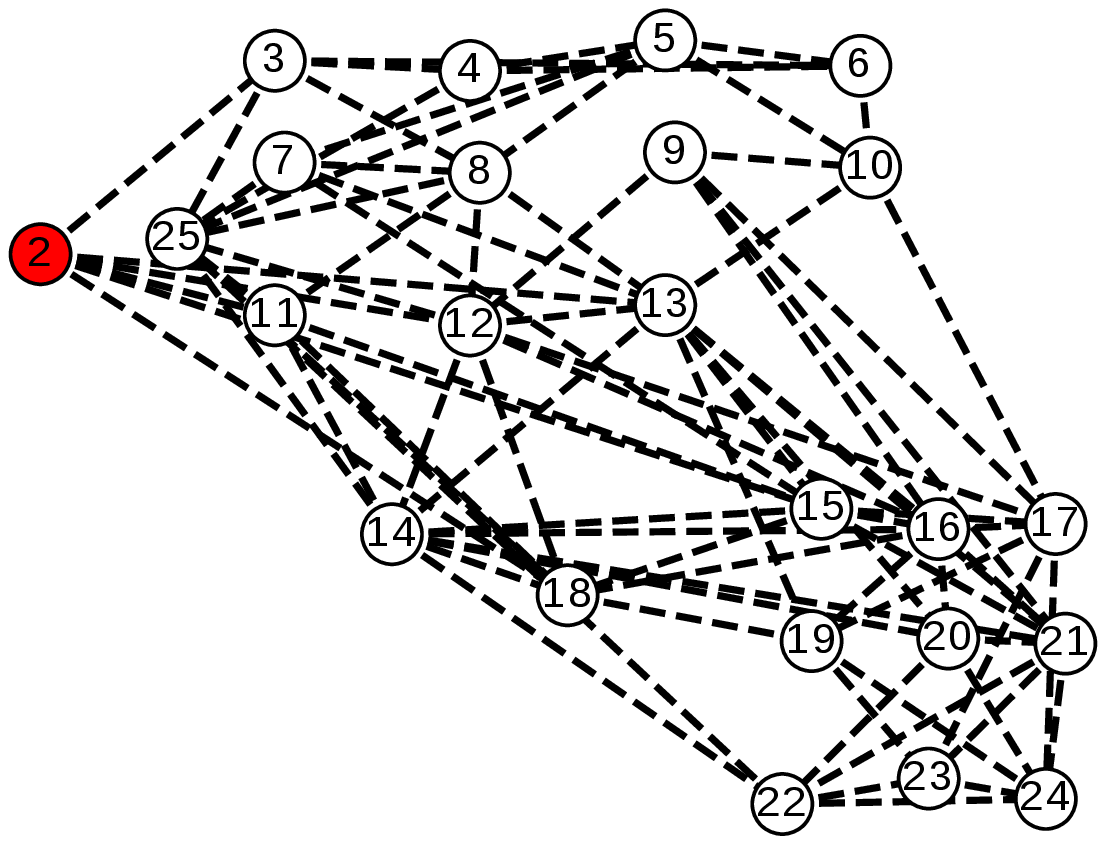}}
\hfil
\subfloat[Data collection tree formed from graph 1]{\includegraphics[scale=0.27]{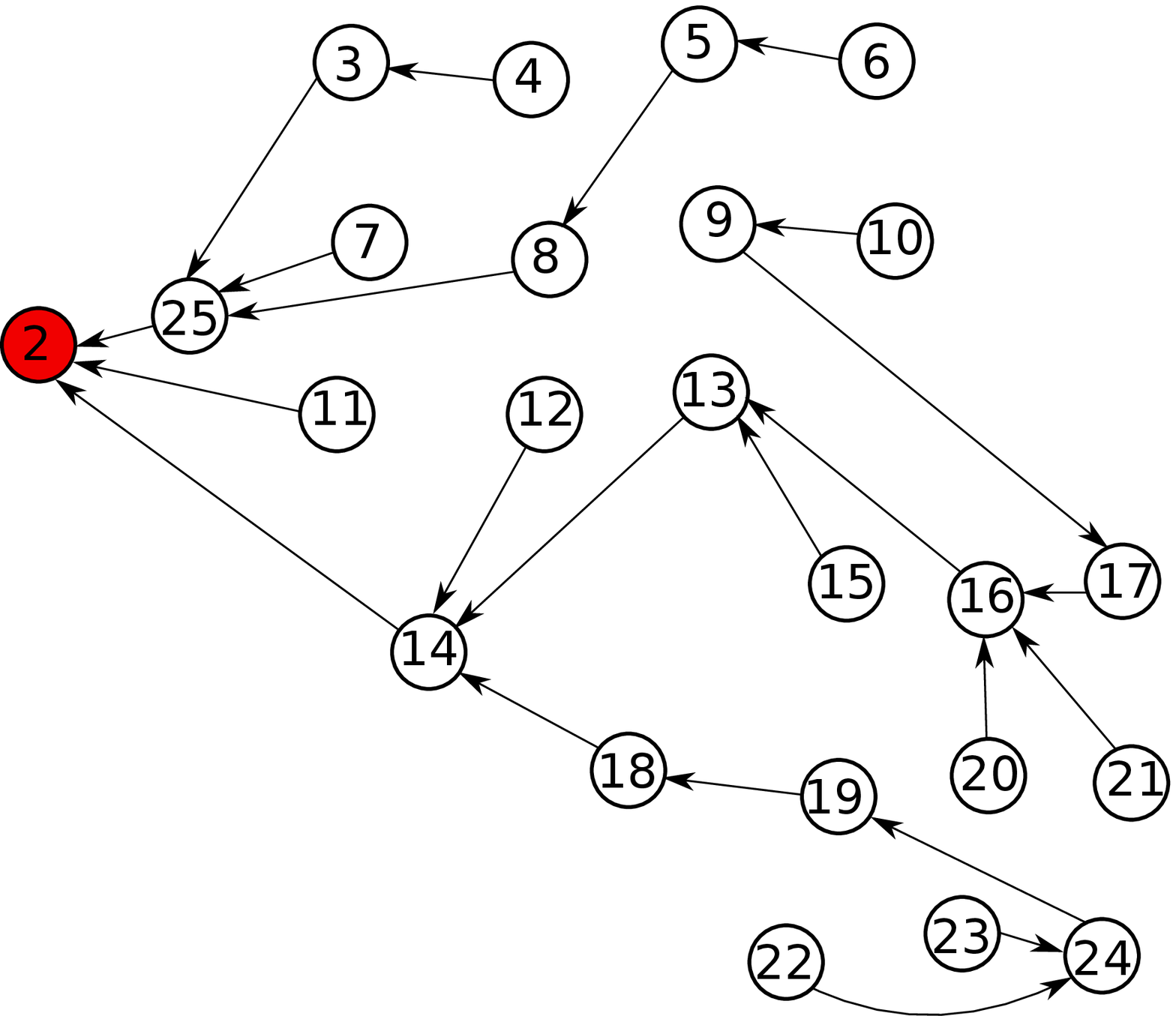}}
\hfil
\subfloat[Data collection tree formed from graph 2]{\includegraphics[scale=0.27]{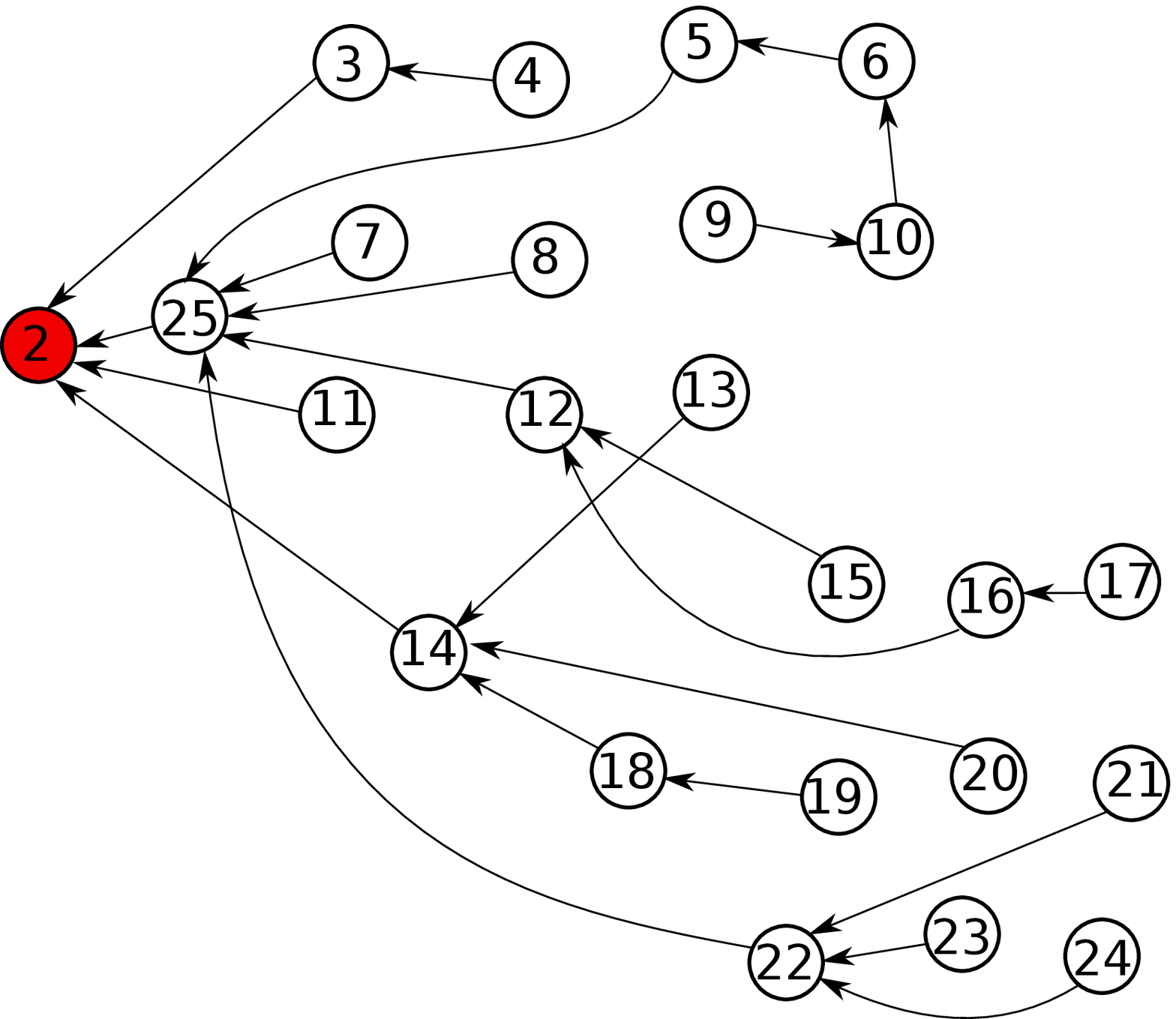}}
\hfil
\subfloat[Data collection tree formed from graph 3]{\includegraphics[scale=0.27]{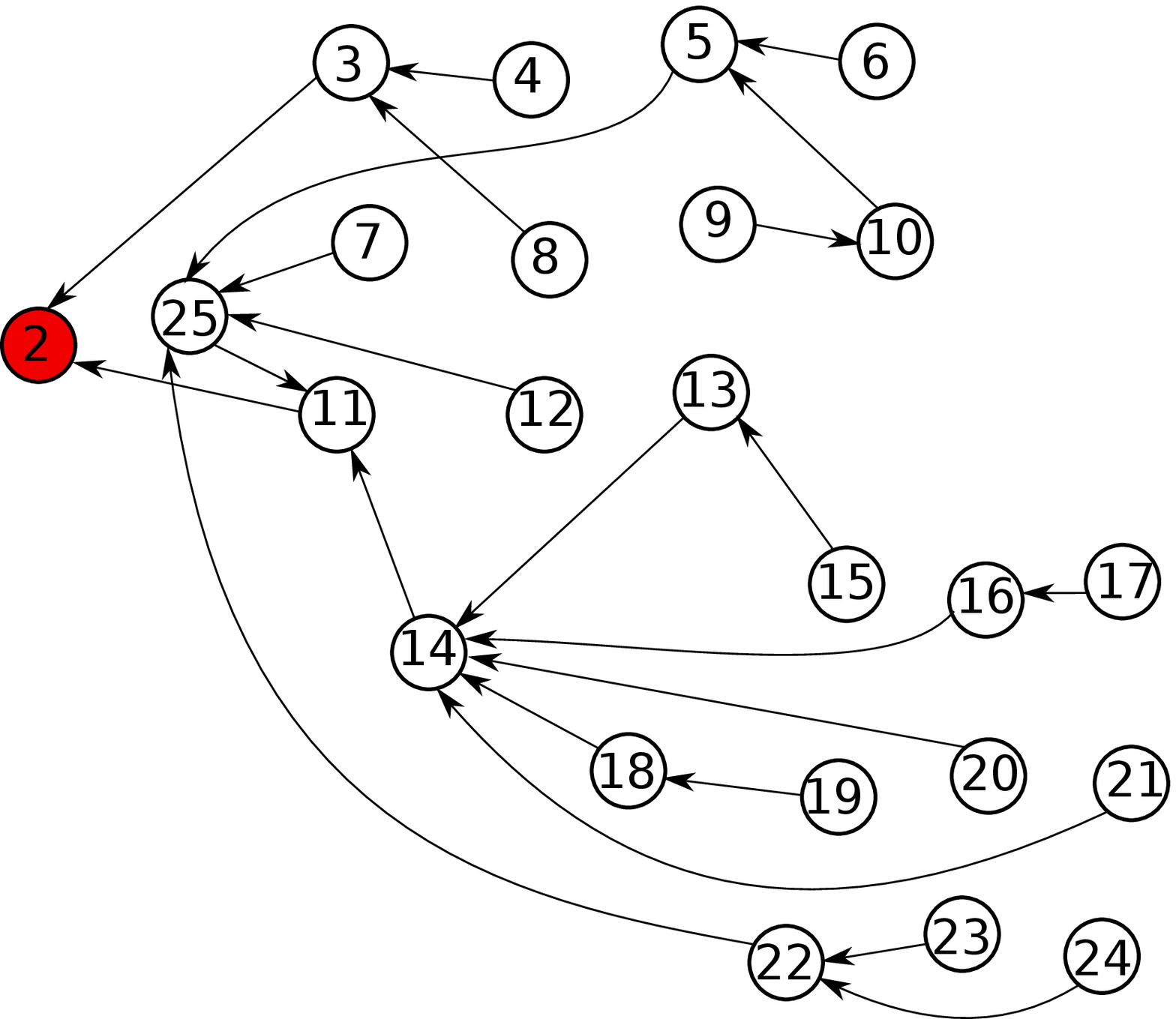}}
\caption{Consecutive graphs and respective data collection trees for test case 2}
\label{consecutive_graphs_1hr}
\end{figure*}

\begin{enumerate}
  \item {Reduction in battery voltages for various nodes varies from 50 mV (NodeID 19) to 151.4 mV (NodeID 21).}
 \item {Reduction in remaining capacity for various nodes varies from 135.7 mAh (NodeID 6) to 764.26 mAh (NodeID 24).}
 \item {If we look at the tree used for the data collection (tree 1), the nodes which spent most of the energy are leaf nodes. This might 
 be because of the fact that they are the ones which went into sleep stage last due to more sleep timeouts. But there are exceptions also. 
 For example Node ID6, finds its position somewhere towards the bottom of the energy expenditure table.}
 
 \item {Node Id 22 is the node which has the least number of sleep message forwards which leads to the fact that its children should have more 
 sleep timeouts and hence should appear somewhere on top of the energy expenditure table.
 It can be seen that the children of Node Id 22-- viz., Node Ids 20, 21, 23, 24 (see Fig. \ref{dct_10mnt_tree1})-- 
 appear near the top in Table \ref{energyexpenditure_10mnt_table}.
 }
 
 \item {Nodes which are closer to the sink are expected to forward more data and thus expected to spend more energy. This effect
 is not clearly visible in the captured energy expenditure profiles of the nodes. This is because, the radio
 which we used consumes almost the same power in both transmit and receive stages. }
 
 \item {The decrease in battery voltages are very small (in milli volt ranges) and hence, the effect of errors in the measured
 voltage due to various factors (like errors due to unstable ADC reference voltage, errors in voltage sampling) might have an impact on
 our observations. Therefore, before coming to strong conclusions,
  we need to verify these observations in other test cases as well. A small error in the sensed voltage may lead to 
 wrong conclusions. }
 
\end{enumerate}

\subsection{Test case 2}
In this test case, the interval between two data collection time slots is 1 hour and the tree rebuilding happens after every 30 time slots,
($30 \times 1 = 30$ hours $ > 1$ day). Based on our previous test case observations, we have introduced two additions in the design:

\begin{enumerate}
 
 \item {An initial small delay is added in each time slot before the data packet transmission. This will help to reduce 
 the increased number of packet transmissions due to errors in time synchronization which happened in the previous test case.}
 
 \item {CC2420 has a receiver sensitivity of $-90$ dBm and in our network architecture design, each node used
 an average RSSI thresholding of $-85$ dBm in neighbour discovery phase. Even with this higher thresholding level, some of the edges
 in the data collection tree could not deliver data packets successfully in a few time slots resulting in a reduced node yield. To address
 this, we have assigned different weighted costs to links based on the RSSI range which is divided into 3 ranges and we called them as 
 good quality links, medium quality links and low quality links as shown in Fig. \ref{rangeweightedrssi}. The weight assignment is done in such a way that
 the nodes will always get added in the tree using any of the possible links in good quality range, followed by any in medium quality range, 
 followed by any in low quality range. This modification 
 in the design is to consider the case of parent node not able to hear the child node after the formation of the data collection tree 
 (one of the cases observed in the earlier test case). 
 }

\end{enumerate}

\begin{figure}
\centering
\includegraphics[scale = 0.3]{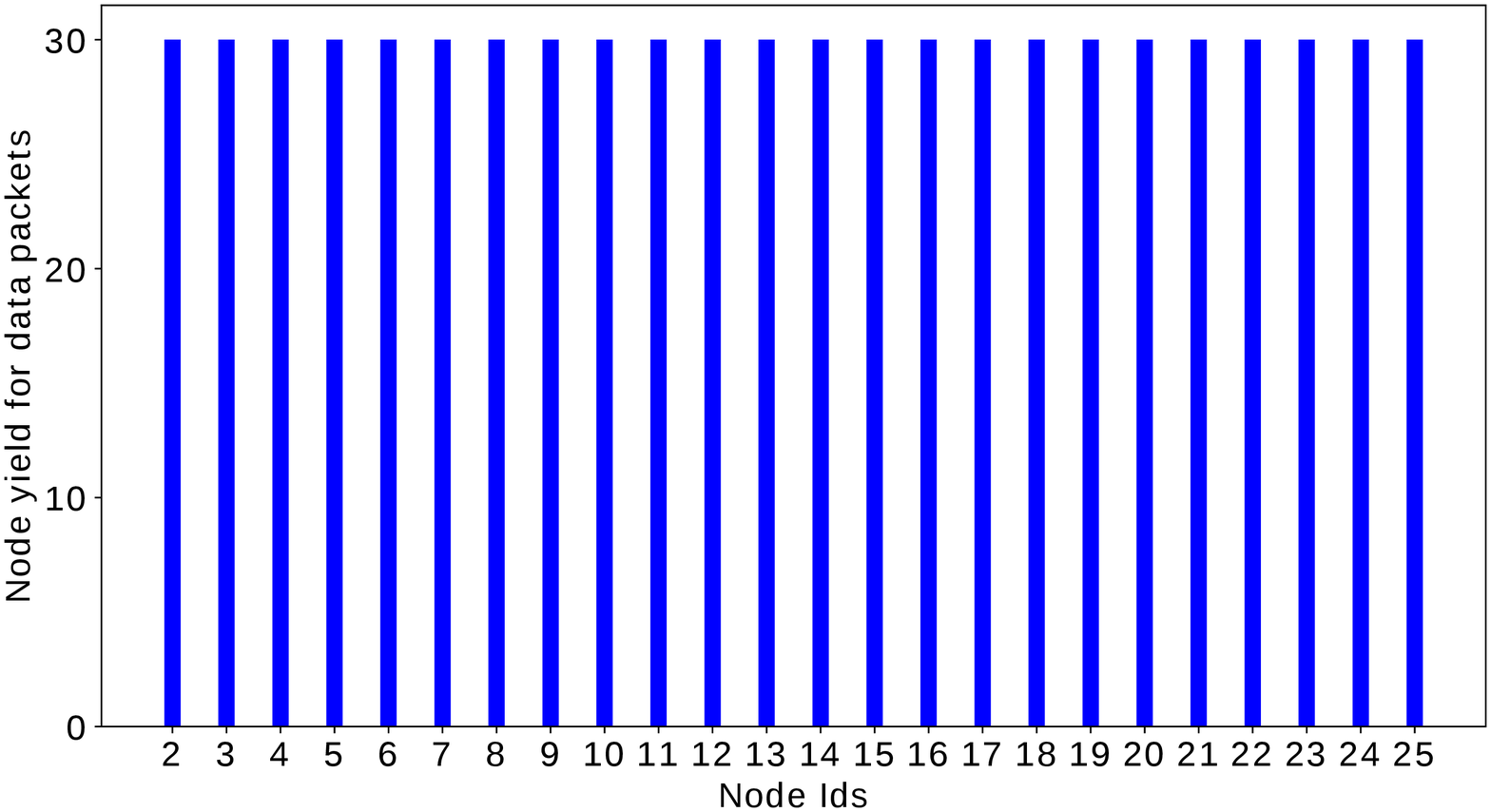}
 \caption{Node yield for test case 2 in each round. Note that the node yield plots for all the three rounds in test case 2 are the same.}

 \label{nodeyield_round2_1hr}
\end{figure}

\begin{figure*}
\centering
\subfloat[packet transmissions in round 1]{\label{combinedtree1_1hr}\includegraphics[scale = 0.55]{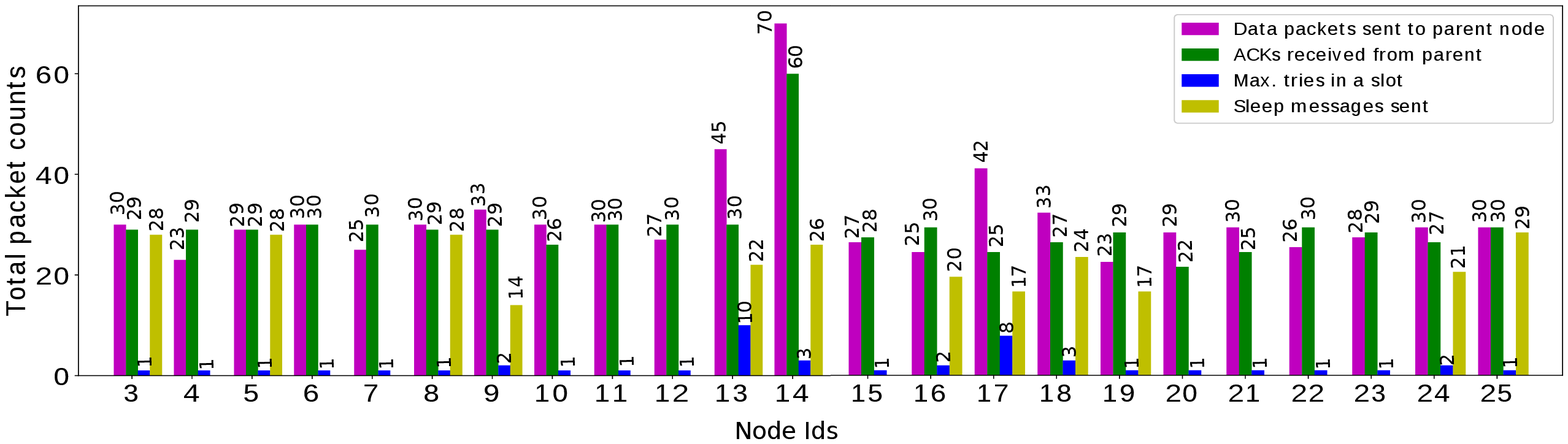}}
\hfill
\subfloat[packet transmissions in round 2]{\label{combinedtree2_1hr}\includegraphics[scale = 0.5]{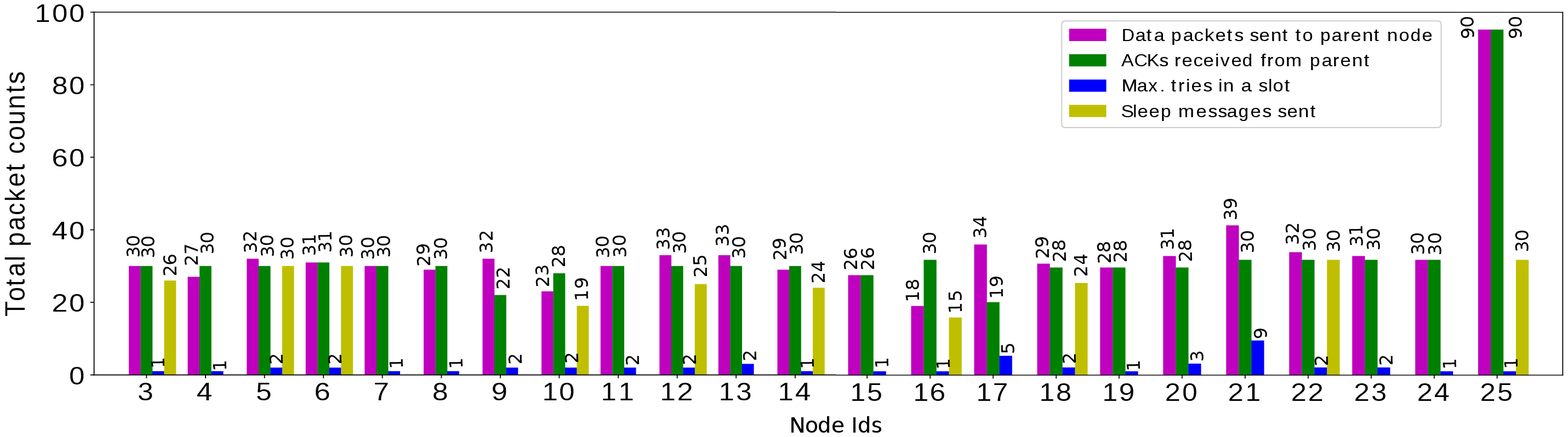}}
\hfill
\subfloat[packet transmissions in round 3]{\label{combinedtree3_1hr}\includegraphics[scale = 0.5]{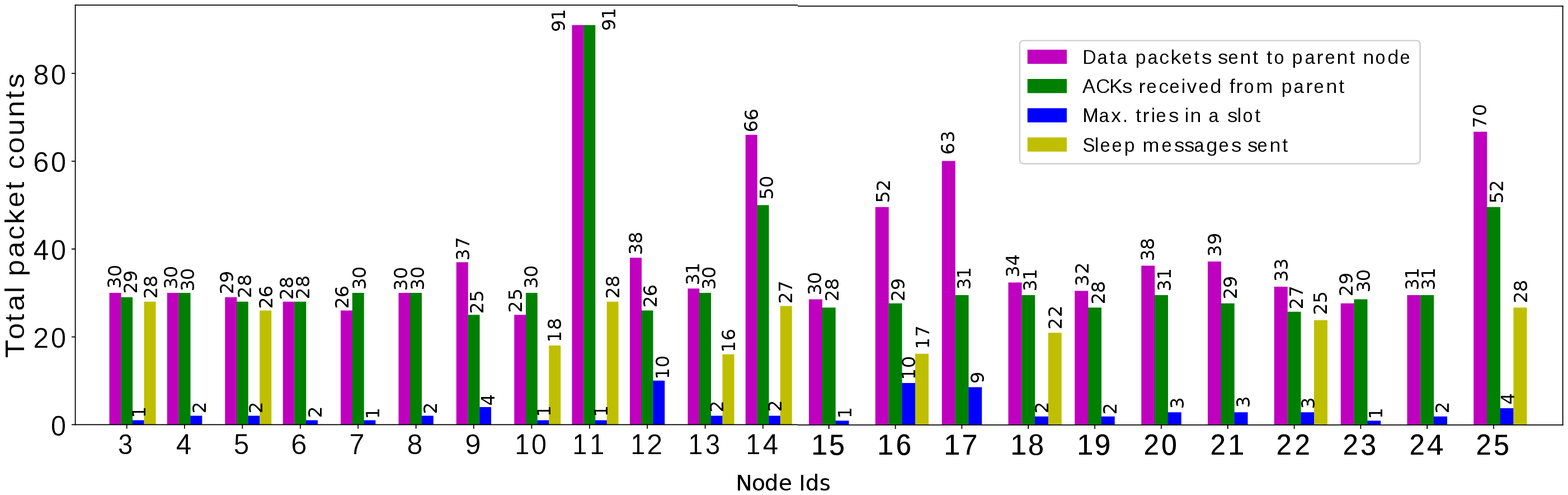}}
\hfill
\caption{Comprehensive packet analysis from the snooper nodes for test case 2}
 \label{snooperanalysis_1hr}
\end{figure*}

Fig. \ref{consecutive_graphs_1hr} shows different network graphs and the corresponding data collection trees and Fig. \ref{nodeyield_round2_1hr} shows the network yield for 
three consecutive data collection rounds corresponding to this test case. All the packet transmissions that happened in the network and which are captured
in the snooper nodes are shown in Fig. \ref{snooperanalysis_1hr}.

\subsubsection{Observations from the snooper data analysis}

\begin{itemize}
 \item {Initial delay in a time slot eliminated the extra packet transmissions due to time synchronization errors which were observed in 
test case 1.}
 \item {There are packet transmissions in the network which the snooper nodes could not capture.}
 \item {ACK-SLEEP dualloss were observed in this test case also (very few times)}
 \item {All the nodes have 100\% node yield in all the rounds.}
 \end{itemize}

\subsubsection{Energy expenditure of nodes}

\begin{figure}
\centering
\subfloat[During round 1 data collection]{\includegraphics[scale=0.28]{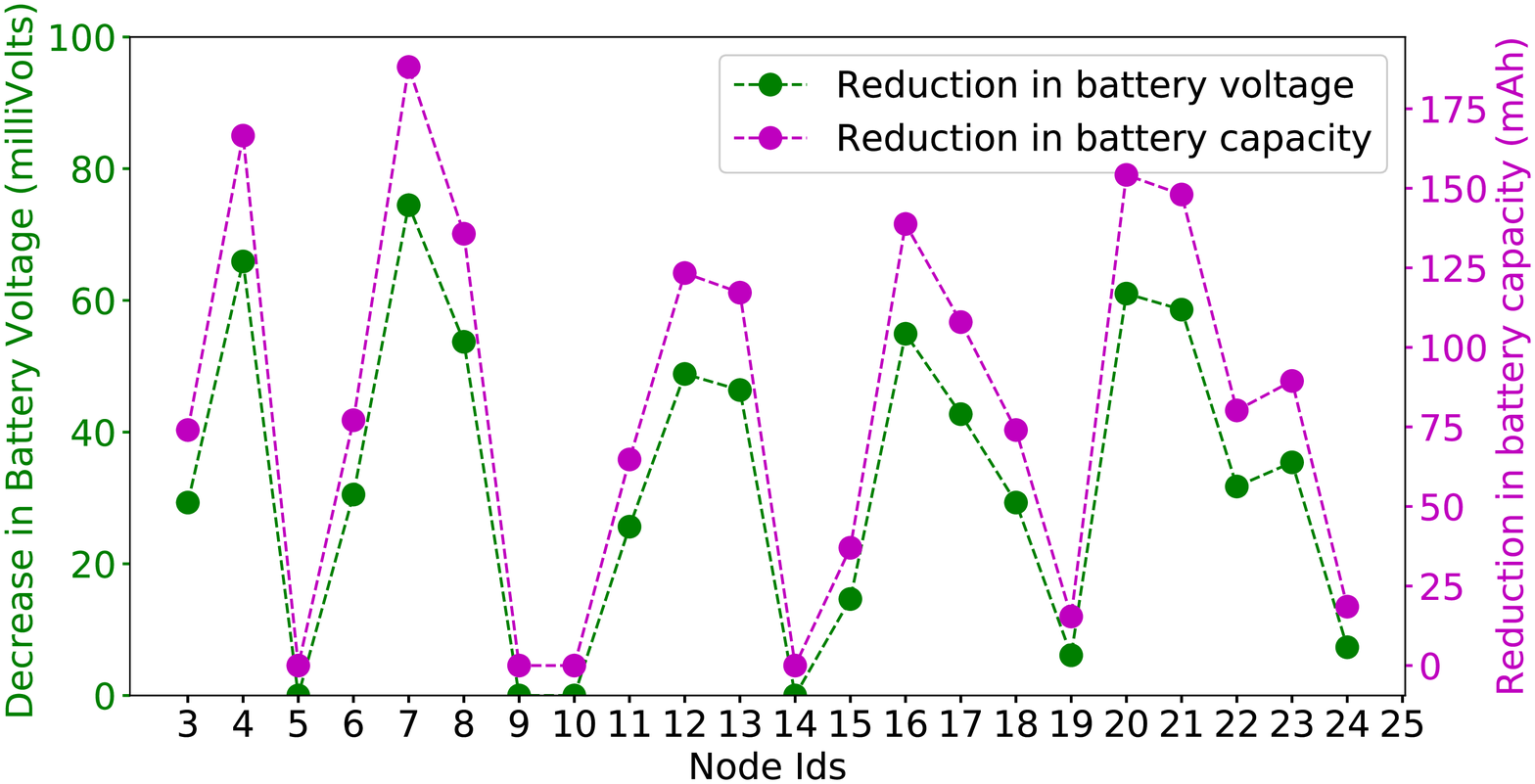}}
\hfil
\subfloat[During round 2 data collection]{\includegraphics[scale=0.28]{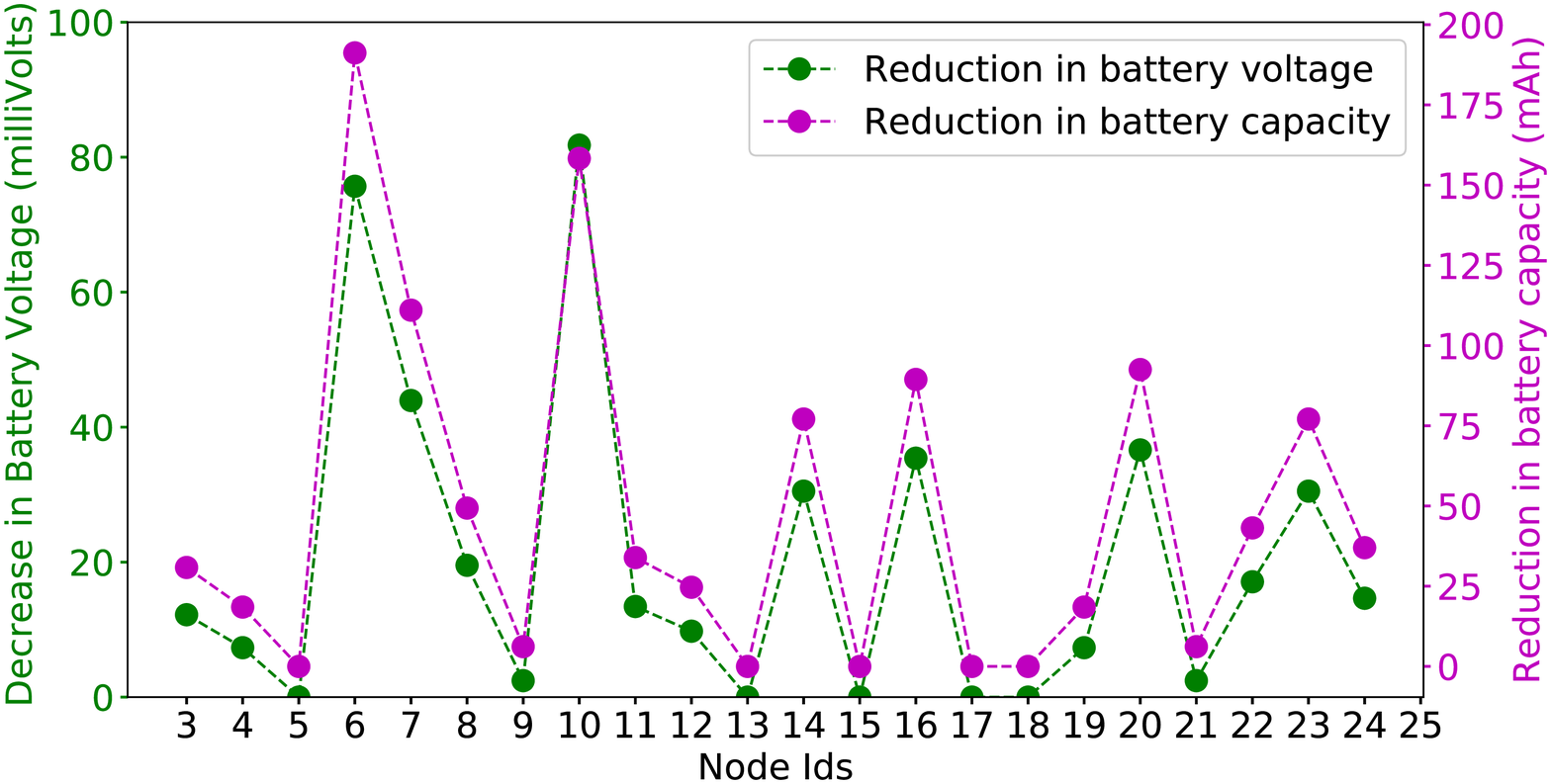}}
\hfil
\subfloat[During round 3 data collection]{\includegraphics[scale=0.28]{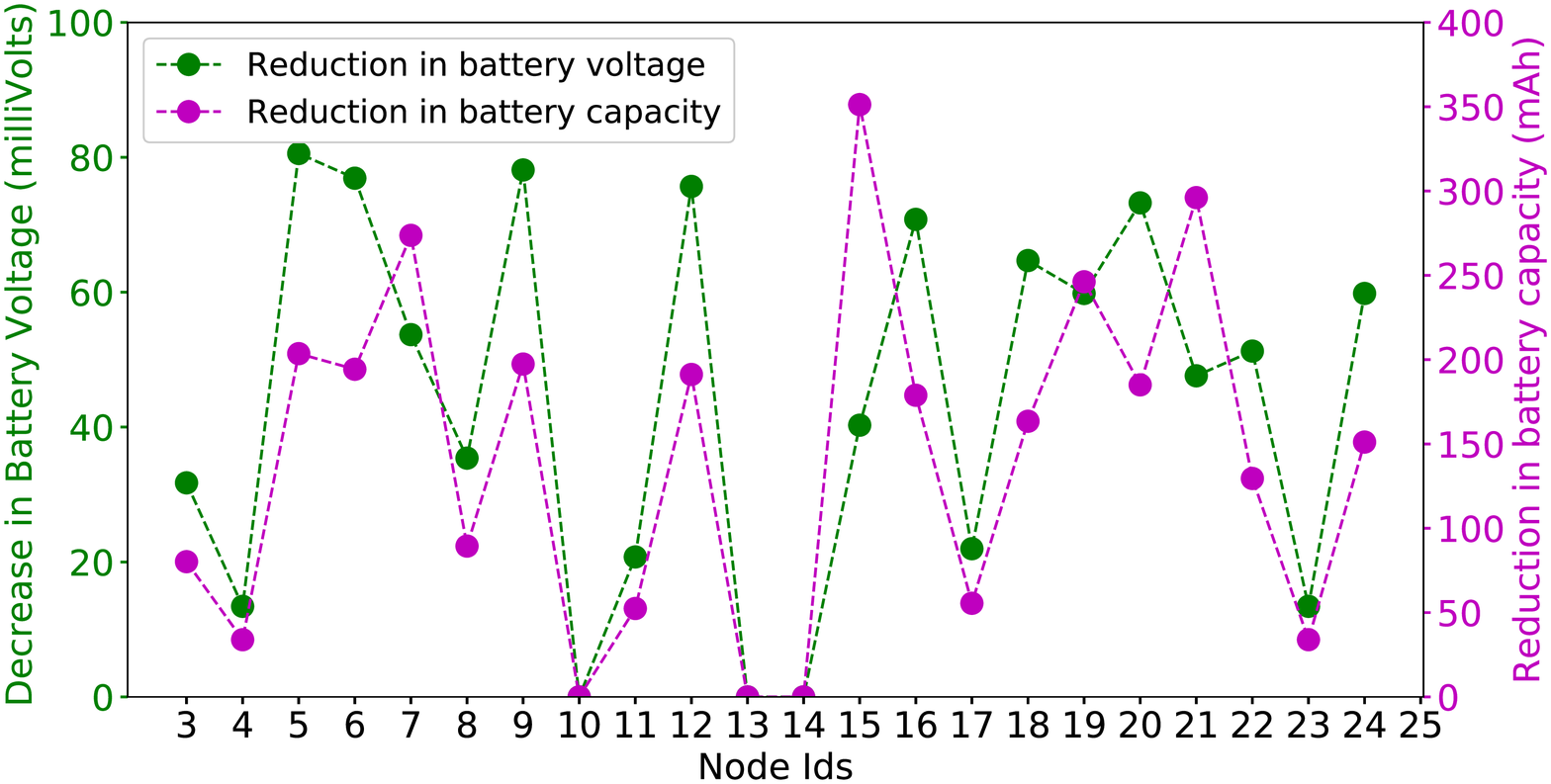}}
\hfil
\caption{Decrease in battery voltage and the corresponding energy expenditure of nodes - test case 2}
\label{energyprofile_1hr}
\end{figure}

In this test case, the data collection interval between two consecutive time slots is 1 hour and the data collection continued for more than 24 hours 
in each round.
The experiment was initiated in such a way that the tree building phase will happen during the day time for the three rounds and 
hence, every node will have continuous data collection occurring throughout the night time. We use the difference between 
the battery voltage reported during the evening slot and the next day morning slot for calculating the energy consumed during the respective 
night. Fig. \ref {energyprofile_1hr} shows the discharge profiles of different nodes in each round.

\subsubsection* {Observations}
\begin{itemize}
 \item {Zeros in Fig. \ref{energyprofile_1hr} indicate that either the difference in voltages was observed as negative (can be because of ADC sensing error) or the sensed voltage is 
 wrong (junk data).}
 
 \item {The decreases in voltages are too small and hence, we are not able to confirm the earlier findings that the sleep message timeouts
 create more energy discharge in the leaf nodes. Also it can be because of larger data collection interval of 1 hour compared to 10 mnt in the previous case
 and hence, the total number of data collection time slots under consideration will be less.
 }
 
 \item { \textit {Qualitatively, we can say that the imbalance in the energy consumption of nodes which can be created due to various factors like
more sleep timeouts for the nodes farther from the sink or more data forwards by the nodes nearby the sink are quite small in a day. Therefore, we do not need to
give unwanted importance to these factors for low duty cycled sensor network design for outdoor applications where solar energy is available.}}
 
\item {These factors become more critical in environments where a rechargeable energy source is not available.}
 
\end{itemize}

\begin{figure*}
\centering
\subfloat[Graph 1]{\includegraphics[scale=0.43]{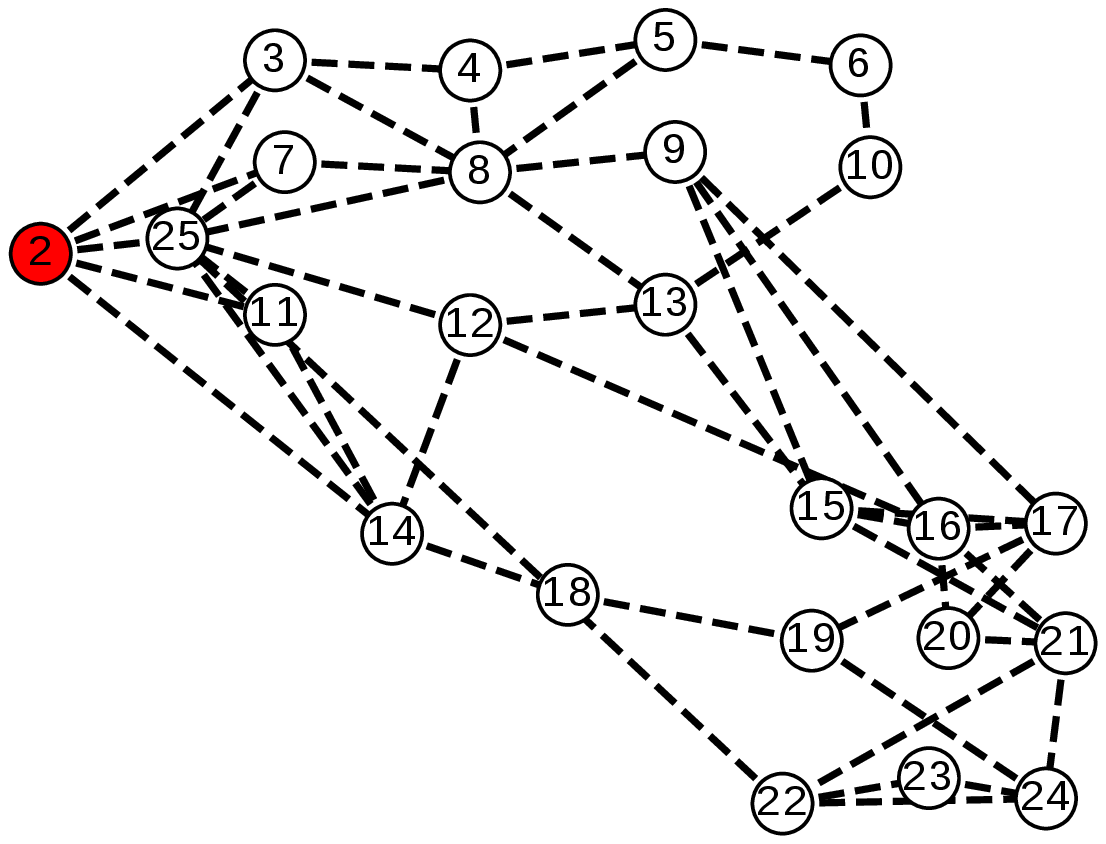}}
\hfil
\subfloat[Data collection tree formed from graph 1]{\includegraphics[scale=0.235]{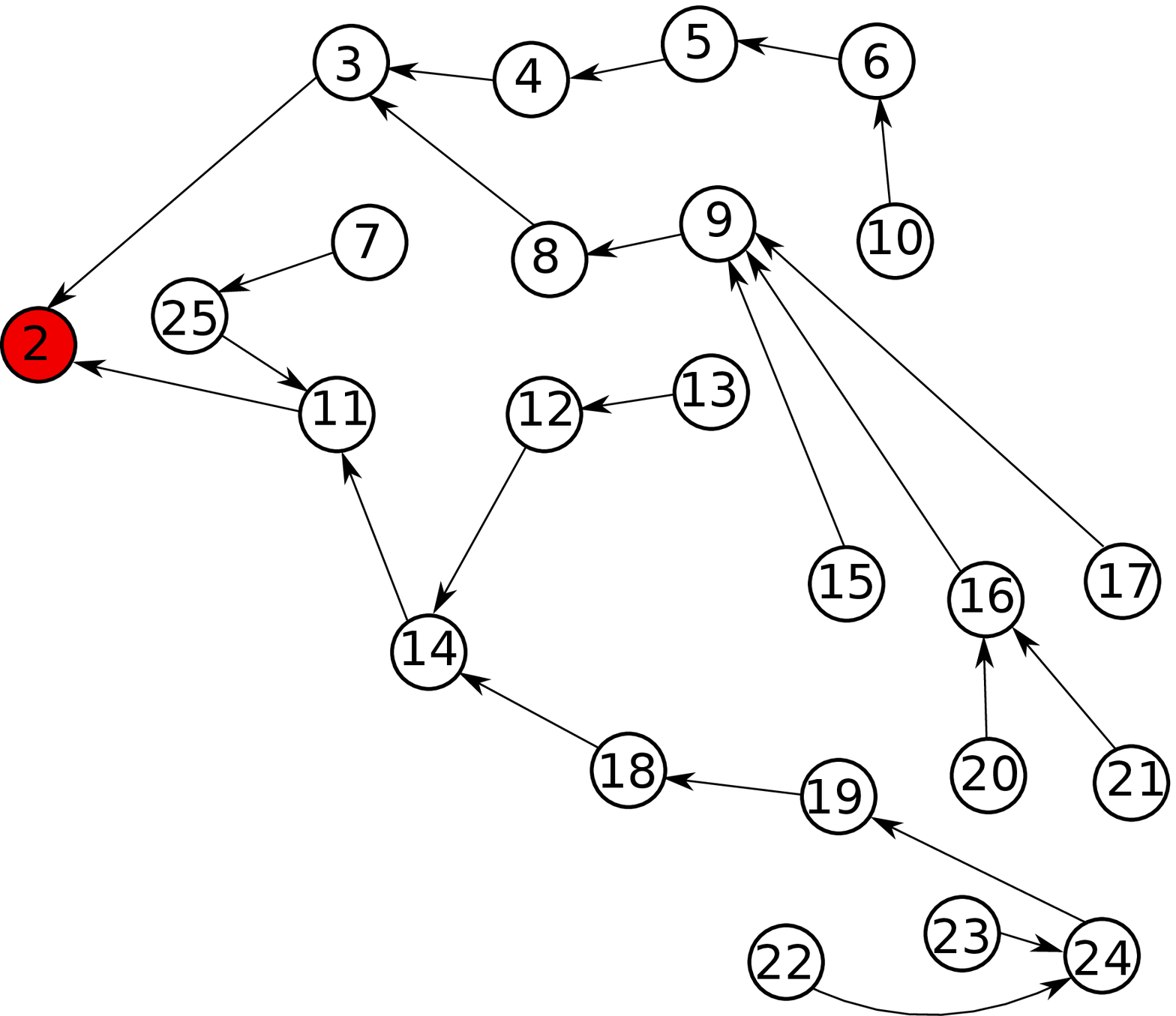}}
\hfil
\subfloat[Node yield at the sink node]{\label{nodeyield_3hr_tree1}\includegraphics[scale=0.25]{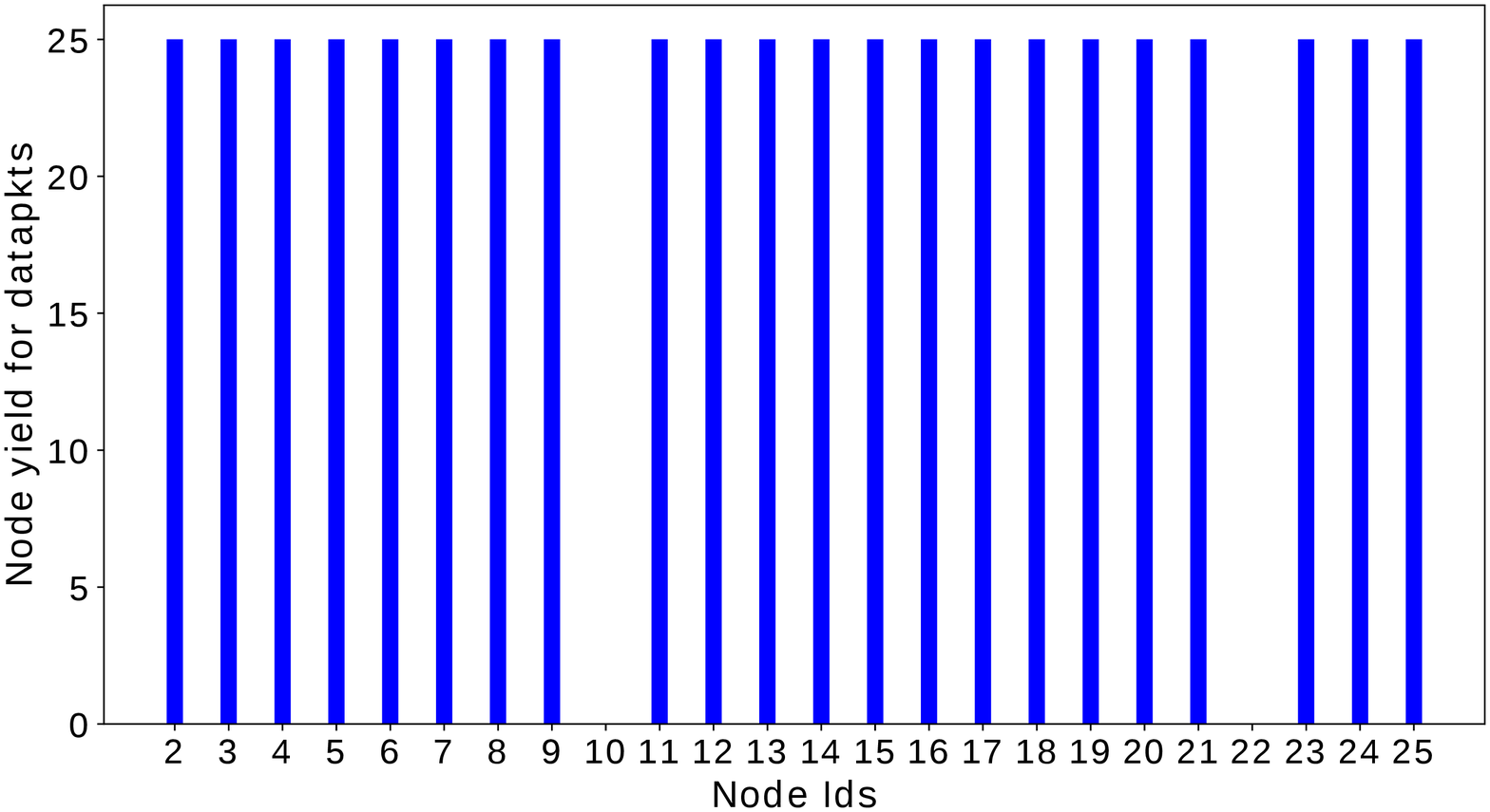}}
\hfil
\subfloat[Snooper data analysis]{\label{nodedataanalysis_3hr_tree1}\includegraphics[scale=0.6]{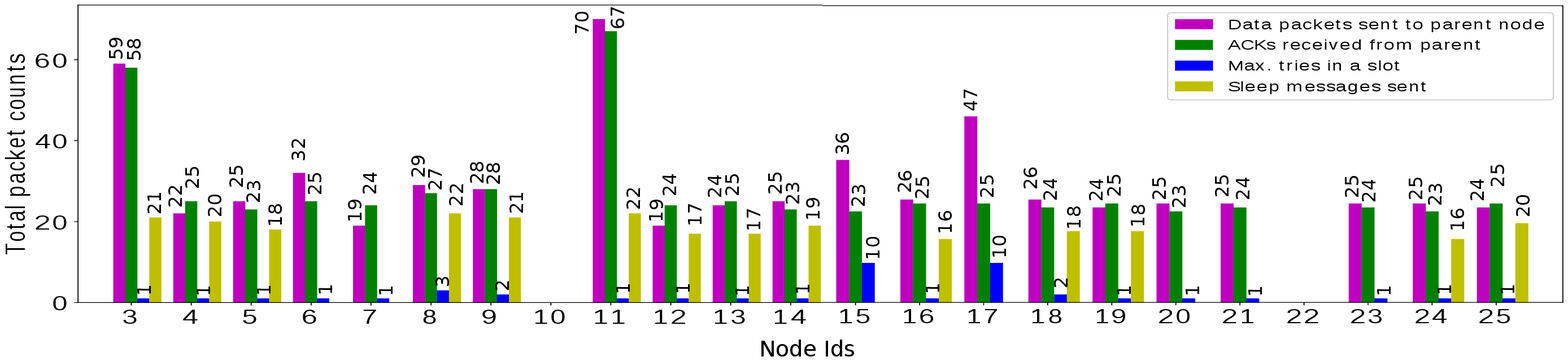}}
\hfil
\caption{Graph, respective data collection tree and the packet transmissions in the network - round 1, test case 3}
\label{datacollectionround1_3hr}
\end{figure*}

\subsection {Test case 3}

In this test case, we have increased the interval between two data collection time slots to 3 hours; also, the sink node collects the network details and
rebuilds the tree after every 25 time slots, $(25 \times 3) = 75 $ hours (3 days). Many of the monitoring applications need to sense and 
collect data from sensors only once every few hours. Through this test case we aim to validate the performance of our proposed simple 
synchronized tree based data collection architecture to collect data through the same tree for a long time.

We have also verified the behaviour of node failure in these trials by switching off the nodes manually.

During the round 1 data collection, node Ids 10 and 22 were switched off after the tree formation, to verify how the data collection 
tree responds to the node failures. As the failed nodes are leaf nodes it is observed that the data collection continued 
through the same tree, just after the removal of the failed leaf nodes. The complete details of round 1 data collection are shown in Fig.
\ref{datacollectionround1_3hr}.

\begin{figure*}
\centering

\subfloat[Graph 2]{\includegraphics[scale=0.5]{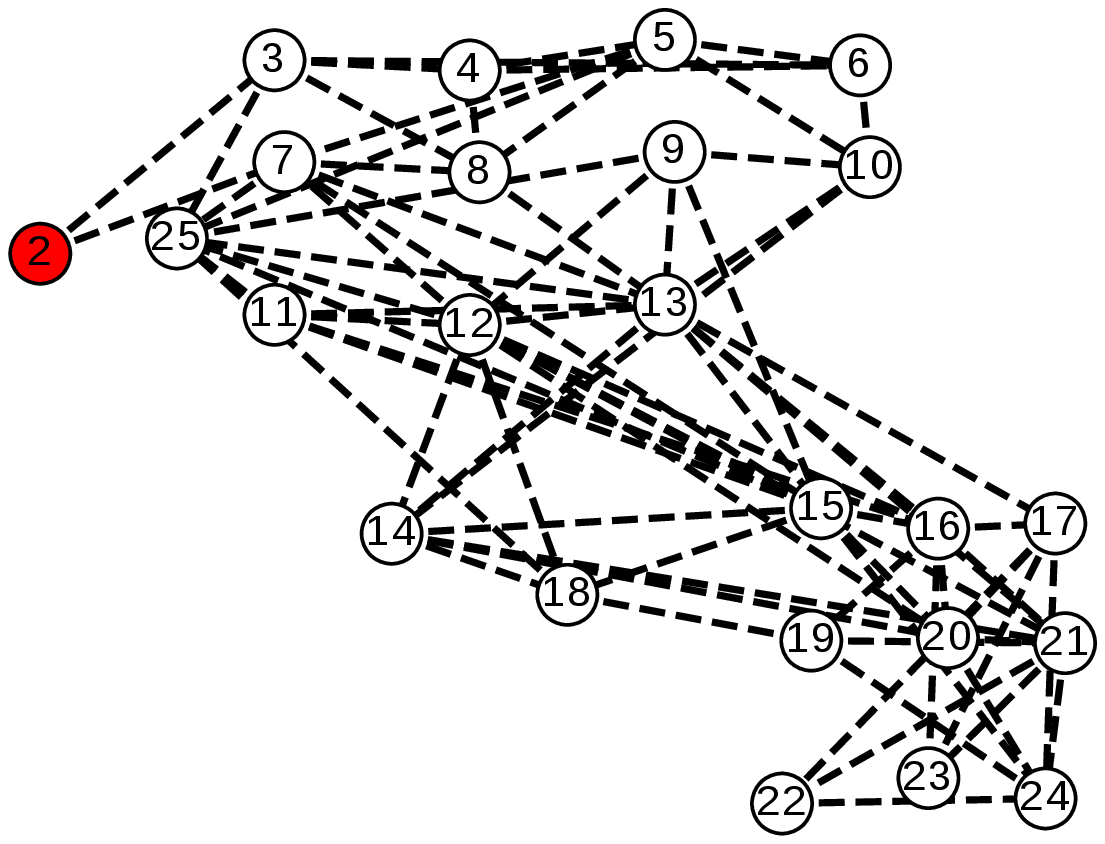}}
\hfill
\subfloat[Data collection tree formed from graph 2]{\includegraphics[scale=0.23]{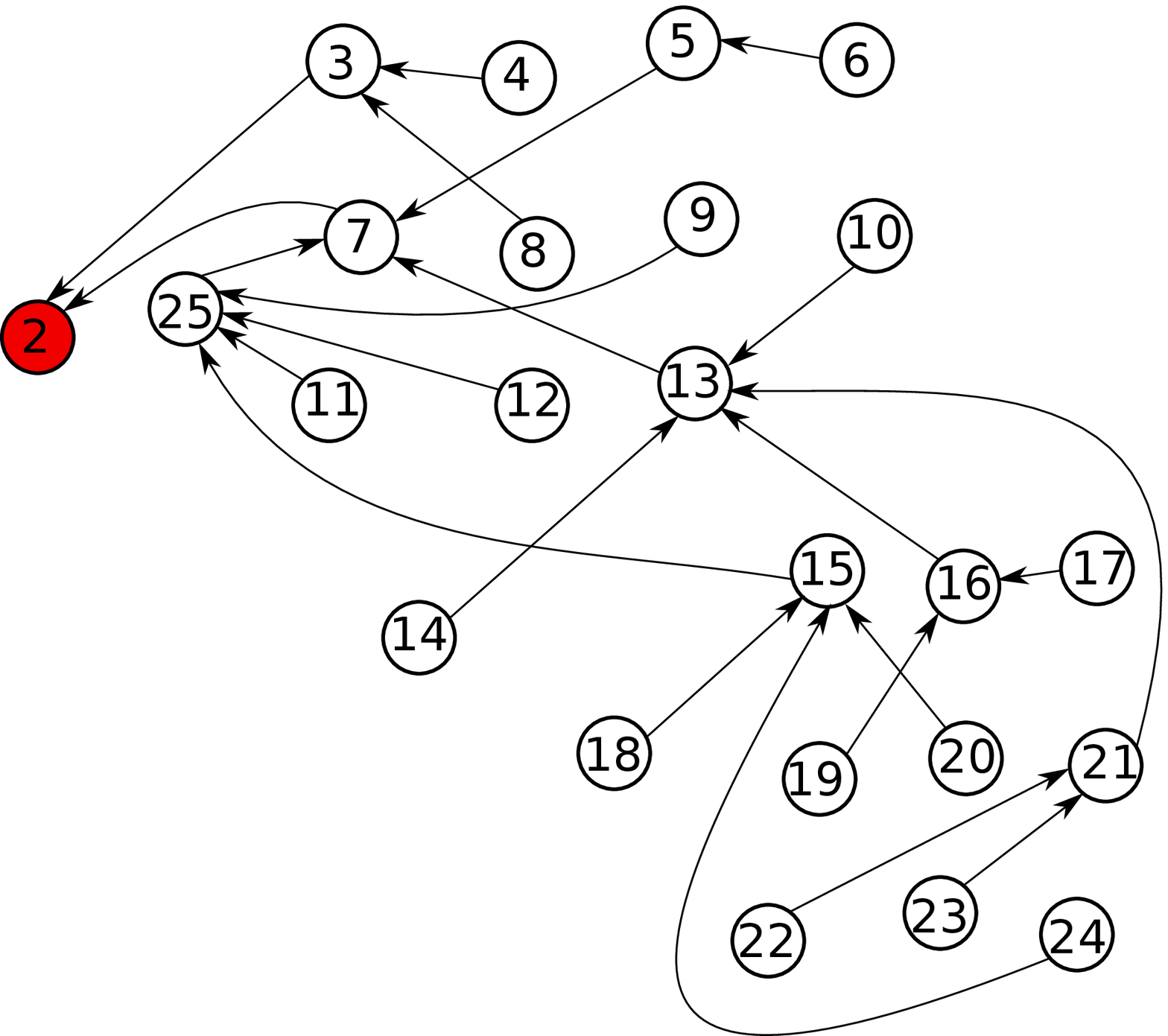}}
\hfill
\subfloat[Data collection tree after removal of node 13]{\includegraphics[scale = 0.25]{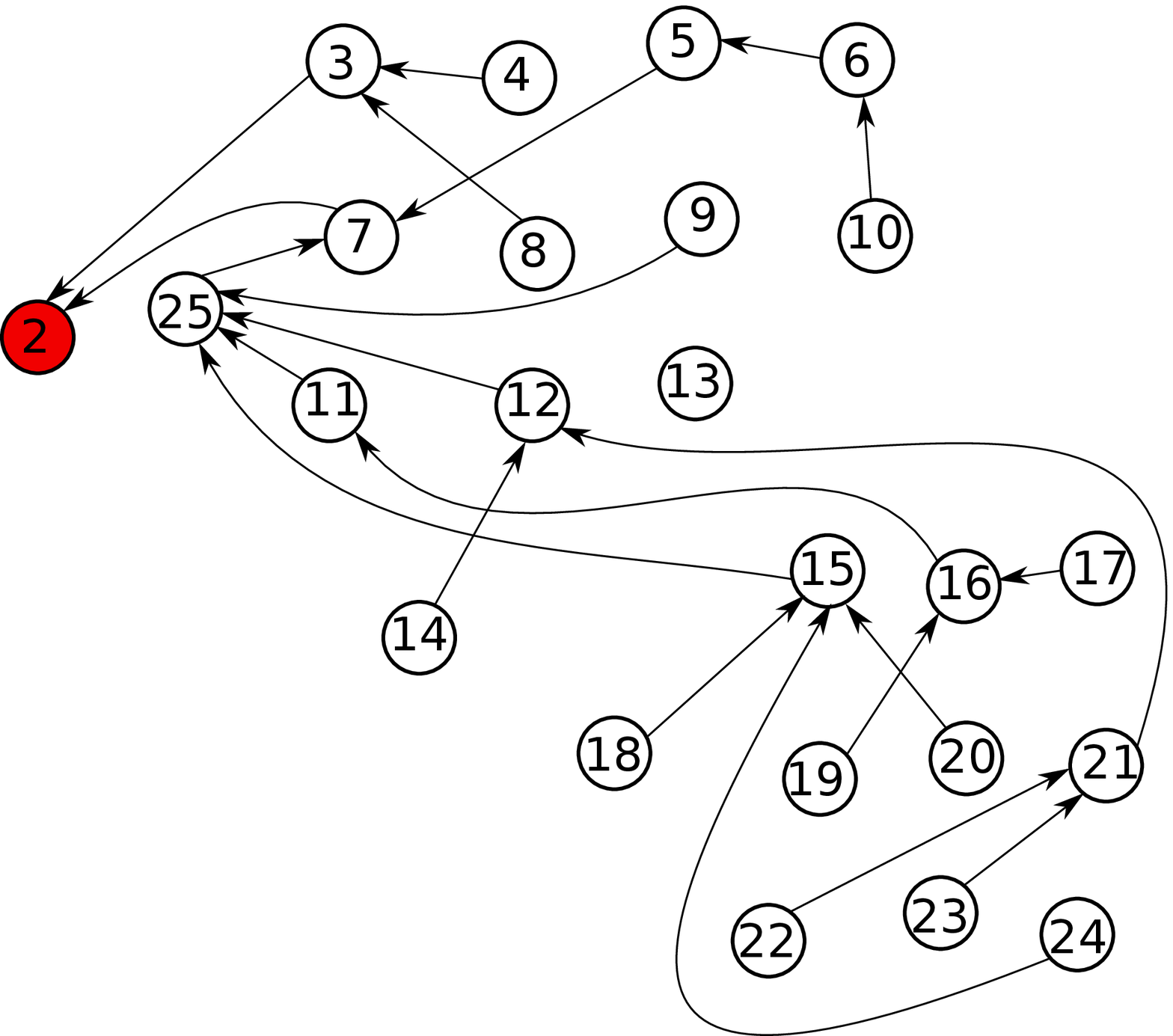}}
\hfill
\subfloat[Snooper data analysis - round 2]{\label{nodedataanalysis_3hr_tree2_nodefailure}\includegraphics[scale = 0.55]{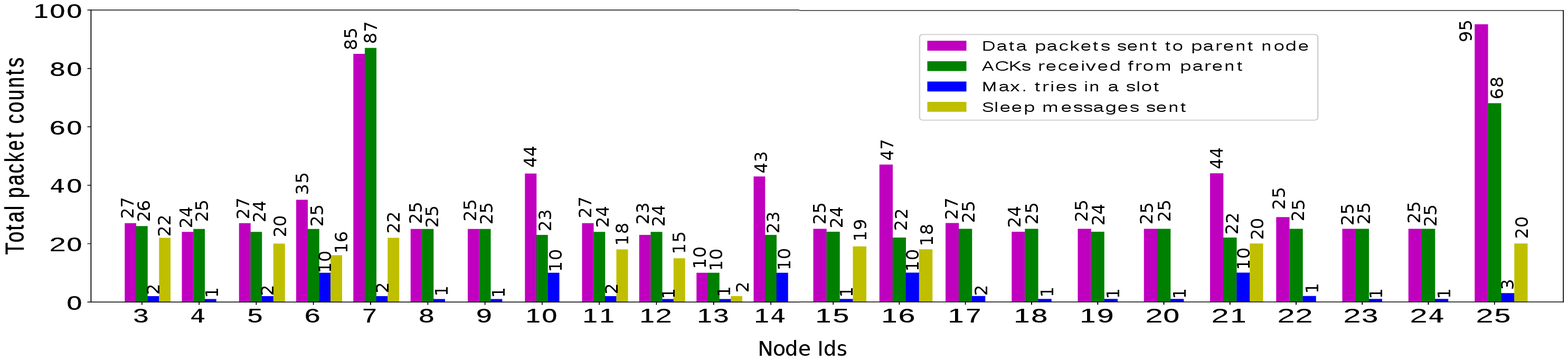}}
\caption{Graphs and the data collection trees - round 2, test case 3}
\label{datacollectionround2_3hr}
\end{figure*}

\subsubsection{Observations from the snooper data analysis of round~1 data collection}
\begin{itemize}
 \item {Node 10 and 22 have zero node yield as these nodes are switched off as a part of node failure testing.
 Also they do not have packet transmissions reflected in the snooper nodes.}
  \item {Nodes 3 and 11 have more data transmissions as they have to forward data from more descendant nodes which does not fit into one packet.}
 \item {ACK-SLEEP dualloss occurred in Node 15 and Node 17. But it happened just once in node 15 and two times in node 17 out of 25 time slots. }
\end{itemize}

During the data collection round 2, after collecting the data through the tree for the first 3 time slots, node Id 13 was switched off 
intentionally to understand how the tree will be adjusted and the new tree is shown in Fig. \ref{nodedataanalysis_3hr_tree2_nodefailure}. 
Please note that this tree is constructed after the removal of node Id 13,
from the previously collected network graph, in accordance 
with the node failure handling mechanisms detailed in Section \ref{nodefailure_schmes}.

\begin{figure}
\centering
 \includegraphics[scale = 0.3]{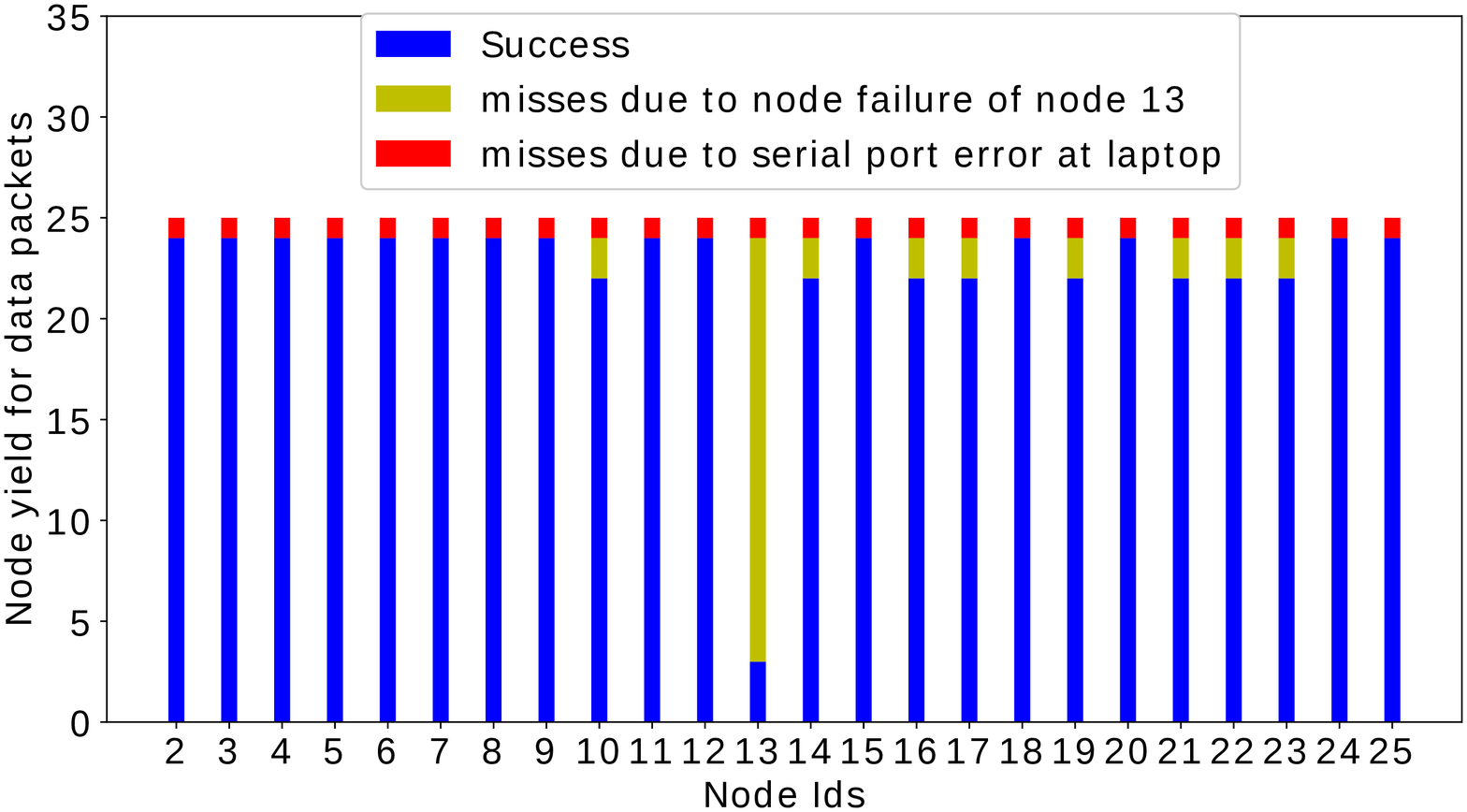}
 \caption{Node yield at the sink node for data collection round 2, test case 3}
 \label{nodeyield_round2_3hr}
\end{figure}

\subsubsection{Observations from the snooper data analysis of round~2 data collection}
\begin{itemize}
\item {Nodes 10,14,16,21 have more packet transmissions with max tries in a data collection time slot as 10 because of the failure of their parent node 13.}
\item {Nodes 7 and 25 have more packet transmissions as they have to forward more data from its descendants which cannot be fitted into
a single data packet.}
\item {There are packet transmissions in the network which are not captured in the snooper nodes (inferred from the fact that some nodes have more ACKs from their parent
than their data packet transmissions).}
\item {Node 13 shows ten data packet transmissions to its parent node even though it was alive only for 3 data collection time slots. 
This is because it was not able to fit the data from its children into a single packet and hence, transferred it as multiple packets.}
\end{itemize}

\begin{figure*}

\centering

\subfloat[Graph 3]{\includegraphics[scale=0.45]{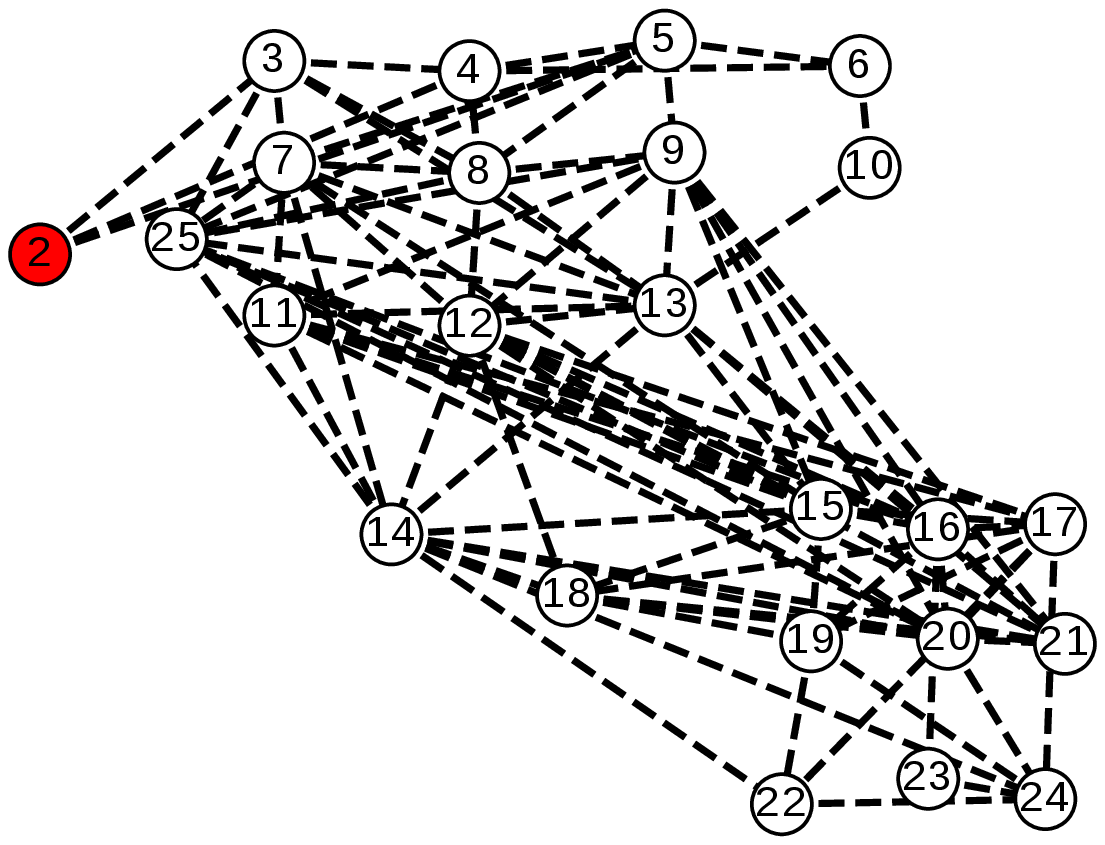}}
\hfil
\subfloat[Data collection tree formed from graph 3]{\includegraphics[scale=0.27]{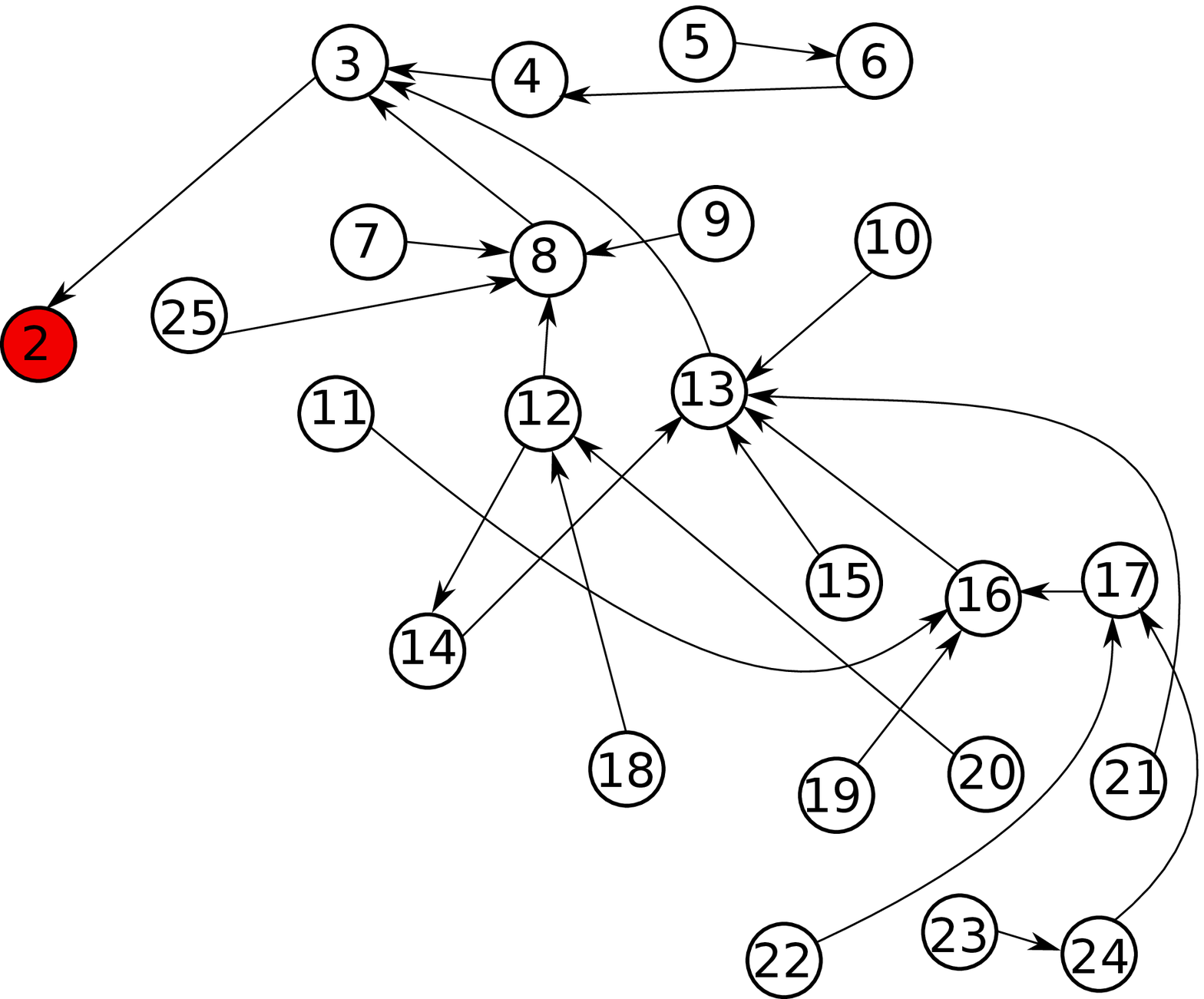}}
\hfil
\subfloat[Node yield at the sink node]{\label{nodeyield_3hr_tree3}\includegraphics[scale=0.25]{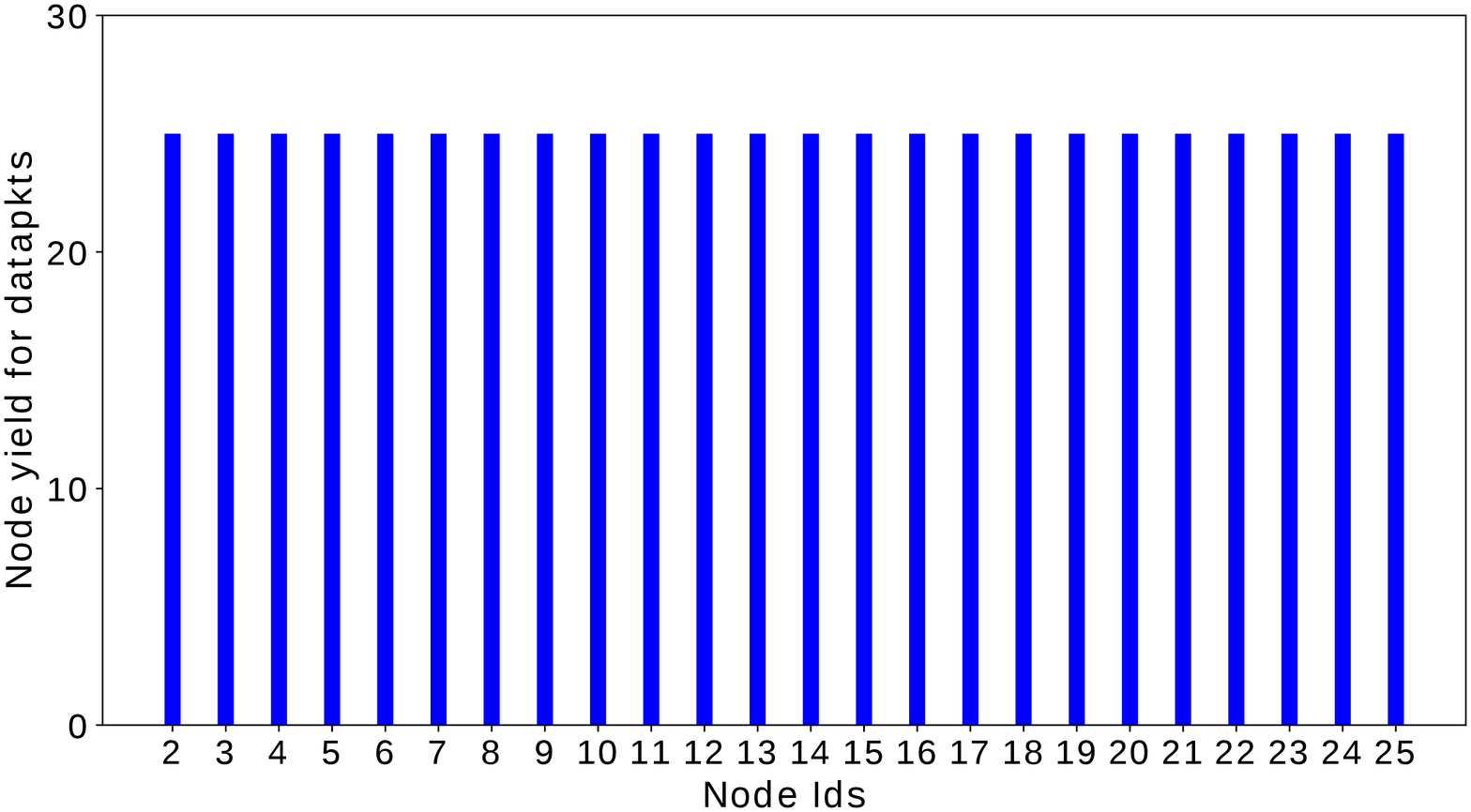}}
\hfil
\subfloat[Snooper data analysis]{\label{nodedataanalysis_3hr_tree3}\includegraphics[scale=0.7]{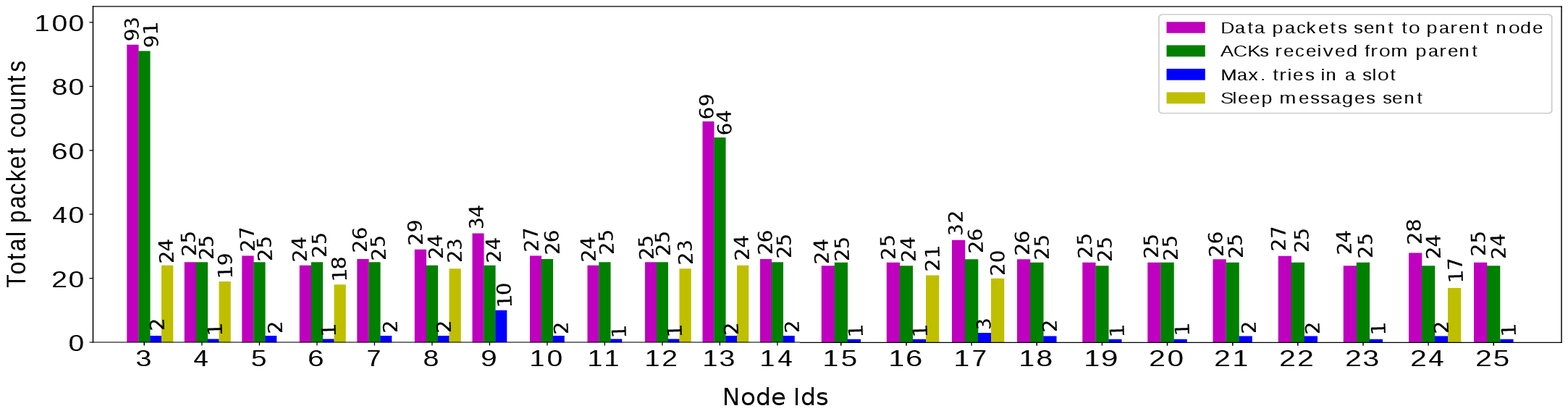}}
\hfil
\caption{Graph, respective data collection tree and the packet transmissions in the network - round 3, test case 3}
\label{datacollectionround3_3hr}
\end{figure*}

\begin{figure}
\centering
\subfloat[Battery voltage of node-Id4]{\includegraphics[scale=0.22]{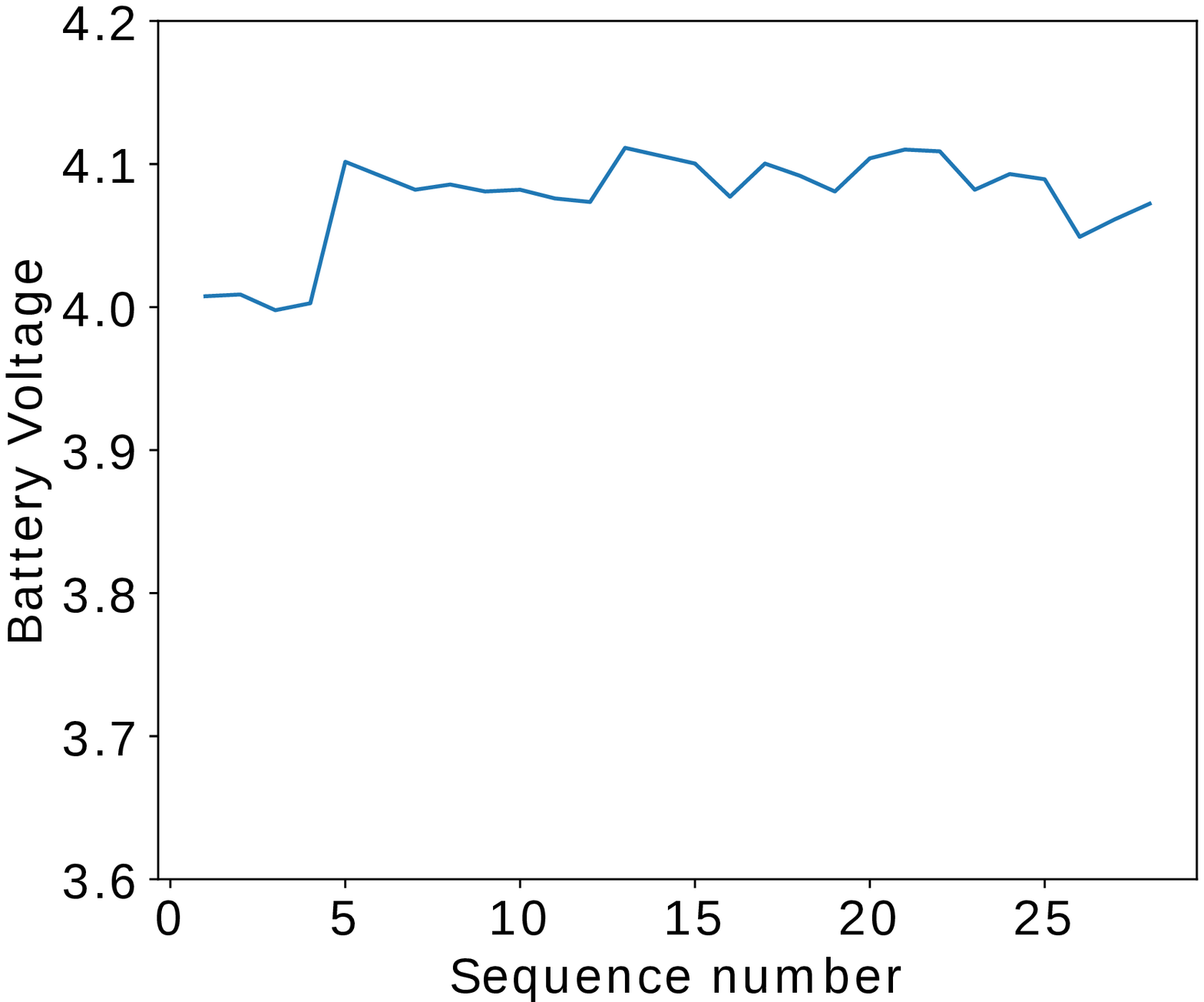}}
\hfil
\subfloat[All parameters of node-Id4]{\includegraphics[scale=0.19]{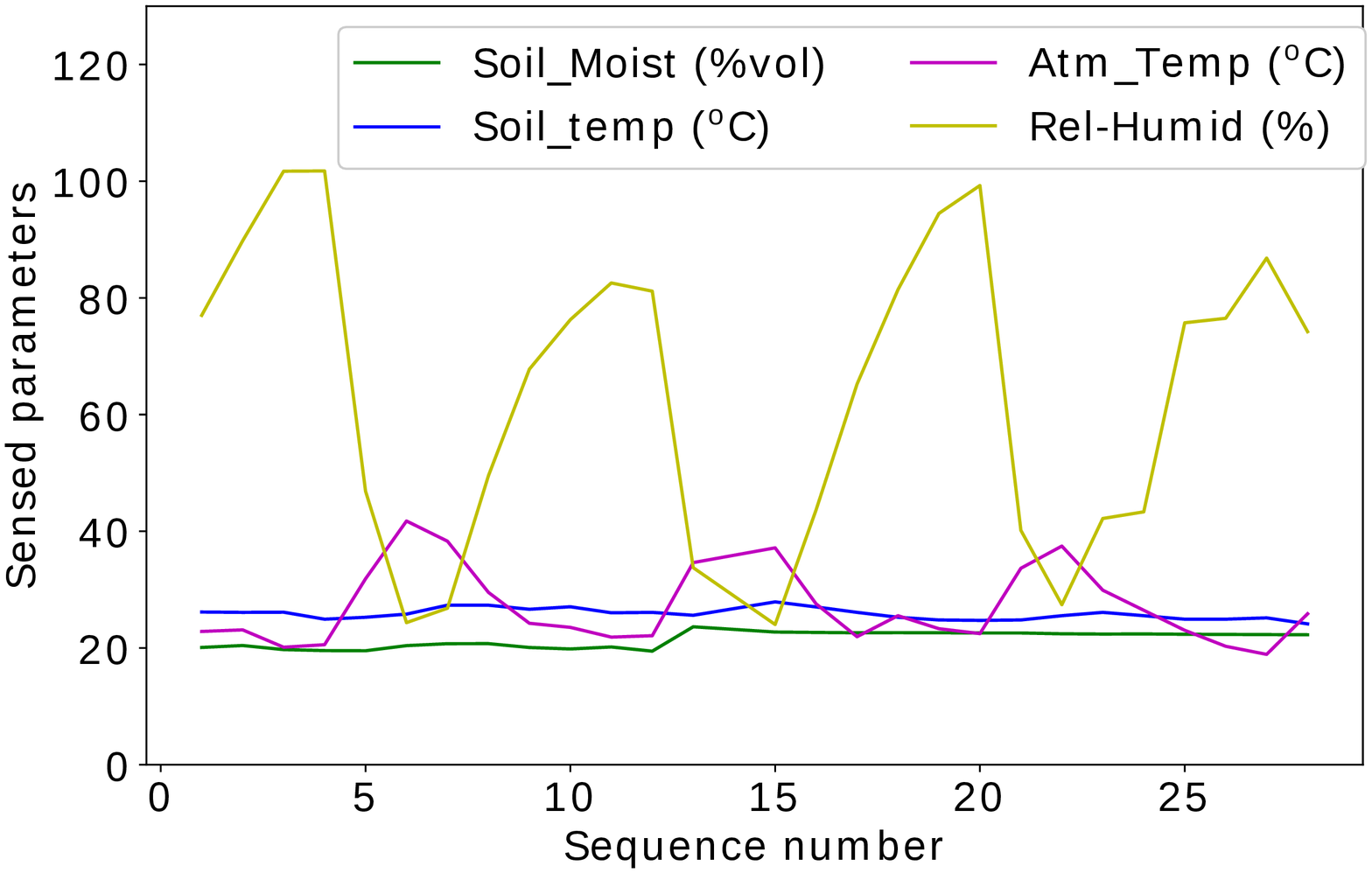}}
\hfil
\subfloat[Battery voltage of node-Id5]{\includegraphics[scale=0.22]{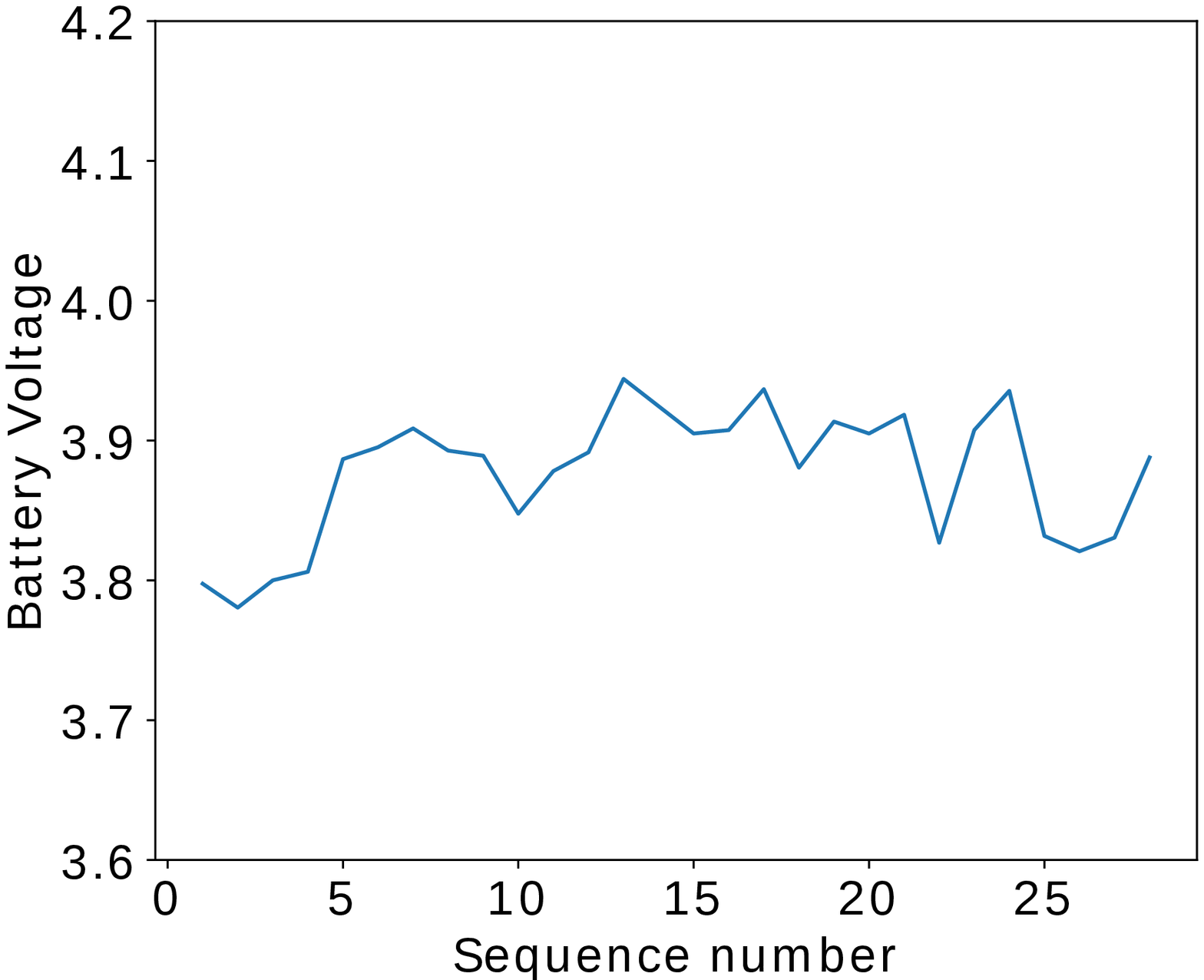}}
\hfil
\subfloat[All parameters of node-Id5]{\includegraphics[scale=0.19]{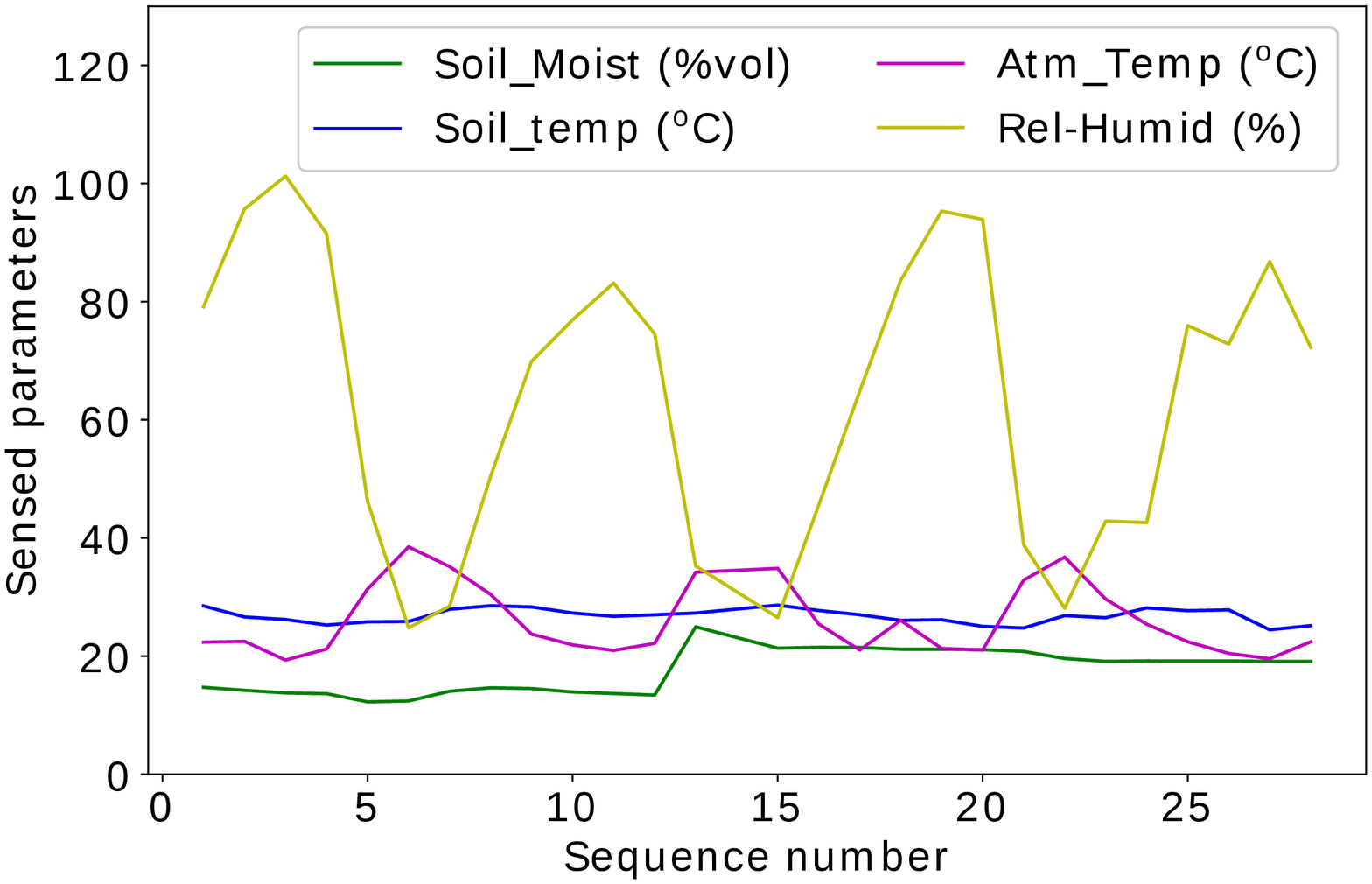}}
\caption{Sensed parameters - data collection round 2, test case 3. Time interval between two consecutive sequence numbers is 3 hours.}
\label{sennsordata_3hr_tree2}
\end{figure}

Data collection in round 3 which is captured in Fig. \ref{datacollectionround3_3hr} has almost the same 
characteristics as that of 1-hour interval case. Node 3 has to forward multiple data packets in a data collection time slot since 
it could not fit its data as well as that of its descendants into a single data packet. ACK SLEEP timeout occurred just once in node 9.

The sensor data variations for one of the data collection rounds is shown in Fig. \ref{sennsordata_3hr_tree2} for two different nodes. 
For optimum irrigation, the soil moisture in agricultural fields needs to be maintained between two limits, the field capacity (FC) and the permanent
wilting point (PWP) \cite{veihmeyer1931moisture} \cite{sim2015assessment}. The sensor systems were deployed in red soil having silt loam soil texture which has 22\% as the FC and 12\% 
as the PWP \cite{veeramalluramulu}. The soil moisture was maintained in the required window for most of the time and the rise in the soil moisture
profile in Fig. \ref{sennsordata_3hr_tree2} corresponds to an irrigation.
The variation in atmospheric humidity was observed as approximately 75\%
in a day while the atmospheric temperature has a variation of 20\textdegree{}C. The soil temperature which was measured
at a depth of 15 cm from the surface showed a variation of 5\textdegree{}C in a day. 

\section{Conclusions}
\label{conclusion}
This paper describes the design and implementation of a shortest path tree based data collection network architecture for low duty-cycled monitoring
applications. The proposed scheme is implemented in a wireless sensor network of 24 nodes in an agricultural field of 3 acre size with various
sensors to monitor soil moisture, soil temperature, atmospheric temperature and relative humidity. The performance of the synchronized
tree based data collection approach is characterized with the help of snooper nodes. It is observed that wireless links in
agricultural fields are quite stable in nature and hence, the same data collection tree can be used for collecting the sensor data for a
long time period, which reduces the energy overhead required for data collection tree building. The energy expenditure profiles of the
sensor nodes reveal that the imbalance in the energy consumption of nodes due to various factors are quite small in a day. Hence, complicated 
design approaches to balance loads in the sensor nodes may not be required in low duty-cycled applications where nodes can 
recharge their batteries using solar energy. Also we could observe that simple time synchronization approaches were good enough to ensure that the nodes
follow a periodic time synchronized sleep-wakeup schedule. The network could collect the sensor data from all the nodes at the sink node for
most of the data collection time slots except for a few temporary errors which  happened during the testing. A very simple battery voltage based model 
was used to find out the remaining capacity of the Li-ion cells. We presented the results of extensive tests conducted on our implementation and their analysis in this paper, which provide significant insights.

\subsection{Acknowledgements}
The  authors  would  like  to  acknowledge  DeitY,  MCIT,
Government of India for the financial support through Centre
of Excellence in Nanoelectronics, IIT-Bombay. We also would like to 
thank Water Technology Center, PJTSAU, Hyderabad for providing us the
support to test the sensor network in their agricultural fields. We also thank
all the members of Integrated Systems Lab, IIT Bombay for the fruitful
discussions and suggestions.

\bibliographystyle{EAI_num}
\bibliography{references}

\begin{thebibliography}{10}
\providecommand{\url}[1]{\texttt{#1}}
\providecommand{\urlprefix}{URL }
\expandafter\ifx\csname urlstyle\endcsname\relax
  \providecommand{\doi}[1]{doi:\discretionary{}{}{}#1}\else
  \providecommand{\doi}{doi:\discretionary{}{}{}\begingroup
  \urlstyle{rm}\Url}\fi
\providecommand{\eprint}[2][]{\url{#2}}

\bibitem{john2015design}
\textsc{John, J.} \emph{et~al.} (2015) Design and implementation of a soil
  moisture wireless sensor network. In \emph{NCC} (IEEE): 1--6.

\bibitem{chong2003sensor}
\textsc{Chong, C.Y.} and \textsc{Kumar, S.P.} (2003) Sensor networks:
  evolution, opportunities, and challenges. \emph{Proceedings of the IEEE}
  \textbf{91}(8): 1247--1256.

\bibitem{akyildiz2002survey}
\textsc{Akyildiz, I.F.}, \textsc{Su, W.}, \textsc{Sankarasubramaniam, Y.} and
  \textsc{Cayirci, E.} (2002) A survey on sensor networks. \emph{IEEE
  Communications magazine} \textbf{40}(8): 102--114.

\bibitem{zheng2009wireless}
\textsc{Zheng, J.} and \textsc{Jamalipour, A.} (2009) \emph{Wireless sensor
  networks: a networking perspective} (John Wiley \& Sons).

\bibitem{pakzad2008design}
\textsc{Pakzad, S.N.}, \textsc{Fenves, G.L.}, \textsc{Kim, S.} and
  \textsc{Culler, D.E.} (2008) Design and implementation of scalable wireless
  sensor network for structural monitoring. \emph{Journal of infrastructure
  systems} \textbf{14}(1): 89--101.

\bibitem{mainwaring2002wireless}
\textsc{Mainwaring, A.} \emph{et~al.} (2002) Wireless sensor networks for
  habitat monitoring. In \emph{Proceedings of the 1st ACM international
  workshop on Wireless sensor networks and applications} (Acm): 88--97.

\bibitem{lombardo2017wireless}
\textsc{Lombardo, L.} \emph{et~al.} (2017) Wireless sensor network for
  distributed environmental monitoring. \emph{IEEE Transactions on
  Instrumentation and Measurement} .

\bibitem{lazarescu2013design}
\textsc{Lazarescu, M.T.} (2013) Design of a wsn platform for long-term
  environmental monitoring for iot applications. \emph{IEEE Journal on emerging
  and selected topics in circuits and systems} \textbf{3}(1): 45--54.

\bibitem{yang2010design}
\textsc{Yang, J.} and \textsc{Li, X.} (2010) Design and implementation of
  low-power wireless sensor networks for environmental monitoring. In
  \emph{Wireless Communications, Networking and Information Security (WCNIS),
  2010 IEEE International Conference on} (IEEE): 593--597.

\bibitem{werner2006fidelity}
\textsc{Werner-Allen, G.} \emph{et~al.} (2006) Fidelity and yield in a volcano
  monitoring sensor network. In \emph{Proceedings of the 7th symposium on
  Operating systems design and implementation} (USENIX Association): 381--396.

\bibitem{greenorbs}
\textsc{Liu, Y.} \emph{et~al.} (2013) Does wireless sensor network scale? a
  measurement study on greenorbs. \emph{IEEE Transactions on Parallel and
  Distributed Systems} \textbf{24}(10): 1983--1993.

\bibitem{VibhaDhawan}
\textsc{Dhawan, V.} (2017) Water and agriculture in india. \emph{Background
  paper for the South Asia expert panel GFFA 2017} .

\bibitem{fishman2015can}
\textsc{Fishman, R.}, \textsc{Devineni, N.} and \textsc{Raman, S.} (2015) Can
  improved agricultural water use efficiency save india’s groundwater?
  \emph{Environmental Research Letters} \textbf{10}(8): 084022.

\bibitem{mashnik2017increasing}
\textsc{Mashnik, D.} \emph{et~al.} (2017) Increasing productivity through
  irrigation: Problems and solutions implemented in africa and asia.
  \emph{Sustainable Energy Technologies and Assessments} \textbf{22}: 220--227.

\bibitem{palaparthy2013review}
\textsc{Palaparthy, V.S.}, \textsc{Baghini, M.S.} and \textsc{Singh, D.N.}
  (2013) Review of polymer-based sensors for agriculture-related applications.
  \emph{Emerging Materials Research} \textbf{2}(4): 166--180.

\bibitem{hatfield2015temperature}
\textsc{Hatfield, J.L.} and \textsc{Prueger, J.H.} (2015) Temperature extremes:
  effect on plant growth and development. \emph{Weather and climate extremes}
  \textbf{10}: 4--10.

\bibitem{ojha2015wireless}
\textsc{Ojha, T.}, \textsc{Misra, S.} and \textsc{Raghuwanshi, N.S.} (2015)
  Wireless sensor networks for agriculture: The state-of-the-art in practice
  and future challenges. \emph{Computers and Electronics in Agriculture}
  \textbf{118}: 66--84.

\bibitem{keshtgary2012efficient}
\textsc{Keshtgary, M.} and \textsc{Deljoo, A.} (2012) An efficient wireless
  sensor network for precision agriculture. \emph{Canadian Journal on
  Multimedia and Wireless Networks} \textbf{3}(1): 1--5.

\bibitem{mafuta2013successful}
\textsc{Mafuta, M.} \emph{et~al.} (2013) Successful deployment of a wireless
  sensor network for precision agriculture in malawi--wipam. In \emph{3rd IEEE
  International Conference On Networked Embedded Systems For Every
  Application}.

\bibitem{zhou2009wireless}
\textsc{Zhou, Y.}, \textsc{Yang, X.}, \textsc{Wang, L.} and \textsc{Ying, Y.}
  (2009) A wireless design of low-cost irrigation system using zigbee
  technology. In \emph{Networks Security, Wireless Communications and Trusted
  Computing, 2009. NSWCTC'09. International Conference on} (IEEE), ~\textbf{1}:
  572--575.

\bibitem{gutierrez2014automated}
\textsc{Guti{\'e}rrez, J.} \emph{et~al.} (2014) Automated irrigation system
  using a wireless sensor network and gprs module. \emph{IEEE transactions on
  instrumentation and measurement} \textbf{63}(1): 166--176.

\bibitem{kim2008remote}
\textsc{Kim, Y.}, \textsc{Evans, R.G.} and \textsc{Iversen, W.M.} (2008) Remote
  sensing and control of an irrigation system using a distributed wireless
  sensor network. \emph{IEEE transactions on instrumentation and measurement}
  \textbf{57}(7): 1379--1387.

\bibitem{raman2008censor}
\textsc{Raman, B.} and \textsc{Chebrolu, K.} (2008) Censor networks: A critique
  of sensor networks from a systems perspective. \emph{ACM SIGCOMM Computer
  Communication Review} \textbf{38}(3): 75--78.

\bibitem{van2003adaptive}
\textsc{Van~Dam, T.} and \textsc{Langendoen, K.} (2003) An adaptive
  energy-efficient mac protocol for wireless sensor networks. In
  \emph{Proceedings of the 1st international conference on Embedded networked
  sensor systems} (ACM): 171--180.

\bibitem{lu2004adaptive}
\textsc{Lu, G.}, \textsc{Krishnamachari, B.} and \textsc{Raghavendra, C.S.}
  (2004) An adaptive energy-efficient and low-latency mac for data gathering in
  wireless sensor networks. In \emph{Parallel and Distributed Processing
  Symposium, 2004. Proceedings. 18th International} (IEEE): 224.

\bibitem{el2003wisemac}
\textsc{El-Hoiydi, A.}, \textsc{Decotignie, J.D.}, \textsc{Enz, C.} and
  \textsc{Le~Roux, E.} (2003) Wisemac, an ultra low power mac protocol for the
  wisenet wireless sensor network. In \emph{Proceedings of the 1st
  international conference on Embedded networked sensor systems} (ACM):
  302--303.

\bibitem{rajendran2006energy}
\textsc{Rajendran, V.}, \textsc{Obraczka, K.} and \textsc{Garcia-Luna-Aceves,
  J.J.} (2006) Energy-efficient, collision-free medium access control for
  wireless sensor networks. \emph{Wireless networks} \textbf{12}(1): 63--78.

\bibitem{sohrabi2000protocols}
\textsc{Sohrabi, K.}, \textsc{Gao, J.}, \textsc{Ailawadhi, V.} and
  \textsc{Pottie, G.J.} (2000) Protocols for self-organization of a wireless
  sensor network. \emph{IEEE personal communications} \textbf{7}(5): 16--27.

\bibitem{rhee2008z}
\textsc{Rhee, I.} \emph{et~al.} (2008) Z-mac: a hybrid mac for wireless sensor
  networks. \emph{IEEE/ACM Transactions on Networking (TON)} \textbf{16}(3):
  511--524.

\bibitem{ahn2006funneling}
\textsc{Ahn, G.S.} \emph{et~al.} (2006) Funneling-mac: a localized,
  sink-oriented mac for boosting fidelity in sensor networks. In
  \emph{Proceedings of the 4th international conference on Embedded networked
  sensor systems} (ACM): 293--306.

\bibitem{sun2014energy}
\textsc{Sun, W.}, \textsc{Yang, Z.}, \textsc{Zhang, X.} and \textsc{Liu, Y.}
  (2014) Energy-efficient neighbor discovery in mobile ad hoc and wireless
  sensor networks: A survey. \emph{IEEE Communications Surveys \& Tutorials}
  \textbf{16}(3): 1448--1459.

\bibitem{TPSN}
\textsc{Ganeriwal, S.}, \textsc{Kumar, R.} and \textsc{Srivastava, M.B.} (2003)
  Timing-sync protocol for sensor networks. In \emph{Proceedings of the 1st
  international conference on Embedded networked sensor systems} (ACM):
  138--149.

\bibitem{FTSP}
\textsc{Mar{\'o}ti, M.}, \textsc{Kusy, B.}, \textsc{Simon, G.} and
  \textsc{L{\'e}deczi, {\'A}.} (2004) The flooding time synchronization
  protocol. In \emph{Proceedings of the 2nd international conference on
  Embedded networked sensor systems} (ACM): 39--49.

\bibitem{gnawali2009collection}
\textsc{Gnawali, O.} \emph{et~al.} (2009) Collection tree protocol. In
  \emph{Proceedings of the 7th ACM conference on embedded networked sensor
  systems} (ACM): 1--14.

\bibitem{liu2013citysee}
\textsc{Liu, Y.} \emph{et~al.} (2013) Citysee: Not only a wireless sensor
  network. \emph{IEEE Network} \textbf{27}(5): 42--47.

\bibitem{barrenetxea2008sensorscope}
\textsc{Barrenetxea, G.} \emph{et~al.} (2008) Sensorscope: Out-of-the-box
  environmental monitoring. In \emph{Proceedings of the 7th international
  conference on Information processing in sensor networks} (IEEE Computer
  Society): 332--343.

\bibitem{vicaire2009achieving}
\textsc{Vicaire, P.} \emph{et~al.} (2009) Achieving long-term surveillance in
  vigilnet. \emph{ACM Transactions on Sensor Networks (TOSN)} \textbf{5}(1): 9.

\bibitem{langendoen2006murphy}
\textsc{Langendoen, K.}, \textsc{Baggio, A.} and \textsc{Visser, O.} (2006)
  Murphy loves potatoes: Experiences from a pilot sensor network deployment in
  precision agriculture. In \emph{Parallel and Distributed Processing
  Symposium, 2006. IPDPS 2006. 20th International} (IEEE): 8--pp.

\bibitem{woo2003taming}
\textsc{Woo, A.}, \textsc{Tong, T.} and \textsc{Culler, D.} (2003) Taming the
  underlying challenges of reliable multihop routing in sensor networks. In
  \emph{Proceedings of the 1st international conference on Embedded networked
  sensor systems} (ACM): 14--27.

\bibitem{sky2007ultra}
\textsc{Sky, T.} (2007), Ultra low power ieee 802.15. 4 compliant wireless
  sensor module humidity, light, and temperature sensors with usb.

\bibitem{tinyos}
\textsc{Levis, P.} \emph{et~al.} (2005) Tinyos: An operating system for sensor
  networks. In \emph{Ambient intelligence} (Springer),  115--148.

\bibitem{cc2420datasheet}
\textsc{Datasheet, C.} (2006) 2.4 ghz ieee 802.15. 4/zigbee-ready rf
  transceiver. \emph{Chipcon products from Texas Instruments} .

\bibitem{fastcollection}
\textsc{Incel, O.D.} \emph{et~al.} (2012) Fast data collection in tree-based
  wireless sensor networks. \emph{IEEE Transactions on Mobile computing}
  \textbf{11}(1): 86--99.

\bibitem{baccour2012radio}
\textsc{Baccour, N.} \emph{et~al.} (2012) Radio link quality estimation in
  wireless sensor networks: A survey. \emph{ACM Transactions on Sensor Networks
  (TOSN)} \textbf{8}(4): 34.

\bibitem{RSSI_underappreciated}
\textsc{Levis, K.} (2006) Rssi is under appreciated. In \emph{Proceedings of
  the Third Workshop on Embedded Networked Sensors, Cambridge, MA, USA},
  \textbf{3031}: 239242.

\bibitem{temp_effects_RSSI}
\textsc{Boano, C.A.}, \textsc{Tsiftes, N.}, \textsc{Voigt, T.}, \textsc{Brown,
  J.} and \textsc{Roedig, U.} (2010) The impact of temperature on outdoor
  industrial sensornet applications. \emph{IEEE Transactions on Industrial
  Informatics} \textbf{6}(3): 451--459.

\bibitem{floodingcomparison}
\textsc{Williams, B.} and \textsc{Camp, T.} (2002) Comparison of broadcasting
  techniques for mobile ad hoc networks. In \emph{Proceedings of the 3rd ACM
  international symposium on Mobile ad hoc networking \& computing} (ACM):
  194--205.

\bibitem{cormen2009introduction}
\textsc{Cormen, T.H.} (2009) \emph{Introduction to algorithms} (MIT press).

\bibitem{Dijkstra}
\textsc{Bertsekas, D.P.}, \textsc{Gallager, R.G.} and \textsc{Humblet, P.}
  (1987) \emph{Data networks}, ~\textbf{2} (Prentice-hall Englewood Cliffs,
  NJ).

\bibitem{SPTpaper}
\textsc{Upadhyayula, S.} and \textsc{Gupta, S.K.} (2007) Spanning tree based
  algorithms for low latency and energy efficient data aggregation enhanced
  convergecast (dac) in wireless sensor networks. \emph{Ad Hoc Networks}
  \textbf{5}(5): 626--648.

\bibitem{karl2007protocols}
\textsc{Karl, H.} and \textsc{Willig, A.} (2007) \emph{Protocols and
  architectures for wireless sensor networks} (John Wiley \& Sons).

\bibitem{TinyosTEP_packetleveltimesync}
\textsc{Maroti, M.} and \textsc{Sallai, J.} (2008) Packet-level time
  synchronization. \emph{TinyOS Core Working Group, Technical Report, May} .

\bibitem{Elapsed_timeonarrival}
\textsc{Kus{\`y}, B.} \emph{et~al.} (2006) Elapsed time on arrival: a simple
  and versatile primitive for canonical time synchronization services.
  \emph{International Journal of Ad Hoc and Ubiquitous Computing}
  \textbf{2}(1): 239--251.

\bibitem{rootzone_maize}
\textsc{Gao, Y.} \emph{et~al.} (2010) Distribution of roots and root length
  density in a maize/soybean strip intercropping system. \emph{Agricultural
  water management} \textbf{98}(1): 199--212.

\bibitem{du2008water}
\textsc{Du, T.}, \textsc{Kang, S.}, \textsc{Zhang, J.}, \textsc{Li, F.} and
  \textsc{Yan, B.} (2008) Water use efficiency and fruit quality of table grape
  under alternate partial root-zone drip irrigation. \emph{Agricultural water
  management} \textbf{95}(6): 659--668.

\bibitem{capacitivework}
\textsc{Aravind, P.} \emph{et~al.} (2015) A wireless multi-sensor system for
  soil moisture measurement. In \emph{SENSORS, 2015 IEEE} (IEEE): 1--4.

\bibitem{MCP9700A}
Microchip (2016) \emph{Low-Power Linear Active Thermistor}.
  \urlprefix\url{http://ww1.microchip.com/downloads/en/DeviceDoc/20001942G.pdf}.

\bibitem{HIH5031}
Honeywell (2010) \emph{Low-Volatge Humidity sensors}.
  \urlprefix\url{https://sensing.honeywell.com/index.php?ci\_id=49692}.

\bibitem{barsukov2004challenges}
\textsc{Barsukov, Y.} (2004) Challenges and solutions in battery fuel gauging.
  \emph{Consultant to Texas Instruments, Inc} .

\bibitem{lithiumion_datasheet}
 (2006), Panasonic 18650 li-ion cell.
  \urlprefix\url{http://www.omnitron.cz/download/datasheet/NCR-18650PF.pdf}.

\bibitem{veihmeyer1931moisture}
\textsc{Veihmeyer, F.} and \textsc{Hendrickson, A.} (1931) The moisture
  equivalent as a measure of the field capacity of soils. \emph{Soil Science}
  \textbf{32}(3): 181--194.

\bibitem{sim2015assessment}
\textsc{Sim, R.}, \textsc{Plummer, S.} and \textsc{Fellahi, M.} (2015)
  Assessment of potential markets for soil moisture sensor in tanzania.
  \emph{D-Lab, March} .

\bibitem{veeramalluramulu}
\textsc{Rao, P.}, \textsc{Kumar, A.}, \textsc{Ramulu, V.} and \textsc{Devi, U.}
  (2013) \emph{Efficient Irrigation Technologies for Enhancing Water
  Productivity in Major crops of Andhra Pradesh} (Water Technology Centre,
  College of Agriculture - Hyderabad).

\end{thebibliography}

\end{document}